\documentclass[aps,superscriptaddress,preprint,a4]{revtex4-1}
\usepackage{amssymb}
\usepackage{amsmath}
\usepackage{bm}
\usepackage{color}
\usepackage{epsfig}
\usepackage{graphicx}
\usepackage{subeqnarray}

\begin{document}

\title{Efficient low temperature simulations for fermionic reservoirs with 
the hierarchical equations of motion method: Application 
to the Anderson impurity model}

\author{Xiaohan Dan}
\affiliation{Beijing National Laboratory for
Molecular Sciences, State Key Laboratory for 
Structural Chemistry of Unstable and Stable 
Species, Institute of Chemistry, Chinese Academy
of Sciences, Zhongguancun, Beijing 100190, China}
\affiliation{University of Chinese Academy of 
Sciences, Beijing 100049, China}

\author{Meng Xu}
\affiliation{Institute for Complex Quantum Systems and IQST,
Ulm University - Albert-Einstein-Allee 11, D-89069 Ulm, Germany}

\author{J. T. Stockburger}
\affiliation{Institute for Complex Quantum Systems and IQST,
Ulm University - Albert-Einstein-Allee 11, D-89069 Ulm, Germany}

\author{J. Ankerhold}
\affiliation{Institute for Complex Quantum Systems and IQST,
Ulm University - Albert-Einstein-Allee 11, D-89069 Ulm, Germany}

\author{Qiang Shi}\email{qshi@iccas.ac.cn}
\affiliation{Beijing National Laboratory for
Molecular Sciences, State Key Laboratory for 
Structural Chemistry of Unstable and Stable 
Species, Institute of Chemistry, Chinese Academy
of Sciences, Zhongguancun, Beijing 100190, China}
\affiliation{University of Chinese Academy of 
Sciences, Beijing 100049, China}

\begin{abstract}
The hierarchical equations of motion (HEOM) approach is an accurate
method to simulate open system quantum dynamics, which allows for 
systematic convergence to numerically exact results. To represent
effects of the bath, the reservoir correlation functions are usually 
decomposed into summation of multiple exponential terms in the HEOM
method. Since the reservoir correlation functions become highly 
non-Markovian at low temperatures or when the bath has complex 
band structures, a present challenge is to obtain accurate 
exponential decompositions that allow efficient simulation with 
the HEOM. In this work, we employ the barycentric representation 
to approximate the Fermi function and hybridization functions 
in the frequency domain. The new method, by approximating 
these functions with optimized rational decomposition, greatly 
reduces the number of basis functions in decomposing the 
reservoir correlation functions, which further allows the HEOM 
method to be applied to ultra-low temperature and general 
band structures. We demonstrate the
efficiency, accuracy, and long-time stability of the new
decomposition scheme by applying it to the Anderson impurity 
model (AIM) in the low temperature regime with the Lorentzian 
and tight-binding hybridization functions. 

\end{abstract}
\maketitle

\section{Introduction}
A paradigmatic setting in solid state physics is related to a 
quantum system coupled to a continuum of electronic states, 
where a typical example is the charge transport in single-molecule 
junctions or through quantum dots \cite{galperin08,zimbovskaya13,
ratner13, xiang16,thoss18}. Such systems usually consist of a 
molecule or quantum dot attached to two metal leads, forming an 
open quantum system where the molecule can exchange 
electrons and energy with the leads. Owing to the advances in 
experimental techniques \cite{chen99,park00,park02,xu03,qiu04}, 
it is now possible to measure a variety of transport properties. 
The experimental observations, including Coulomb blockade \cite{park02},  
spin blockade \cite{heersche06},  Kondo effect \cite{liang02,parks10},  
negative differential resistance \cite{osorio10,chen99,gaudioso00,xu15},  
switching \cite{blum05,choi06}, and hysteresis \cite{ballmann12}, 
have shown the potential of molecular junctions in the field of 
molecular electronics and stimulated the development of 
transport theory and simulation techniques.

Different methods are now available to solve the transport 
problem in molecular junctions and likewise for arrays of 
quantum dots. Approximate methods such as the 
quantum master equations (QMEs) \cite{mitra04,donarini06,
timm08,leijnse08,esposito09b,hartle11,levy19}, although being able to 
provide many useful physical insights, introduce significant 
approximations and are often limited to certain parameter regimes.
For example, in a previous study, we have shown that QMEs based 
on perturbation theory fail in the strong coupling regime \cite{dan22}.
Many numerically exact methods have also been developed, 
which include the numerical renormalization 
group (NRG) \cite{anders05,anders06, bulla08,nghiem18,de19}, 
density-matrix 
renormalization group (DMRG) \cite{schollwock05,
heidrich09,he19,kohn21}, and multilayer multi-configuration 
time-dependent Hartree (ML-MCTDH) in the second quantization 
representation (SQR) \cite{wang09,wang11,wang18}. These methods 
usually require discretization of the lead hybridization functions, 
such that special treatments are needed to avoid discretization 
artifacts at long simulation times \cite{nishimoto04,zitko09,wang18}.

Other approaches utilize the non-interacting nature of 
the leads and do not require discretization. These include
the path integral (PI) \cite{weiss08b,segal10,weiss13},  
continuous-time quantum Monte-Carlo (CT-QMC) \cite{muhlbacher08,
werner09,schiro09, gull11,antipov16, krivenko19}, 
Inchworm Monte Carlo \cite{cohen15,antipov17}, and the 
hierarchical equations of motion (HEOM) approach \cite{jin08,li12,
hartle13,hartle15b,shi18,erpenbeck19,zhang20,dan22}.

In this work, 
we focus on the HEOM method originally developed by Tanimura and 
Kubo \cite{tanimura89,tanimura06}. Jin \emph{et al} \cite{jin07,jin08} 
developed the HEOM method for the Fermionic bath problem, 
which has later been applied to the charge transport 
problem in molecular junctions \cite{zheng13,hartle13,
hartle15b,wang16,wang16c,schinabeck16,erpenbeck19}.
Despite this popularity, it is known that 
the applicability of HEOM is often limited to moderate or high 
temperature regimes \cite{hartle15b,zhang20}.
The reason is that, in the HEOM, the reservoir correlation 
function $C(t)$ that describes the fluctuation and dissipation 
effects of the environment, are represented using a finite 
set of basis functions to construct the hierarchical structures.
In the literature, the most widely used scheme in the moderate to 
high temperature regime is based on exponential decomposition of 
the reservoir correlation function with the aid of Matsubara 
expansion of the Bose/Fermi function \cite{tanimura90,tanimura06,
jin08}. At low temperatures though, the long memory time of $C(t)$ 
leads to rapid growth in the number of basis functions, 
rendering HEOM simulations very expensive and in many cases not feasible.

To resolve this problem, several methods have been developed 
based on more efficient decomposition schemes
of $C(t)$, which include the Pad\'{e} spectrum 
decomposition (PSD) \cite{ozaki07,hu10,hu11}, 
logarithmic discretization scheme \cite{ye17},
orthogonal functions decomposition based on Chebyshev 
polynomials \cite{tian12,popescu15,nakamura18,rahman19},  
Hermite polynomials \cite{tang15}, Bessel functions \cite{nakamura18,
ikeda20}, or product of oscillatory and exponential 
functions \cite{ikeda20}, re-summation over poles 
(RSHQME) \cite{erpenbeck18}, and more recently, the Fano spectrum 
decomposition (FSD) \cite{cui19,zhang20}, and the Prony fitting 
decomposition (PFD) \cite{chen22} schemes.
These important developments have now allowed the HEOM
to reach the experimentally very low temperature 
regimes \cite{li17,ye17,han18,zhang20,chen22}, 
and to deal with more complex band structures of 
the lead \cite{zheng09b,xie13b,hou14,erpenbeck18,chen22}. 
However, many of these new methods still have severe 
limitations. For example, the RSHQME approach becomes 
increasingly expensive with the simulation time \cite{erpenbeck18}, 
the FSD method may suffer from asymptotic instability problem in 
certain parameter regimes \cite{zhang20}.  
The time domain decomposition approaches
also require a large number of basis functions when 
increasing the simulation time \cite{tian12,tang15,duan17}.

Recently, the free-pole HEOM (FP-HEOM) method was proposed 
by our groups \cite{xu22} based on the barycentric 
representation \cite{nakatsukasa18} of the spectrum of 
the reservoir correlation function $C(t)$ as a very efficient 
tool to cure the above deficiencies. It shows high accuracy 
and computational efficiency for a broad class of bosonic 
reservoirs including those with sub-ohmic and bandgap 
spectral densities. Further, it applies to the complete 
temperature range down to zero 
temperature. In this work, we extend this framework to 
the charge transport problem in molecular junctions or quantum dots, 
where the metal leads are described by non-interacting 
Fermionic baths. We name this scheme the barycentric 
spectrum decomposition (BSD) method henceforth.
The BSD method is easy to implement with available packages 
and the accuracy of the decomposition can be controlled 
by a single predefined parameter \cite{nakatsukasa18}.
To demonstrate the performance of the BSD scheme for fermionic baths, we apply it here
to simulate charge transport in the Anderson 
impurity model (AIM), one of the benchmark models in solid state physics.

We first use the BSD scheme to decompose the Fermi distribution
function. Compared with traditional methods such as PSD, 
the BSD scheme is more accurate in the full fitting 
range, with errors not exceeding a predefined precision 
criteria. Moreover, the BSD scheme is superior in the low 
temperature regime. It is found that, to achieve a given 
precision, the number of BSD basis functions 
increases only linearly when the temperature drops 
exponentially, compared to the exponential increase of 
the number of PSD basis functions.

To demonstrate the capability to perform simulations at low 
temperatures and the numerical stability of the new method, 
the voltage-driven dynamics and the Kondo resonance in the AIM 
with the Lorentzian hybridization function are simulated.
The observed long time hysteresis behavior at low temperature 
shows the numerical stability of the BSD scheme.
The Kondo resonance at low temperatures is investigated by 
calculating the retarded Green's function directly using the 
HEOM, and the results show that the efficiency and accuracy are 
at least comparable with the most recent PFD method \cite{chen22}.

Previous frequency domain decomposition schemes usually depends 
on writing the hybridization functions into 
forms where the poles can be obtained 
analytically \cite{hu11,liu14,wang16c,zhang20}.
The BSD scheme also has the advantage that it does not 
rely on these analytical forms, and is capable to treat
general forms of spectral density or hybridization functions.
In the last example showing its applicability to 
arbitrary band structures, the charge transport 
dynamics of the AIM with a tight-binding 
hybridization function are studied. 
It is shown that the BSD-based HEOM can produce 
reliable zero temperature dynamics, in agreement with 
previously results from the ML-MCTDH-SQR method \cite{wang13b}. 
Moreover, quantum dynamics at different temperatures can also be explored
using the BSD scheme.

The outline of this paper is as follows. 
In Sec.~\ref{sec:HEOM-AIM}, we introduce the model system 
and the HEOM method. In Sec.~\ref{sec:PhyQuant}, 
we present how to use the HEOM to calculate the transport 
current and the impurity spectral function. 
In Sec.~\ref{sec:BSD}, we present details of the BSD 
scheme applied to the reservoir correlation function for 
the Fermionic bath. Numerical results are presented in 
Sec.~\ref{sec:Numerical}, where we analyze the efficiency of 
the BSD scheme in decomposing the Fermi distribution function, 
and show that the HEOM method combined with the BSD scheme 
can be efficiently applied to simulate charge transport in 
the AIM at very low temperature, and with general band 
structures. Conclusions and discussions are made in
Sec.~\ref{sec:conclusion}.

%%%%%%%%%%%%%%%%%
\section{theory}
\label{sec:theory}
\subsection{Model Hamiltonian and the HEOM method}
\label{sec:HEOM-AIM}
Within the open quantum system framework, which consists of a 
molecule (referred to as the ``system") attached to two metal 
leads (representing the ``bath"), the total Hamiltonian of 
the AIM can be written as:
\begin{equation}\label{Eq:totimpH}
\hat H_T = \hat H_{S} + \hat H_{B} + \hat H_{I} \;\;.
\end{equation}
where $\hat H_{S}$, $\hat H_{B}$, and $\hat H_{I}$ 
correspond to the Hamiltonian of the system, the leads, and 
the coupling between them, respectively. Their explicit forms 
are given by: 
\begin{subequations}
\begin{equation}
\hat H_{S} = \sum\limits_{\alpha}\varepsilon_{\alpha} 
\hat a_{\alpha}^{\dag}
\hat a_{\alpha} + U \hat a_{\uparrow}^{\dag} \hat a_{\uparrow}
\hat a_{\downarrow}^{\dag}\hat a_{\downarrow} \;\;,
\end{equation}
\begin{equation}\label{Eq:impHB}
\hat H_{B} =  \sum\limits_{\alpha n,l}
\varepsilon_{\alpha n,l}
\hat c_{\alpha n,l}^{\dag}\hat c_{\alpha n,l} \;\;,
\end{equation}
\begin{equation}\label{Eq:impHI}
\hat{ H_{I} } = \sum\limits_{\alpha n, l} V_{\alpha n, l}
\hat c_{\alpha n,l}^{\dag} \hat a_{\alpha} +
{{h.c.}} \;\;.
\end{equation}
\end{subequations}
Here, $\alpha=\{\uparrow,\downarrow\}$ denotes the two spin states, 
which are degenerate when there is no external magnetic field.
$l=L,R$ represents the left or right lead.
$\hat c^{(\dag)}_{\alpha n,l}$ and $\hat a_{\alpha}^{(\dag)}$ 
are the annihilation (creation) operators of the lead and 
molecule electrons, with the energy $\varepsilon_{\alpha n,l}$ 
and $\varepsilon_{\alpha}$, respectively. $U>0$ represents 
the repulsive Coulomb interaction when both spin states on 
the molecule are occupied. $V_{\alpha n, l}$ is the coupling between 
the molecule and the $l$th lead. The molecule-lead coupling 
can be characterized by the hybridization function defined as:
\begin{equation}
\Gamma_{\alpha, l}(\varepsilon)=2\pi\sum\limits_{n}
|V_{\alpha n, l}|^{2} \delta(\varepsilon-\varepsilon_{\alpha n, l}) \;\;.
\end{equation}

The reservoir correlation functions are related to the corresponding 
hybridization function through \cite{jin07,jin08,hartle15b}:
\begin{equation}\label{Eq:bathFDT}
C^\sigma_{\alpha,l}(t)=\int_{-\infty}^{+\infty}
\frac{d\varepsilon}{2\pi\hbar} 
{e^{\sigma i\varepsilon t/\hbar }\Gamma_{\alpha,l}(\varepsilon)}
f_l^\sigma(\varepsilon)\;\;,
\end{equation}
where $\sigma=\pm$, and $f_l^\sigma(\varepsilon) = [1
+e^{\sigma \beta_{l}(\varepsilon-\mu_l)}]^{-1}$ 
is the Fermi function for the electrons/holes.
Qualitatively, at sufficiently elevated temperatures the correlation 
function decays exponentially while at very low temperatures algebraic 
decay gives rise to long time tails. The latter depends on the details 
of the hybridization and Fermi function in close proximity to the 
Fermi level, i.e. low-frequency properties. It is this behavior 
that renders any numerical simulation to be highly non-trivial for 
low temperature Fermionic reservoirs. 

We may also encounter situations 
where a time-dependent bias voltage is applied to the leads. 
In such cases, for the time-dependent voltage $V_l(t)$ 
applied on lead $l$, the non-stationary reservoir 
correlation functions $\tilde{C}^{\sigma}_{\alpha,l}(t,\tau)$ 
are described by \cite{jin08,zheng13,zhang20}:
\begin{equation}
\tilde{C}^{\sigma}_{\alpha,l}(t,\tau)={\exp}  
\Bigg[ -i\frac{e}{\hbar}\sigma  \int_\tau^t dt' V_l(t') \Bigg] 
C^{\sigma}_{\alpha,l}(t-\tau) \;\;.
\end{equation}

In the HEOM, the reservoir correlation function 
$C^\sigma_{\alpha,l}(t)$ is represented by using a finite 
set of basis functions \cite{erpenbeck18,cui19,chen22}. 
Usually, when applying a sum-over-poles expansion of 
$\Gamma_{\alpha,l}(\varepsilon) f_l^\sigma(\varepsilon)$, 
the basis is a set of exponential functions
\begin{equation}\label{Eq:SOP_Bcorr}
C^\sigma_{\alpha,l}(t) \simeq \sum_{k=1}^K 
d^\sigma_{\alpha lk}e^{-\nu^\sigma_{\alpha lk}t} \;\;. 
\end{equation}

The HEOM for the Fermionic bath is then given 
by \cite{jin08,hartle13,schinabeck16}:
\begin{equation}
\label{Eq:drho1}
\frac{\partial }{\partial t}{\hat{\rho}}_{{\bf{J}}}(t)
= -\Big[\frac{i}{\hbar}\mathcal{L}_S +\gamma_{\bf{J}}(t)
\Big]{\hat{\rho}}_{{\bf{J}}}
-\frac{i}{\hbar}\sum\limits_{m=1}^{\Omega} (-1)^{\Omega-m} 
\mathcal{C}_{j_m}\hat{\rho}_{{{\bf J}_{m}^{-}}}
- i\sum\limits_{j_{{\Omega+1}}} 
\mathcal{A}_{\bar{j}_{{\Omega+1}}} 
{\hat{\rho}}_{_{{\bf J}^{+}}} \;\;,
\end{equation}
where the subscript ${\bf J}$ denotes $\{j_{1}j_{2},...,j_{\Omega}\}$, 
${{{\bf J}^{+}}}$ denotes $\{j_{1}j_{2},...,j_{\Omega},j_{\Omega+1}\}$, 
and ${{{\bf J}_{m}^{-}}}$ denotes 
$\{j_{1},...,j_{m-1},j_{m+1},...,j_{\Omega}\}$, 
where the combined-label index $j=(\alpha, l, k, \sigma )$, 
$\bar{j}=(\alpha, l, k, \bar{\sigma})$, $\bar{\sigma}=-\sigma$.
The reduced density operator (RDO) is $\hat{\rho}_{\bf 0}$, 
with ${\bf 0}$ means that $\bf J$ does not contain any terms.
When ${\bf J}\neq{\bf 0}$, $\hat{\rho}_{\bf{J}}$s are the 
auxiliary density operators (ADOs) that contain the system-bath 
correlations. $\Omega$ is the tier of $\hat{\rho}_{\bf{J}}$, 
which is the total number of terms in ${\bf J}$. 
$\mathcal{L}_S\hat{\rho}_{\bf{J}}=[\hat{H}_S,\hat{\rho}_{\bf{J}}]$. 
$\gamma_{\bf{J}}(t)$, the only term that contains the 
effect of time-dependent bias voltage, 
is $\gamma_{\bf{J}}(t) = \sum\limits_{( \alpha, l, k, \sigma)
\in {\bf{J}}}[\nu^\sigma_{\alpha lk} + ie\sigma V_l(t)/\hbar  ]$.
The operators $\mathcal{C}_j$ and 
$\mathcal{A}_j$ 
couples the $\Omega th$ tier to the $(\Omega-1)th$ 
and the $(\Omega+1)th$ tiers, respectively. 
Their actions on the RDO/ADOs are given by:
\begin{subequations}\label{Eq:ACoperator}
\begin{equation}
\mathcal{C}_j\hat{\rho}_{\bf{J}} = 
d_{j}\hat{a}_j \hat{\rho}_{\bf{J}} - 
(-1)^{\Omega} d^*_{\bar{j}}\hat{\rho}_{\bf{J}}\hat{a}_j \;\;, 
\end{equation}
\begin{equation}
\mathcal{A}_j\hat{\rho}_{\bf{J}} = 
\hat{a}_j \hat{\rho}_{\bf{J}} + 
(-1)^{\Omega} \hat{\rho}_{\bf{J}} \hat{a}_j \;\;.
\end{equation}
\end{subequations}
For $j = ( \alpha, l, k, \sigma)$, 
$d_{j}=d^\sigma_{\alpha lk}$, 
$\hat a_j = \hat a^\sigma_\alpha$ 
with $\hat a^{-(+)}_\alpha\equiv \hat a^{(\dag)}_\alpha$,
representing the system annihilation (creation) operators.
More details of the above HEOM for Fermionic bath 
can be referred to Refs.~\cite{jin08,hartle13,xu17}.

In practical calculations, except for the number of 
basis functions $K$ in Eq. (\ref{Eq:SOP_Bcorr}), the number of 
tiers $\Omega$ also needs to be truncated in conventional HEOM.
We denote the terminal tier level as $N_{\rm{trun}}$, 
i.e., we set $\hat{\rho}_{\bf{J}}=0$ for all ADOs 
with $\Omega>N_{\rm{trun}}$. The computational cost 
of the HEOM in Eq. (\ref{Eq:drho1}) increases rapidly as 
$N_{\rm{trun}}$ increases, especially when $K$ is large.
To solve this problem, we employ two different approaches:
the first one is based on the on-the-fly filtering 
method \cite{shi09b,hartle15b}, and the second one is 
the matrix product state (MPS) method to propagate
the HEOM \cite{shi18,ke22a}. By using the MPS method, the 
computational and memory costs increase 
linearly as the number of basis functions $K$ increases. 
Moreover, the precision control parameter 
in the MPS method is the bond dimension, rather than truncation 
tier $N_{\rm{trun}}$. For more details on the MPS-HEOM, 
we refer to the earlier work in Ref.~\cite{shi18}. 
In our later numerical simulations, the HEOM is 
propagated by using the on-the-fly filtering 
method unless the MPS method is explicitly mentioned.

\subsection{Calculation of the physical quantities}
\label{sec:PhyQuant}
To study the transport property of the AIM, we need to calculate 
the transport current which is defined as
$I_{l}(t)=-e\frac{d \hat{N}_l}{dt}$, with $\hat{N}_l$ being 
the electron number operator on lead $l$ and $e$ being the 
elementary charge.  In the HEOM, $I_{l}(t)$ can be calculated 
from the first-tier ADOs $\hat{\rho}_{\alpha k,l}^{\sigma}
(t)$ as \cite{jin08}:
\begin{equation}\label{Eq:expressI}
I_{l}(t) = ie \sum_{\alpha k}{\rm Tr}_S\{ 
 \hat{\rho}_{\alpha k,l}^{+}(t)\hat{a}_\alpha 
-\hat{a}_\alpha^\dag \hat{\rho}_{\alpha k,l}^{-}(t)\}, 
\quad l=L,R \;\;,
\end{equation}
where ${\rm Tr}_S$ denotes the trace over system DOFs.

When the electron-electron interaction is strong in AIM, 
Kondo resonance arises at low-temperature,
which can be quantitatively described by the Kondo peak 
of the spectral function near the Fermi energy \cite{li12,hewson97}.
Thus to verify the efficiency of BSD at low temperatures, 
we are interested in calculating
the spectral function $A_\alpha (\omega)$ given by
the Fourier transform of the retarded Green's 
function $G^R_\alpha (t)$ \cite{mahan00,kohn21}:
\begin{equation}\label{Eq:Aome}
A_\alpha (\omega) = -\frac{1}{\pi} {\rm Im}
\int_{-\infty}^{\infty} 
dt e^{i\omega t} G^R_\alpha (t)\;\; ,
\end{equation}
\begin{equation}\label{Eq:Greenfunc}
G^R_\alpha (t) = -i \Theta (t) \langle \{ \hat{a}_\alpha(t),
\hat{a}^\dag_\alpha \} \rangle \;\; ,
\end{equation}
where $\langle \hat{O} \rangle = {\rm Tr}
[\hat{\rho}^{eq} \hat{O}]$ 
with $\hat{\rho}^{eq}=e^{-\beta\hat{H}_T}/{\rm Tr}
[e^{-\beta\hat{H}_T}]$ 
describing the thermal equilibrium of the total system,
$\{\cdot,\cdot\}$ is the anti-commutator.

When using the HEOM to calculate $G^R_\alpha (t)$, 
the equations of motion are slightly different 
due to the anti-commuting property of the Fermion 
operators. As an example, we first consider 
the correlation function 
$\langle \hat{a}_\alpha(t) \hat{a}^\dag_\alpha \rangle$:
\begin{eqnarray}
\label{Eq:corr-c}
\langle \hat{a}_\alpha(t) \hat{a}^\dag_\alpha \rangle &= &
{\rm Tr}[ e^{i\hat{H}_T t/\hbar} \hat{a}_\alpha 
e^{-i\hat{H}_T t/\hbar} 
\hat{a}^\dag_\alpha \hat{\rho}^{eq}]  \nonumber \\ 
& = & {\rm Tr}_S\{\hat{a}_\alpha {\rm Tr}_B
[e^{-i\hat{H}_T t/\hbar} \hat{a}^\dag_\alpha 
\hat{\rho}^{eq} e^{i\hat{H}_T t/\hbar}]\} \;\;.
\end{eqnarray}
In this work, the equilibrium density operator $\hat{\rho}^{eq}$ 
of the total system is represented by all the RDO/ADOs after 
propagating the initial state till equilibrium. 
We note that there are also other methods to obtain the 
equilibrium state, via the imaginary time 
propagation \cite{tanimura14,song15b,xing22,ke22b}, 
or iterative solver for linear equations \cite{ye16,kaspar21}.
After obtaining the equilibrium RDO/ADOs, 
$\hat{a}^\dag_\alpha \hat{\rho}^{eq}$ is obtained by 
multiplying $\hat{a}^\dag_\alpha$ on the left 
to all equilibrium RDO/ADOs. The resulted new RDO/ADOs 
are then propagated to time $t$.

Due to the anti-commutation relation between 
$\hat{a}^\dag_\alpha$ 
and the RDO/ADOs, the resulting HEOM for the evolution of 
$\hat{a}^\dag_\alpha \hat{\rho}^{eq}$ have the same form as 
Eq.~(\ref{Eq:drho1}), but with the operators $\mathcal{C}_j$ and 
$\mathcal{A}_j$ replaced by $\mathcal{C}^L_j$ and 
$\mathcal{A}^L_j$:
\begin{subequations}
\label{Eq:ACoperator-l}
\begin{equation}
\mathcal{C}^L_j\hat{\rho}^L_{\bf{J}} = 
(-1)d_{j}\hat{a}_j \hat{\rho}^L_{\bf{J}} - 
(-1)^{\Omega} d^*_{\bar{j}}\hat{\rho}^L_{\bf{J}}\hat{a}_j \;\; , \\
\end{equation}
\begin{equation}
\mathcal{A}^L_j\hat{\rho}^L_{\bf{J}} = 
(-1)\hat{a}_j \hat{\rho}^L_{\bf{J}} + 
(-1)^{\Omega} \hat{\rho}^L_{\bf{J}} \hat{a}_j \;\;  .
\end{equation}
\end{subequations}
The superscript $L$ in the above Eq. (\ref{Eq:ACoperator-l}) 
indicates that the original equilibrium RDO/ADOs 
$\hat{\rho}_{\bf{J}}$ is multiplied by 
$\hat{a}^\dag_\alpha$ on the left (i.e., $\hat{\rho}^L_{\bf{J}} 
\equiv \hat{a}^\dag_\alpha \hat{\rho}_{\bf{J}}$). 
Note that the additional minus sign when $\hat{a}_j$ 
acts on the left of $\hat{\rho}^L_{\bf{J}}$, compared 
to Eq.\ (\ref{Eq:ACoperator}). 
It comes from constructing the equation 
of motion of $\hat{\rho}^L_{\bf{J}}$. After taking the derivative 
of $\hat{\rho}^L_{\bf{J}}$, 
we obtain terms like 
$\hat{a}^\dag_\alpha \mathcal{C}_j \hat{\rho}_{\bf{J}}$ 
and $\hat{a}^\dag_\alpha \mathcal{A}_j \hat{\rho}_{\bf{J}}$.
To obtain the final equation of motion, 
$\hat{a}_j$ should be switched to the 
left or right side of $\hat{\rho}^L_{\bf{J}}$, 
such that the left acting term (i.e. $\hat{a}_j 
\hat{\rho}^L_{\bf{J}}$) in Eq.~(\ref{Eq:ACoperator-l}) 
has an additional minus sign due to the interchange 
of $\hat{a}_j$ with $\hat{a}^\dag_\alpha$ in $\hat{\rho}^L_{\bf{J}}$.
Further details can be referred to Refs. \cite{jin08,li12}.

After propagating $\hat{a}^\dag_\alpha \hat{\rho}^{eq}$ to time $t$ 
using the above equations, the $\langle \hat{a}_\alpha(t) 
\hat{a}^\dag_\alpha \rangle$ term in Eq.\ (\ref{Eq:corr-c}) can be 
obtained by multiplying $\hat{a}_\alpha$ with the RDO at time $t$
(i.e.\ ${\rm Tr}_B[e^{-i\hat{H}_T t/\hbar} \hat{a}^\dag_\alpha
\hat{\rho}^{eq} e^{i\hat{H}_T t/\hbar}]$), then taking the 
trace over the system DOFs.
For $\langle \hat{a}^\dag_\alpha \hat{a}_\alpha(t)\rangle$, 
one needs to calculate $e^{-i\hat{H}_T t/\hbar} 
\hat{\rho}^{eq}\hat{a}^\dag_\alpha e^{i\hat{H}_T t/\hbar}$. 
Similarly, $\hat{\rho}^{eq} \hat{a}^\dag_\alpha$ is realized by 
multiplying $\hat{a}^\dag_\alpha$ to the right of the
equilibrium RDO/ADOs. And the HEOM for the evolution of 
$\hat{\rho}^{eq} \hat{a}^\dag_\alpha$, 
following the same procedure as described above, is Eq.\ (\ref{Eq:drho1}) 
with $\mathcal{C}_j$ and $\mathcal{A}_j$ replaced by $\mathcal{C}^R_j$ and 
$\mathcal{A}^R_j$
\begin{subequations}
\label{Eq:ACoperator-r}
\begin{equation}
\mathcal{C}^R_j\hat{\rho}^R_{\bf{J}} = 
d_{j}\hat{a}_j \hat{\rho}^R_{\bf{J}} + 
(-1)^{\Omega} d^*_{\bar{j}}\hat{\rho}^R_{\bf{J}}\hat{a}_j \;\;,
\end{equation}
\begin{equation}
\mathcal{A}^R_j\hat{\rho}^R_{\bf{J}} = 
\hat{a}_j \hat{\rho}^R_{\bf{J}} - 
(-1)^{\Omega} \hat{\rho}^R_{\bf{J}} \hat{a}_j \;\;  .
\end{equation}
\end{subequations}
Note that the minus sign appears when $\hat{a}_j$ acts on the 
right of $\hat{\rho}^R_{\bf{J}}$ 
($\hat{\rho}^R_{\bf{J}} 
\equiv \hat{\rho}_{\bf{J}} \hat{a}^\dag_\alpha $), 
due to the same reason 
as in Eq. (\ref{Eq:ACoperator-l}).

%%%%%%%%%%%%%%%%%%%
\subsection{Barycentric spectrum decomposition}\label{sec:BSD}
The sum-over-poles expansion method used in this work 
is based on the availability of high-precision rational 
approximations of general functions $B(z)$ on the real axis. A 
particularly suitable representation for this purpose is the 
barycentric representation \cite{xu22,nakatsukasa18}:
\begin{equation}\label{Eq:BSD} 
\tilde{B}(z) = \sum_{j=1}^{m}\frac{W_jB(S_j)} 
{z-S_j} 
\Bigg{/}\sum_{j=1}^{m}\frac{W_j}{z-S_j} \;\;.
\end{equation}
Here, $m \ge 1$ is an integer that determines 
the order of the rational function.
The function to be approximated by $\tilde B(z)$ is 
$B(z)$, which takes the value $B(z_i)$ for $z_i \in Z$ with 
$Z \subseteq \bold{C}$ as the {\itshape{sample points set}}. 
$\{ S_1,S_2,..,S_m \}$ is a set 
of {\itshape{support points}}, chosen from the sample points set. 
$B(S_j)$ is the value of $B(z)$ at the 
support point $S_j$. The barycentric approximation 
$\tilde{B}(z)$, which uses a rational function to 
interpolate $\tilde{B}(S_j)=B(S_j)$ at the 
support point $S_j$, selects a support points 
set from the sample points set, and calculates the corresponding 
weight $W_j$ \cite{nakatsukasa18}.

In this work, we employ the adaptive Antoulas-Anderson (AAA)
algorithm to obtain the parameters $S_j$ and $W_j$ in 
Eq. (\ref{Eq:BSD}), which is described in detail in 
Ref.~\cite{nakatsukasa18}. The AAA algorithm uses a greedy 
strategy to select the support points, which can be obtained 
with a MATLAB code \cite{nakatsukasa18}, 
and also from the {\it baryrat} Python package \cite{hofreither21}.
After the barycentric representation is obtained, the pole 
structure of the function $\tilde{B}(z)$ can be obtained 
from the calculated support points and weights. Namely, 
$\tilde{B}(z) = \sum_{j=1}^{K}\frac{R_A(j)}{z-P_A(j)}$, 
with the poles $\{P_A(j)\}$ and associated residues $\{R_A(j)\}$.
The poles $\{P_A(j)\}$ and residues $\{R_A(j)\}$ can also be 
obtained directly from the output of the MATLAB or Python packages.
It can be shown that, $\tilde{B}(z)$ is a rational function 
of type $(m-1,m-1)$ \cite{nakatsukasa18}, so the total number 
of poles is $K = m-1$.

BSD for the Bosonic reservoir case was reported previously 
in Ref.~\cite{xu22}, where it is employed to approximate 
the spectrum of the harmonic bath correlation function. 
Significant improvements of the efficiency 
of the HEOM at low temperatures have been observed, and simulations 
can even be performed at zero temperature \cite{xu22}.
For the Fermionic bath considered in this work
using the HEOM in Eq.~(\ref{Eq:drho1}) \cite{jin08}, 
the paths associated with $C^\sigma_{\alpha,l}(t)$ 
and $C^{\bar{\sigma}*}_{\alpha,l}(t)$ are combined to 
define the ADOs by utilizing the relation 
$\nu^\sigma_{\alpha lk}=\nu^{\bar{\sigma}*}_{\alpha lk}$ 
where $\nu^\sigma_{\alpha lk}$ is defined in Eq. (\ref{Eq:SOP_Bcorr}). 
To maintain this conventional 
structure of the Fermionic HEOM in Eq. (\ref{Eq:drho1}), 
expansion of $C^\sigma_{\alpha,l}(t)$ 
and $C^{\bar{\sigma}}_{\alpha,l}(t)$ should be handled 
together to ensure that 
$\nu^\sigma_{\alpha lk}=\nu^{\bar{\sigma}*}_{\alpha lk}$. 

To this end, by utilizing the property 
$f_l^+(\varepsilon)=1-f_l^-(\varepsilon)$, 
we define the symmetric Fermi distribution function as 
\begin{equation}
\label{Eq:sym-fermidis}
f_l^s(\varepsilon) = \frac{1}{1+e^{\beta_{l}(\varepsilon-\mu_l)}}
- \frac{1}{2} \;\;.
\end{equation}
Now, $f_l^+(\varepsilon)=\frac{1}{2}+f_l^s(\varepsilon)$ 
and $f_l^-(\varepsilon)=\frac{1}{2}-f_l^s(\varepsilon)$,
and the reservoir correlation functions can be expressed as:
\begin{subequations}\label{Eq:ct-sym}
\begin{equation}
C^+_{\alpha,l}(t) = \int \frac{d\varepsilon}{2\pi\hbar}
e^{i\varepsilon t/\hbar} \frac{\Gamma_{\alpha,l}(\varepsilon)}{2} 
+ \int \frac{d\varepsilon}{2\pi\hbar}e^{i\varepsilon t/\hbar} 
\Gamma_{\alpha,l}(\varepsilon){f^s_l}(\varepsilon) \;\;,
\end{equation}
\begin{equation}
C^-_{\alpha,l}(t) = \int \frac{d\varepsilon}{2\pi\hbar}
e^{-i\varepsilon t/\hbar} \frac{\Gamma_{\alpha,l}(\varepsilon)}{2} 
- \int \frac{d\varepsilon}{2\pi\hbar}e^{-i\varepsilon t/\hbar} 
\Gamma_{\alpha,l}(\varepsilon){f^s_l}(\varepsilon) \;\;.
\end{equation}
\end{subequations}

From the above equations, we need the pole structures of 
$\Gamma_{\alpha,l}(\varepsilon)$ and 
$\Gamma_{\alpha,l}(\varepsilon){f^s_l}(\varepsilon)$ to 
do the sum-over-poles expansion.
For the pole structure of $\Gamma_{\alpha,l}(\varepsilon)
{f^s_l}(\varepsilon)$, 
it is possible to use the BSD expansion in Eq. (\ref{Eq:BSD}) 
to approximate $\Gamma_{\alpha,l}(\varepsilon)
{f^s_l}(\varepsilon)$ directly as in 
Ref. \cite{xu22}, or to approximate the two functions
$\Gamma_{\alpha,l}(\varepsilon)$ 
and ${f^s_l}(\varepsilon)$ separately.
In practice, the two different choices may result in 
slightly different pole structures for the same accuracy
control parameter. It should be noted that, as long as 
the correlation functions are faithfully reproduced, 
the HEOM will give the same result, although the 
numerical cost could be different depending on the 
pole structure. 
In this work, we choose to approximate the 
hybridization functions and the symmetric Fermi functions 
separately. An advantage of this separate decomposition 
scheme is that, it allows us to discuss the contribution 
to basis functions from the Fermi function, which is 
critical for the low-temperature performance of the 
HEOM (see the discussions later in Sec.~\ref{sec:fermi-BSD}).

To this end the pole structures of the hybridization and Fermi
functions are:
\begin{subequations}\label{Eq:jomefermi_pole}
\begin{equation}
 \tilde{\Gamma}_{\alpha,l}(\varepsilon)  =  \sum_{j=1}^{K_{0\Gamma}}
\frac{R_{\alpha, l}^\Gamma(j)}{\varepsilon-P_{\alpha, l}^\Gamma(j)} \;\;,
\end{equation}
\begin{equation}
 {\tilde{f}^s_l}(\varepsilon) = \sum_{j=1}^{K_{0f}} 
\frac{R_{\alpha, l}^f(j)}{\varepsilon-P_{\alpha, l}^f(j)} \;\;,
\end{equation}
\end{subequations}
where $\tilde{\Gamma}_{\alpha,l}(\varepsilon)$ and 
${\tilde{f}^s_l}(\varepsilon)$ denote 
the barycentric representation of $\Gamma_{\alpha,l}(\varepsilon)$ 
and ${f^s_l}(\varepsilon)$, respectively. $P_{\alpha, l}^{\Gamma/f}(j)$ 
and ${R_{\alpha, l}^{\Gamma/f}(j)}$ are the poles 
and residues, where $K_{0\Gamma}$ and $K_{0f}$ 
(we omit here the subscripts $\alpha$ and $l$ for simplicity, 
same for ${K_\Gamma}$ and ${K_f}$ later)
are the number of poles for each function.

It is noted that $\Gamma_{\alpha,l}(\varepsilon)$ and 
${f^s_l}(\varepsilon)$ are real, so except for those on the 
real axis, the poles and residues should consist of conjugate pairs.
The sum-over-poles expansion of the reservoir correlation functions
is then expressed as the combination of the pole structure 
of the two approximate functions:
\begin{equation}\label{Eq:corr_pole}
C^\sigma_{\alpha,l}(t) \simeq \sum_{j=1}^{K_\Gamma} 
d^{\Gamma \sigma}_{\alpha l j} 
e^{-\nu^{\Gamma \sigma}_{\alpha lj}t} 
+ \sum_{j=1}^{K_f} d^{f \sigma}_{\alpha l j} 
e^{-\nu^{f \sigma}_{\alpha lj}t} \;\;,
\end{equation}
where $\sigma$ denotes $+$ or $-$, ${K_\Gamma}$ and ${K_f}$ 
denote the number of basis functions that come from the poles of 
$\tilde{\Gamma}_{\alpha,l}(\varepsilon)$ and 
${\tilde{f}^s_l}(\varepsilon)$, respectively.
The frequency $\nu^{\Gamma/f \sigma}_{\alpha lj}$ and
coupling strength $d^{\Gamma/f \sigma}_{\alpha l j}$ are
\begin{subequations}\label{Eq:SOPfreq}
\begin{equation}
\nu^{\Gamma \sigma}_{\alpha lj} = 
-i\sigma P_{\alpha, l}^{\Gamma \sigma}(j),\quad 
d^{\Gamma \sigma}_{\alpha l j} = 
i R_{\alpha, l}^{\Gamma \sigma}(j)
[\sigma\frac{1}{2}+{\tilde{f}^s_l}
(P_{\alpha, l}^{\Gamma \sigma}(j))] \;\; ,
\end{equation}
\begin{equation}
\nu^{f \sigma}_{\alpha lj} = -i\sigma 
P_{\alpha, l}^{f \sigma}(j),\quad 
d^{f \sigma}_{\alpha l j} = i R_{\alpha, l}^{f \sigma}(j) 
\tilde{\Gamma}_{\alpha,l}(P_{\alpha, l}^{f \sigma}(j)) \;\; .
\end{equation}
\end{subequations}
Here, the superscript $\sigma$ in $P_{\alpha, l}^{\Gamma/f}(j)$ 
and ${R_{\alpha, l}^{\Gamma/f}(j)}$ distinguishes the poles or 
residues in the upper or lower half plane, with 
$\sigma = +$ for the upper half plane, and
$\sigma = -$ for the lower half plane. Thus,
all the constants in Eq. (\ref{Eq:corr_pole}) can be obtained 
using the barycentric decomposition in Eq. (\ref{Eq:BSD}), 
which is used later in the HEOM.

Depending on the choice 
of the sample points, sometimes there are poles on the real axis. 
They correspond to oscillating terms 
without decay ($e^{i\gamma t}$) in the correlation 
function $C(t)$, and are unphysical for the 
two types of hybridization functions considered 
later in Eq. (\ref{eq:Lorentz}) and Eq. (\ref{Eq:tightbjome}). 
Two types of spurious real poles are observed: 
poles away from the sample point domain $\mathcal{D}$ and 
those at the sharp edge of the original function.
The first type of real poles is due to incompletely chosen 
sample points and can be eliminated by
choosing a larger sample point set. The second type is related to 
the property of the original function. 
When the function (e.g. the $T$=0 fermi function) 
or its first derivative is discontinuous (e.g. near the band edge)
at a certain point, real poles may occur at the discontinuity point. 

There are also some tricks to eliminate the second type of 
real poles, such as by defining $f(x=a) = [f(x\to a^-) + 
f(x\to a^+)]/2$ for a discontinuous function at point $a$.
It is also noted that the residue of the second type of pole 
tends to zero when the resolution of the sample points near 
the discontinuity point becomes finer. 
So, throwing away the unphysical poles is another way to 
treat the above problem. In this case,
we have checked carefully that neglecting the spurious
poles would not affect the accuracy of the fitted function, 
and the final bath correlation functions $C(t)$.
As a consequence, ${K_\Gamma}\leq K_{0\Gamma}/{2}$ and 
${K_f}\leq K_{0f}/{2}$. 
Since $P_{\alpha, l}^{\Gamma/f+}(j)$ and 
$P_{\alpha, l}^{\Gamma/f-}(j)$ are complex conjugates, 
the relation $\nu^{\Gamma/f\sigma}_{\alpha lj}
=\nu^{{\Gamma/f}\bar{\sigma}*}_{\alpha lj}$ 
still holds in the BSD scheme.
Hence, the structure of traditional HEOM will not change when 
applying the BSD scheme for the reservoir correlation functions.

The sample points do not need to be equally spaced. 
This allows us to choose a point set that can focus on the 
regime where the function changes rapidly. 
In this work, logarithmic discretization \cite{bulla08} is 
used for Fermi distributions to ensure that there are enough 
points near the Fermi level. A simple example 
is $Z = \{z_n~|~z_n = \mu \pm D\Lambda^{-n}, n=0,...,N \}\cup\{\mu\}$, 
which is discretized in the domain $\mathcal{D}=[-D+\mu,D+\mu]$ 
and concentrated at $\mu$, with the minimum interval 
$\delta = D\Lambda^{-N}$ controlling the fineness of the 
discretization. For the hybridization functions, we choose 
uniform discretization. In both cases, the minimum interval 
$\delta$ should be chosen to achieve a good resolution 
at the Fermi level or the hard band edges (if any). 
As long as the domain $\mathcal{D}$ is 
suitable and discrete points are dense enough, 
slight changes in the sample point set almost do not affect the BSD result. 

In summary, application of the BSD scheme to obtain 
the sum-over-poles expansion of the reservoir correlation 
function $C^\sigma_{\alpha,l}(t)$
can be achieved through the following steps:

(1): Choose the sample points sets $Z_F$ and $Z_\Gamma$ to 
discretize the Fermi distribution and the hybridization functions,
respectively. The point set is in the domain $\mathcal{D}=[D_{min},
D_{max}]$ covering the main scope of the hybridization function.
For simplicity, we use the same domain for point 
sets $Z_F$ and $Z_\Gamma$. The sample points do not need 
to be equally spaced and should focus on the 
regime where the function changes rapidly.
The minimum interval between adjacent sample points 
is denoted as $\delta_F$ and $\delta_\Gamma$, 
which shows the fineness of the discretization.

(2): By using the sample point set $Z_\Gamma$, 
the value set $\{ \Gamma_{\alpha,l}(z), z\in Z_\Gamma \}$ as input,
and with a predefined precision control parameter $tol_A$ 
that measures the accuracy of the barycentric 
approximation and has been integrated into the AAA 
algorithm \cite{nakatsukasa18}, obtain the 
barycentric representation of $\tilde{\Gamma}_{\alpha,l}(\varepsilon)$. 
The same procedure is applied to obtain ${\tilde{f}^s_l}(\varepsilon)$. 

(3): Calculate the poles and residues of 
$\tilde{\Gamma}_{\alpha,l}(\varepsilon)$ 
and ${\tilde{f}^s_l}(\varepsilon)$. Then use them to 
construct $\nu^{\Gamma/f \sigma}_{\alpha lj}$ 
and $d^{\Gamma/f \sigma}_{\alpha l j}$ in Eqs. (\ref{Eq:corr_pole}) 
through (\ref{Eq:SOPfreq}), 
which are the $\nu_{j_m}$ and $d_{j_m}$ coefficients in the 
HEOM in Eq. (\ref{Eq:drho1}).

\section{Results}\label{sec:Numerical}

In this section, we apply the BSD method to simulate the 
charge transport of AIM at different temperatures with 
different band structures. To do this, we first use 
the AAA algorithm to approximate the Fermi function 
to show the low temperature performance of BSD. The real 
time dynamics and the Kondo resonance of AIM with the
Lorentzian hybridization functions in the low temperature 
regime are then explored. 
In the end, we demonstrate the performance of the BSD scheme 
for arbitrary band structures by considering AIM with the
tight-binding hybridization function.

\subsection{Efficient decomposition of the Fermi function 
at low temperature}\label{sec:fermi-BSD}
\begin{figure}
\centering
\includegraphics[width=14cm]{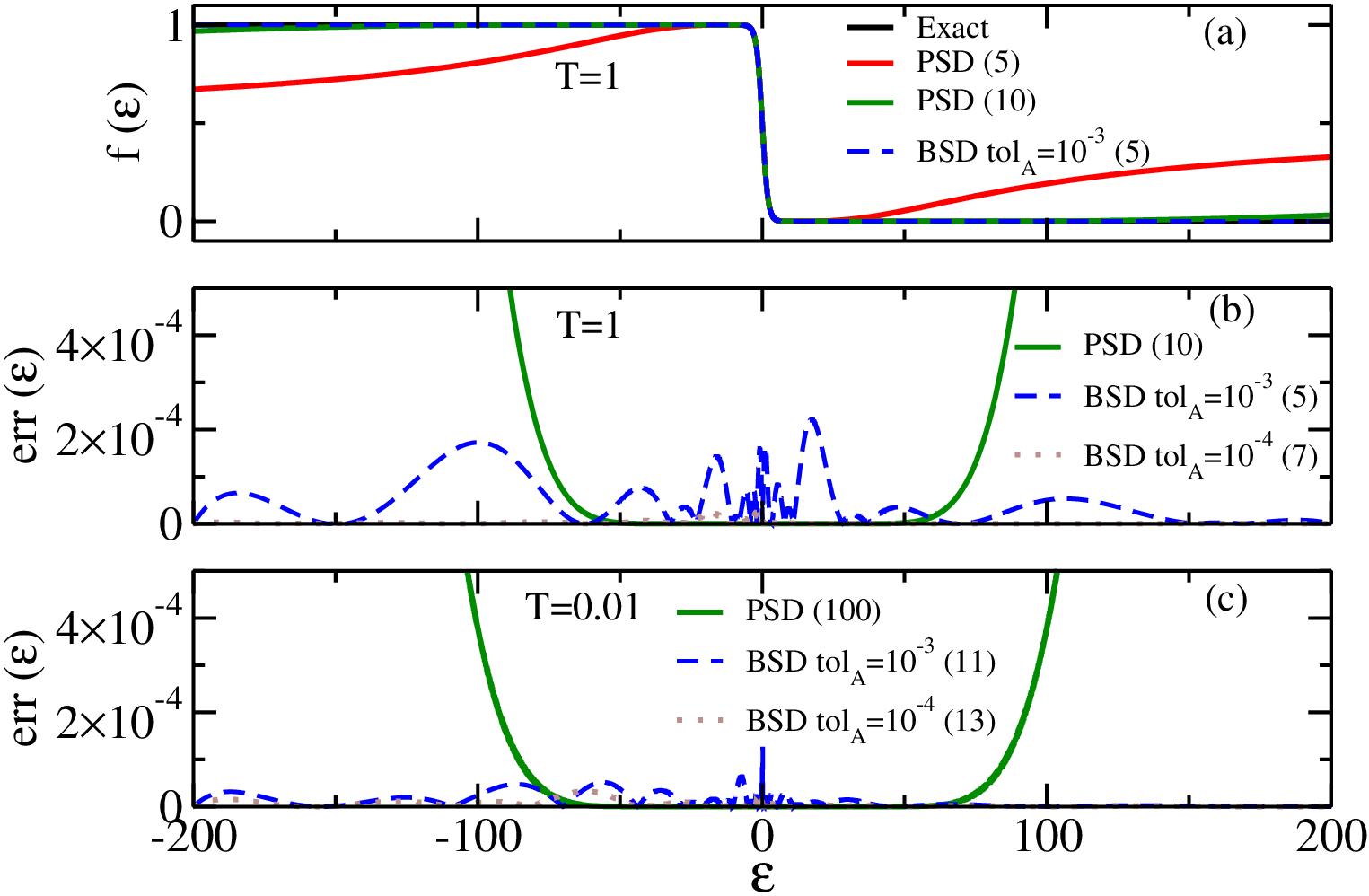}
\caption{Performance of the BSD scheme to approximate the 
Fermi function $f(\varepsilon) = 1/(1+e^{\beta\varepsilon})$ 
(here $k_B\equiv1,\hbar\equiv 1$), compared with the 
conventional PSD scheme. Panel (a) shows the approximate 
Fermi functions using different schemes at $T=1$. 
The number in the parentheses is the number of basis functions 
$K_f$. Panels (b) and (c) present the error defined 
as  $err(\varepsilon)=|f^{BSD/PSD}(\varepsilon)-f(\varepsilon)|$ 
for different schemes, at $T=1$ and $T=0.01$, respectively.
Here, in the BSD using the AAA algorithm, the sample points
are obtained in the domain $\mathcal{D}=[-200,200]$ with sufficiently 
dense discretization, the accuracy control parameter   
is given by $tol_A$.}
\label{Fig:fermi_padevsbary}
\end{figure}

Fig.~\ref{Fig:fermi_padevsbary} shows the performance of 
the BSD scheme to decompose the Fermi function 
$f(\varepsilon) = 1/(1+e^{\beta\varepsilon})$ 
(here $k_B\equiv1, \hbar\equiv 1$) at different temperatures, 
compared with the conventional PSD scheme.
In Fig.~\ref{Fig:fermi_padevsbary}(a) at $T=1$, the approximate 
function obtained 
from the PSD scheme shows high accuracy at small $\varepsilon$,
but it deviates significantly from the exact curve at 
large $\varepsilon$. As the number of PSD basis functions 
increases, the accuracy improves and the deviation from the 
exact curve occurs at much larger $\varepsilon$. Since 
there is usually a finite range for 
the hybridization function $\Gamma_{\alpha,l}(\varepsilon)$,
it is usually not a problem in decomposing the reservoir 
correlation functions, as the deviation of PSD approximation from 
$\Gamma_{\alpha,l}(\varepsilon)f_l^\sigma(\varepsilon)$ 
becomes smaller and smaller with larger numbers of basis 
function. The corresponding reservoir correlation function 
$C^\sigma_{\alpha l}(t)$ obtained from the PSD scheme then 
approaches the exact result, allowing systematic convergence 
of the HEOM. 

The approximate distribution from the BSD scheme is 
more accurate than the PSD with the same number of basis 
functions. For $tol_A=10^{-3}$, the deviation of the BSD curve 
from the exact curve is barely noticeable on the curves
shown in Fig.~\ref{Fig:fermi_padevsbary}(a). 
Fig.~\ref{Fig:fermi_padevsbary}(b) shows the error at $T=1$, 
the BSD errors oscillates at small $\varepsilon$, and 
the maximum error does not exceed $tol_A$. 
Fig.~\ref{Fig:fermi_padevsbary}(c) 
shows the error at low temperature $T=0.01$.
For the PSD scheme, the convergence slows down as 
the temperature decreases, and the PSD requires nearly 10 
times as many basis functions as the $T=1$ 
case to reach similar accuracy. This severely limits the 
application of the PSD-based HEOM at low 
temperatures. However, for the BSD scheme, the number 
of required basis functions increases only
slightly as the temperature decreases, with $K_f=5$ 
for $T=1$ and $K_f=11$ for $T=0.01$, when $tol_A=10^{-3}$. 
This behavior greatly reduces the number of basis 
functions required for low-temperature simulations, 
indicating the superiority of the BSD scheme at low temperatures.
In addition, both the $T=1$ and $T=0.01$ cases show 
that the number of required basis functions 
only increases slightly as the accuracy increases 
(by decreasing $tol_A$), thus allowing for systematic 
convergence to numerically exact results.

\begin{figure}
\centering
\includegraphics[width=14cm]{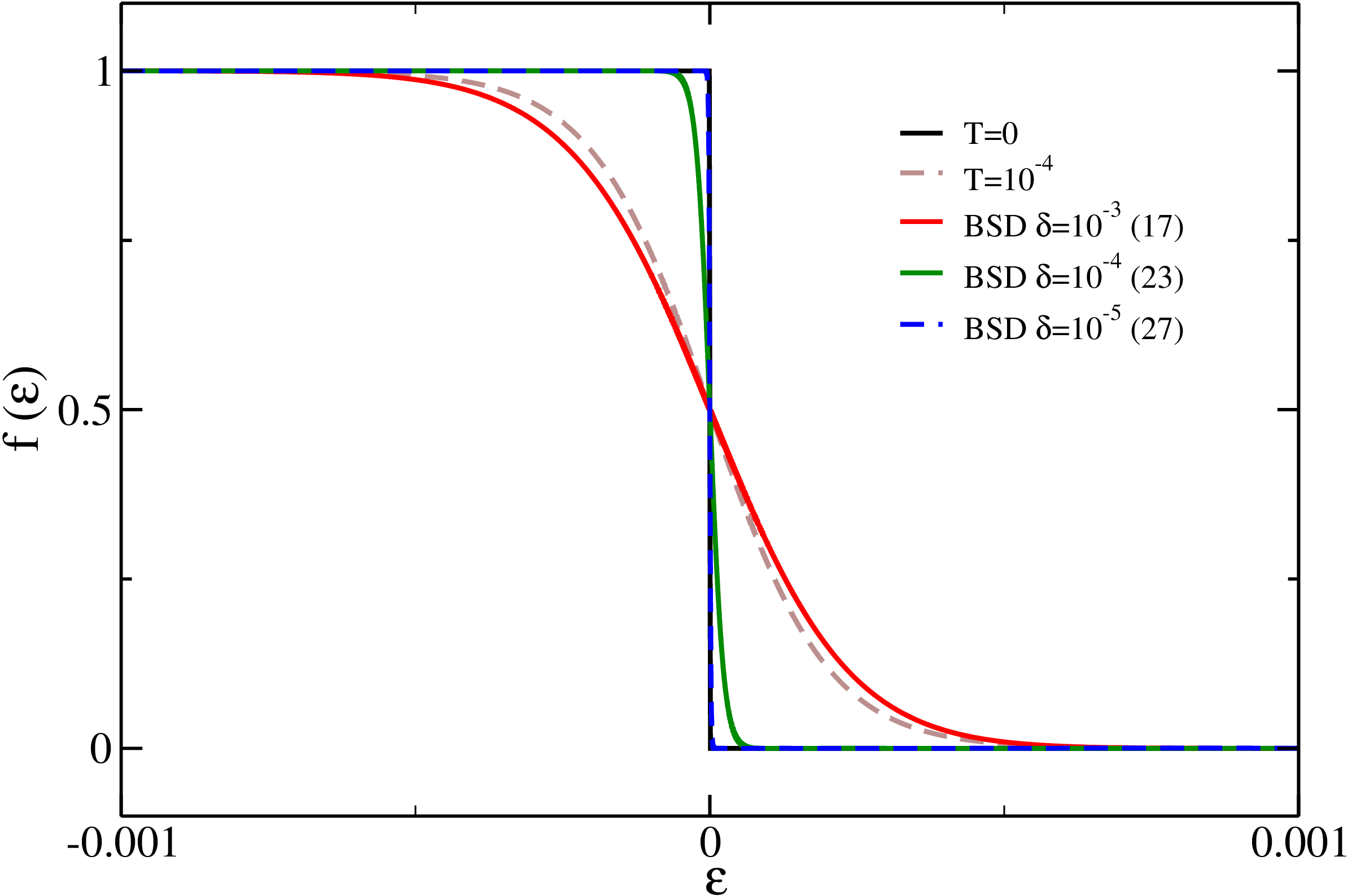}
\caption{
Performance of the BSD scheme to approximate 
the Fermi function $f(\varepsilon) = 1/(1+e^{\beta\varepsilon})$ 
(here, $k_B\equiv1,\hbar\equiv 1$) at $T=0$.
The BSD results are shown for different minimum 
discretization intervals $\delta_F$. 
The exact $T=0$ and $T=10^{-4}$ Fermi functions are also 
shown for comparison. 
The sample points are obtained by 
discreting the domain $\mathcal{D}=[-200,200]$, 
and the accuracy control parameters is $tol_A=10^{-3}$. 
The number of basis functions $K_f$ is given in the 
 parentheses.}
\label{Fig:fermi_T0-1}
\end{figure}

Fig. \ref{Fig:fermi_T0-1} shows the performance of the BSD 
scheme when setting $T=0$. In this case, the Fermi distribution 
becomes a step function.  We present the BSD results for 
different minimum discretization intervals $\delta_F$, 
where the discretization domain is $\mathcal{D}=[-200,200]$
with $tol_A=10^{-3}$. It can be seen that, the BSD results are 
affected by the interval between the points where the Fermi 
function jumps. And $K_f$ increases when decreasing $\delta_F$: 
$K_f=$17, 23, 27 for $\delta_F=10^{-3},10^{-4},10^{-5}$, 
respectively. We find that the BSD approximate results 
can only reach the step function asymptotically,
and correspond to a temperature approximately one-tenth of the 
discretization interval (see the comparison between 
the $\delta_F=10^{-3}$ curve and the Fermi function 
at $T=10^{-4}$). 

\begin{figure}
\centering
\includegraphics[width=14cm]{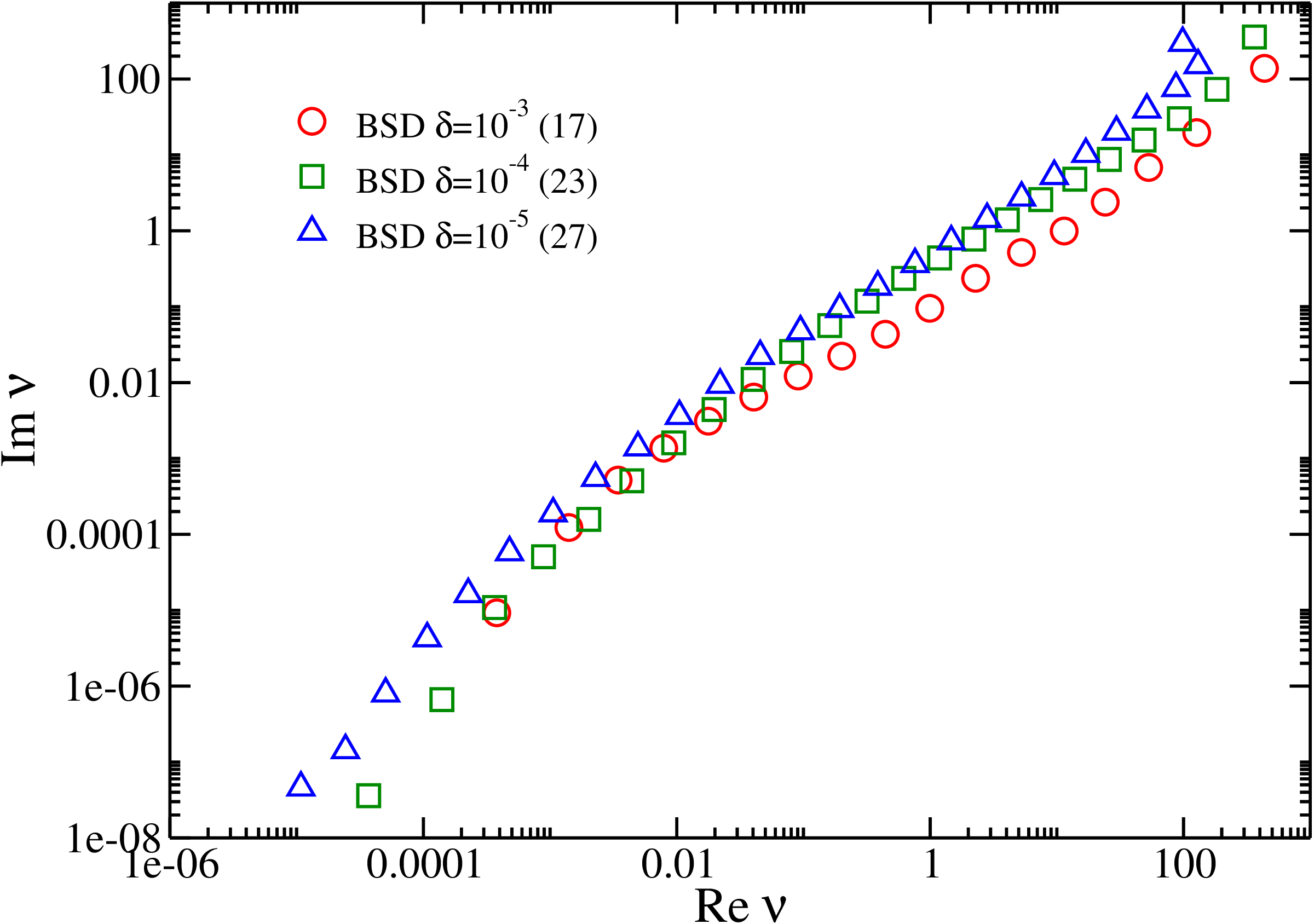}
\caption{
Distribution of real and imaginary parts of the poles in the upper 
half plane, i.e., $\nu^{f +}_{j}$ 
in Eq.\ (\ref{Eq:corr_pole}) (we omit the $\alpha$ or $l$ 
subscripts here), 
for the BSD results in Fig.~\ref{Fig:fermi_T0-1}.
}
\label{Fig:fermi_T0-pole}
\end{figure}

Fig.~\ref{Fig:fermi_T0-pole} shows distribution of the real and 
imaginary parts of the complex frequencies $\nu^{f +}_{\alpha l j}$
with different $\delta_F$. It can be seen that,
the largest frequency originates from the boundary of discretization, 
and the smallest frequency is on the same order of 
the minimum interval, which corresponds to an effective very small 
temperature. The frequencies are nearly uniformly distributed on the
logarithmic scale, confirming the moderate increase of $K_f$ 
when decreasing $\delta_F$.

\begin{figure}
\centering
\includegraphics[width=14cm]{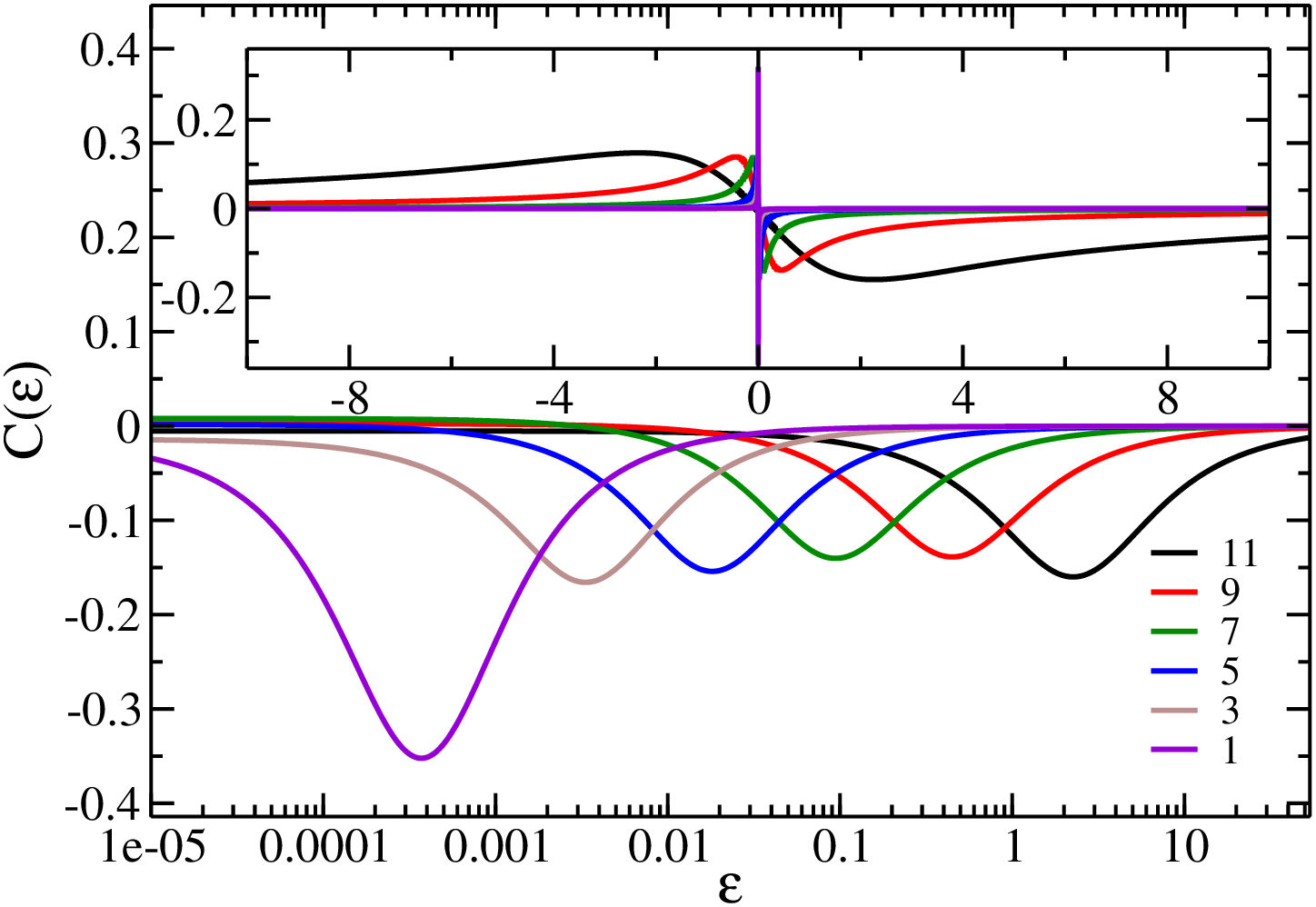}
\caption{
The spectrum of BSD basis functions 
$C_j(\varepsilon)$ 
($2{\rm Re}\frac{d^{f +}_{j}}{i\varepsilon + \nu^{f +}_{j}}$)
for $\delta_F=10^{-3}$ and $T=0$ as in Fig.~\ref{Fig:fermi_T0-1}.
To calculate the coupling strength $d^{f +}_{j}$ using 
Eq.~(\ref{Eq:SOPfreq}b), 
an auxiliary Lorentzian type hybridization function is chosen, 
with parameters $\eta=1$, $\gamma=10$, and $\varepsilon^0=0$ 
(the $\alpha$ and $l$ scripts are omitted).
The basis functions are sorted in ascending  
order of ${\rm Re}\nu^{f +}_{j}$, from $j=1$ to $j=17$, 
the number in the legend indicates the $j$th basis functions.
For simplicity, only the spectra for six low-frequency basis 
functions are shown. The inset shows the spectra in linear scale 
for the $\varepsilon$-axis.}
\label{Fig:Come}
\end{figure}

We further characterize this logarithmic distributed BSD pole structure 
by plotting $C_j(\varepsilon)$ in Fig.~\ref{Fig:Come}, 
where $C_j(\varepsilon) = 2{\rm Re}\frac{d^{f +}_{j}}{i\varepsilon 
+ \nu^{f +}_{j}}$ is the spectrum of different BSD basis functions
[note that $\sum_j C_j(\varepsilon) = \Gamma{(\varepsilon)}
f^+(\varepsilon)$ with scripts $\alpha$ and $l$ omitted].  
By assuming that $\nu^{f +}_{j}$ are arranged in ascending order, 
we show the results for the 1st, 3rd, 5th, 7th, 9th, and 11th poles, 
for $\delta_F=10^{-3}$ and $T=0$.
The results show almost linearly distributed frequencies on 
the logarithmic scale. Compared with the nearly $1/T$ growth for 
Matsubara [i.e. $(2n+1)\pi/\beta$] or Pad\'{e} basis functions, 
the number of basis functions from the BSD increases almost linearly 
as the temperature decreases exponentially. This makes the BSD 
scheme advantageous when performing very low temperature simulations.
This behavior is also similar to the logarithmic discretization 
in the NRG framework \cite{bulla08}, where $C_j(\varepsilon)$s 
replaces the discrete states in the NRG. However, since each 
$C_j(\varepsilon)$ includes a distribution of frequencies, 
the long-time dynamics of HEOM do not suffer from the discretization 
effect in the NRG approach \cite{anders05,zitko09}.

\begin{figure}
\centering
\includegraphics[width=14cm]{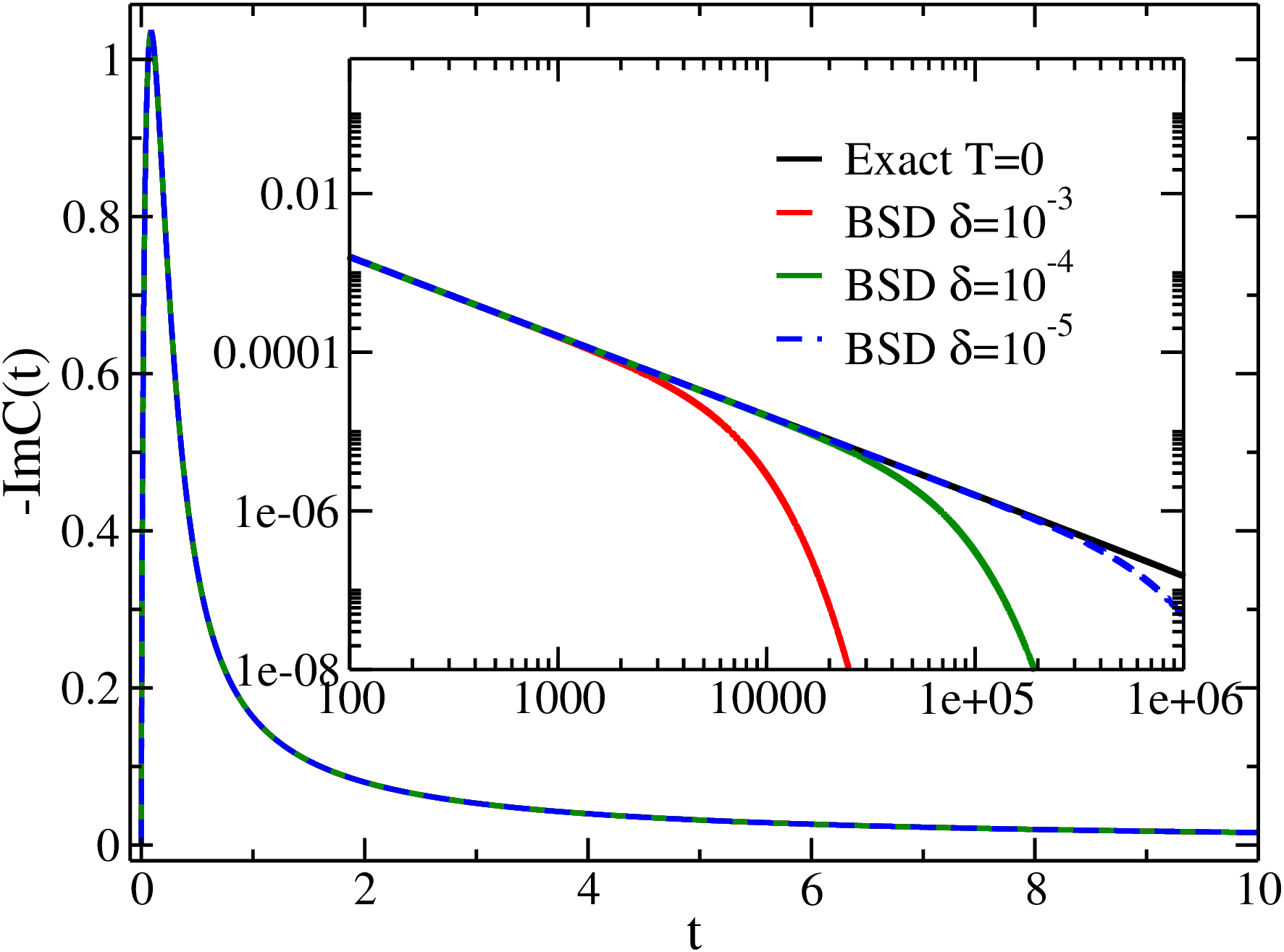}
\caption{
The reservoir correlation function $C(t)$ 
for the approximate BSD Fermi functions in Fig. \ref{Fig:fermi_T0-1}.
An auxiliary Lorentzian type hybridization function is chosen, 
with parameters $\eta=1$, $\gamma=10$, and $\varepsilon^0=0$ 
(the $\alpha$ and $l$ scripts are omitted).
The other parameters are the same as in Fig. \ref{Fig:fermi_T0-1}.
For simplicity only $-{\rm Im} C^+(t)$ is shown.}
\label{Fig:imCorr_T0}
\end{figure}

Though the analytic form of the Fermi function at $T=0$ cannot 
be obtained directly from the BSD, the approximate distribution 
function can approach the $T=0$ results as we decrease the 
discretization interval and introduce more low-frequency terms. 
It is found that the weights [$d^{f \sigma}_{\alpha l j}$ in 
Eq.\ (\ref{Eq:corr_pole})] 
associated with low-frequencies are also very small, so the 
low-frequency term will contribute to the dynamics only 
when $\nu^{f +}_{\alpha l j} t \gg 1$ for all the high-frequency 
terms $\nu^{f +}_{\alpha l j}$. This is illustrated in 
Fig. \ref{Fig:imCorr_T0}, 
where we show the corresponding $C(t)$ for the approximate BSD Fermi 
functions in Fig. \ref{Fig:fermi_T0-1}. To draw the reservoir 
correlation function $C(t)$, a hybridization function 
with Lorentzian type [defined later in Eq. (\ref{eq:Lorentz})] 
is chosen with $\eta=1$, $\gamma=10$, $\varepsilon^0=0$ (here 
we omit the lead label $l$). For this hybridization function 
and with the Fermi energy $\mu=0$, the real part of $C(t)$ shows
single exponential decay and only the imaginary part is affected 
by the temperature \cite{chen22}.
For simplicity, we only show $-{\rm Im}~ C^+(t)$ in 
Fig. \ref{Fig:imCorr_T0}.
It can be seen from Fig. \ref{Fig:imCorr_T0} that, 
in the short time region, all the $C(t)$ curves are the 
same for different BSD discretization intervals $\delta_F$. 
The deviation from the exact $T=0$ curve only occurs 
at longer times, which is shown in the inset. For minimum 
discretization intervals $\delta_F=10^{-3},10^{-4}$,~and 
$10^{-5}$, the deviation appears at approximately $t=10^{3},
10^{4}$,~and $10^{5}$, respectively.

From the above results, one can assume that the $T=0$ results 
can be obtained using the BSD scheme for a given accuracy. 
With the increase of the number of basis functions, 
we find that the weight $d^{f \sigma}_{\alpha l j}$ of the 
introduced low-frequency mode gets very small, 
such that its effect on the dynamics is very small before 
reaching a very long simulation time. 
This observation can also be understood from a different 
perspective: From Fig. \ref{Fig:imCorr_T0}, 
$C(t)$ obtained from the BSD scheme is the same as the zero 
temperature result till a long time $t \sim \delta_F^{-1}$, 
so one can assume that the dynamics represent 
the ``true'' zero temperature result till this time. For 
short-time simulations, or if the dissipation is 
relatively fast such that the system reaches its steady 
state before $\delta_F^{-1}$, the simulated dynamics is 
essentially at $T=0$. In practice, we will show later that, for 
the transport dynamics of the AIM, the result obtained from BSD 
with a small $\delta_F$ can converge and reproduce 
the correct zero temperature dynamics.

\subsection{Low temperature dynamics with 
the Lorentzian hybridization function}
We first consider the Lorentzian type
hybridization function that has been employed in many recent 
studies \cite{jin08,li12,hartle15b,xu17,zhang20,dan22},  where
\begin{equation} \label{eq:Lorentz}
\Gamma_{\alpha,l}(\varepsilon)=\frac{\eta_{l}\gamma_{l}^2}
{(\varepsilon-\varepsilon^0_{l})^2+\gamma_{l}^2} \;\;.
\end{equation}
Here, $\varepsilon^0_{l}$ and $\gamma_{l}$ 
denote the band center and width of the lead $l$,
and $\eta_l$ is the coupling strength between the molecule
and the lead $l$. This form of hybridization function only 
has two simple poles that are symmetric with respect to 
the real axis, leading to $K_\Gamma=1$ in Eq.\ (\ref{Eq:corr_pole}).
The majority of the exponential terms in Eq. (\ref{Eq:SOP_Bcorr})
thus originates from the BSD decomposition of the Fermi function. 

To illustrate the numerical stability of the HEOM combined 
with the BSD scheme, we first calculate the voltage-driven 
dynamics of the AIM. In the low-temperature regime, due to the 
interplay between the quantum coherence and the Kondo resonance, 
the real-time dynamics exhibit nontrivial memory 
effects \cite{zheng13}. As shown by Zheng \emph{et al}. \cite{zheng13}, 
driven by an external periodic voltage, the corresponding $I-V$ 
curve show hysteresis and self-crossing feature. In a later work, 
by using the FSD-based HEOM, it was shown that much stronger 
memory effects can be observed at even lower 
temperatures \cite{zhang20}. However, the FSD has an asymptotic 
instability problem \cite{zhang20},  and the time-dependent current
start to diverge after a certain time.
It was also shown that the instability becomes more severe 
when increasing the truncation tier $N_{\rm{trun}}$ of 
HEOM \cite{zhang20}.

\begin{figure}
\centering
\includegraphics[width=14cm]{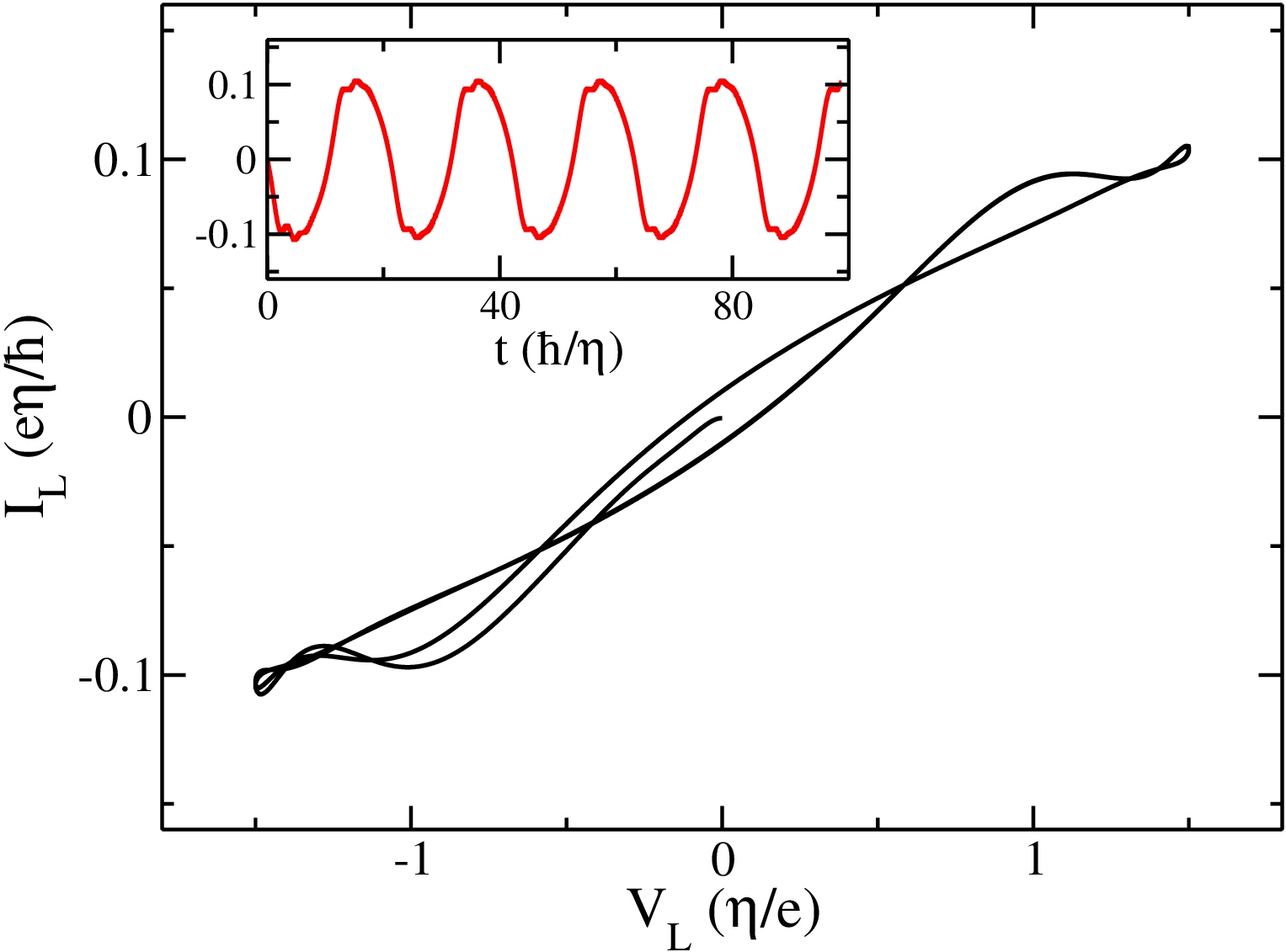}
\caption{Dynamic $I-V$ characteristics for the AIM with the 
Lorentzian hybridization function, with an $ac$ driving voltage 
$V_L(t)=-V_R(t)=V_0 \sin(\omega_0 t)$.
The parameters are (in units of $\eta$), 
$k_B T=0.05$, $\mu_{L}=\mu_{R}=0$, 
$\eta_{L}=\eta_{R}=1$, $\gamma_{L}=\gamma_{R}=20$, 
$\varepsilon^0_{L}=\varepsilon^0_{R}=0$, 
$\varepsilon_\uparrow=\varepsilon_\downarrow=-6$, $U=12$, 
$eV_0=1.5$, and $\hbar\omega_0=0.3$. 
Here, the BSD scheme is used to obtain the optimized 
pole structure of the Fermi function, with discretization 
domain $\mathcal{D}=[-200,200]$ and the accuracy control
parameter $tol_A=10^{-3}$. $K_f=9$ basis functions are used 
to decompose the Fermi function at this temperature.
The inset shows the corresponding transient current flow 
out of left lead $I_L(t)$. All simulations are performed using 
the on-the-fly filtering algorithm with the HEOM
truncation level $N_{\rm{trun}}=5$.}
\label{Fig:ac_volt_current}
\end{figure}

Fig.~\ref{Fig:ac_volt_current} shows the dynamic $I-V$ 
characteristics for the same parameter of Fig. 3
in Ref. \cite{zhang20}, where, in units of $\eta$, 
$k_B T=0.05$, $\mu_{L}=\mu_{R}=0$, 
$\eta_{L}=\eta_{R}=1$, 
$\gamma_{L}=\gamma_{R}=20$, 
$\varepsilon^0_{L}=\varepsilon^0_{R}=0$, 
$\varepsilon_\uparrow=\varepsilon_\downarrow=-6$, $U=12$, 
and the $ac$ voltage is $V_L(t)=-V_R(t)=V_0 \sin(\omega_0 t)$ 
with $eV_0=1.5$, $\hbar\omega_0=0.3 $. 
Here, the truncation tier is set to $N_{\rm{trun}}=5$ that 
ensures convergence. At this temperature, the BSD scheme 
using the AAA algorithm gives $K_f=9$, when discretizing 
the Fermi functions dense enough in the domain 
$\mathcal{D}=[-200,200]$, and setting $tol_A=10^{-3}$.

\begin{figure}
\centering
\includegraphics[width=14cm]{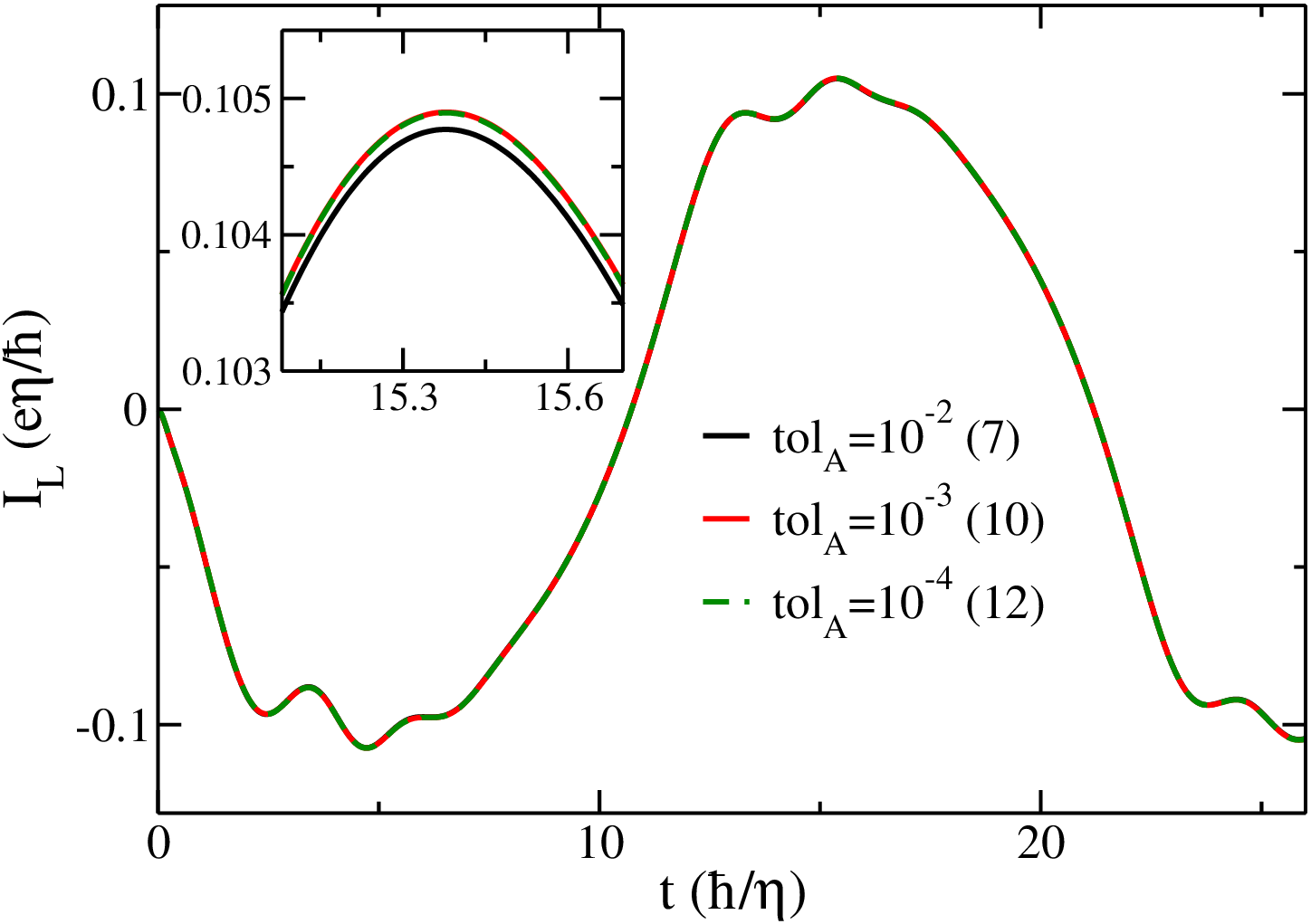}
\caption{Transient current flow out of the left lead $I_L(t)$, 
with different BSD accuracy control parameter $tol_A$. 
All other parameters are the same as those in 
Fig. \ref{Fig:ac_volt_current}. The number in the parentheses
denotes the total number of basis functions $K_\Gamma+K_f$
used in the simulation. The inset shows the
curves near the peak current.}
\label{Fig:baryconv_I}
\end{figure}

To obtain the transport current, we first propagate the 
total system without voltage ($\mu_{L}=\mu_{R}=0$) till 
equilibrium, then apply the $ac$ voltage to the left and 
right leads and propagate the system using 
Eq. (\ref{Eq:drho1}). Finally, the transient current is 
calculated by Eq. (\ref{Eq:expressI}). 
It can be seen in Fig. \ref{Fig:ac_volt_current} 
that, the $I-V$ characteristics are 
in excellent agreement with the results in Ref. \cite{zhang20}.
The inset shows the real-time current dynamics.
For longer simulations, the hysteresis loop is 
almost unchanged, as the $I-V$ curve repeats the earlier cycles.
It can be seen that the current has a phase shift relative 
to the driving voltage, and due to the Kondo resonance, 
there are also overtone responses. These two features 
manifest themselves in the hysteresis behavior with multiple 
self-crossing points of $I-V$ characteristics \cite{zheng13}.
Moreover, the current dynamics obtained from BSD-based 
HEOM runs smoothly over $100~\hbar/\eta$, compared with 
the FSD-based HEOM that diverges at about $3~\hbar/\eta$ 
for $N_{\rm{trun}}=5$ \cite{zhang20}, showing the long-time 
stability of the BSD scheme.
We further use this example to demonstrate the accuracy 
of the BSD scheme. In Fig.~\ref{Fig:baryconv_I},
we show the convergence of the transient current with 
respect to the BSD tolerance parameter $tol_A$. 	
The other parameters are the same as Fig. \ref{Fig:ac_volt_current}.
For $tol_A=10^{-2},10^{-3},10^{-4}$, the number of basis functions 
to decompose the Fermi function is ${K_f}=$ 6, 9, and 11, 
respectively (note that $K_\Gamma=1$ for the 
Lorentzian hybridization function, and the number in the 
figure legend denotes $K_\Gamma+{K_f}$).
It can be seen that, even with a relatively rough precision 
control parameter $tol_A=10^{-2}$, the dynamics are very 
close to convergence. This shows that the BSD scheme 
has high precision and $tol_A$ in the AAA algorithm 
is an effective precision control parameter.
For $tol_A=10^{-3}$, the transient current 
is essentially indistinguishable from those 
with $tol_A=10^{-4}$. Based on these observations, 
we choose $tol_A=10^{-3}$ in later simulations.

\begin{figure}
\centering
\includegraphics[width=12cm]{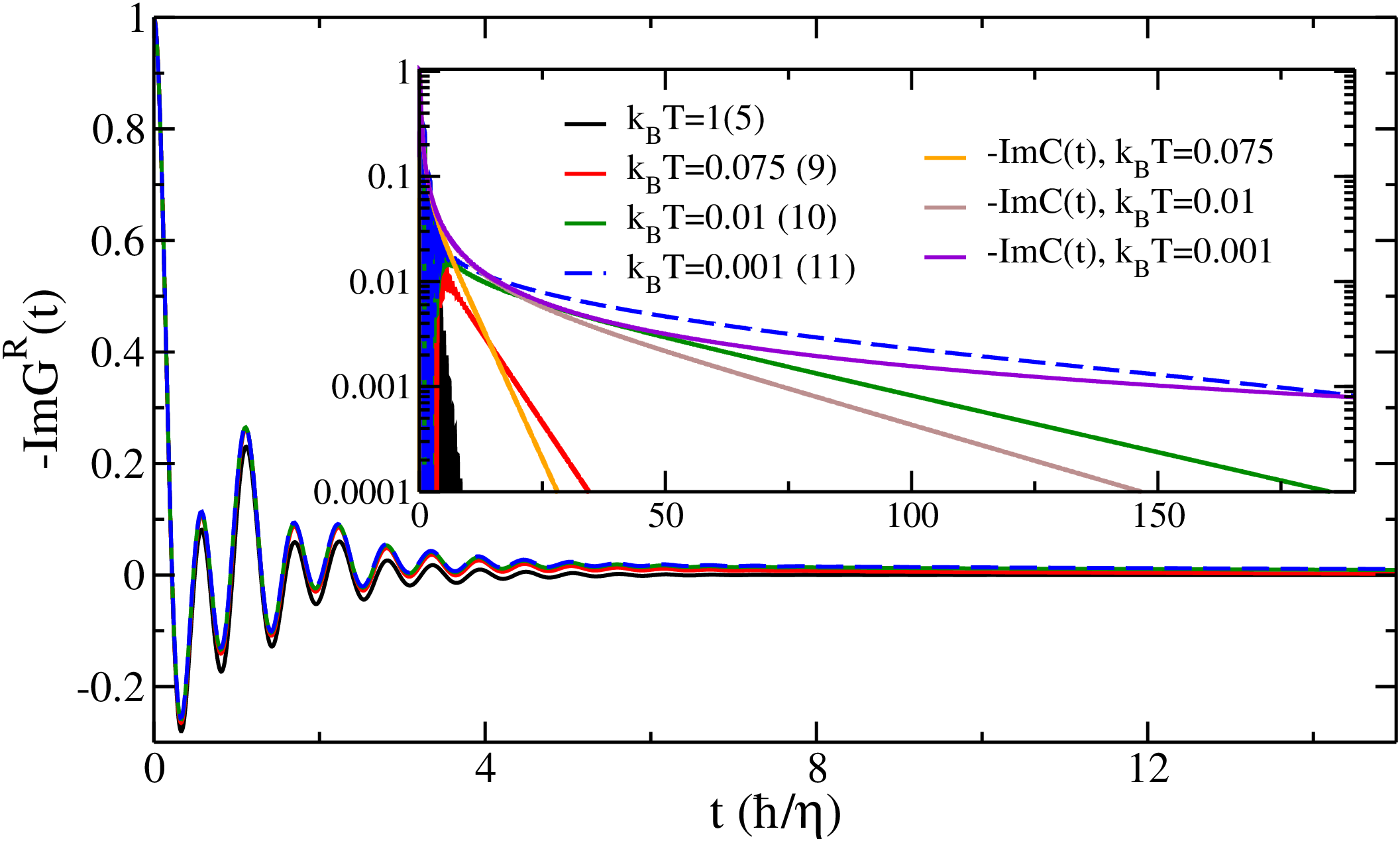}
\vspace{1em}
\caption{
Imaginary part of the retarded Green's function 
$-{\rm Im}G^R(t)$ for the AIM with Lorentzian hybridization 
function at different temperatures.
The inset shows the long-time decay, where 
the imaginary parts of the reservoir correlation function  
$-{\rm Im} C^+_{L}(t)$ (in units of $\eta^2/\hbar$) 
at different temperatures are also shown for comparison. 
The parameters are, in units 
of $\eta$, $\eta_{L}=\eta_{R}=1$, $\gamma_{L}=\gamma_{R}=10$, 
$\varepsilon^0_{L}=\varepsilon^0_{R}=0$, $\mu_{L}=\mu_{R}=0$,
$\varepsilon_\uparrow=\varepsilon_\downarrow=-5$, and $U=15$.
The truncation tier of HEOM is set to $N_{\rm{trun}}=6$, 
and the number of basis functions $K_\Gamma+K_f$ used in 
the simulations is 5, 9, 10, 11 for $k_B T$=1, 0.075, 0.01, 
and 0.001, respectively. These numbers are also shown in the 
parentheses of the legends.
In performing the BSD, the discretization domain for the 
Fermi distribution is $\mathcal{D}=[-200,200]$,
and the accuracy control parameter is $tol_A=10^{-3}$.}
\label{Fig:kondo_s1}
\end{figure}

Fig.~\ref{Fig:kondo_s1} shows the Green's 
function $G^R_\alpha(t)$ in Eq. (\ref{Eq:Greenfunc}), 
due to the similar behavior of ${\rm Im} G_\alpha^R(t)$ 
and ${\rm Re} G_\alpha^R(t)$, 
only $-{\rm Im} G_\alpha^R(t)$ [i.e., ${\rm Re} 
\langle \{ \hat{a}_\alpha(t) ,\hat{a}_\alpha^\dag \} \rangle$] 
is shown for simplicity. We choose the same system parameters 
studied previously with the PSD- \cite{li12} and FSD-based 
HEOM \cite{zhang20}.
That is, in units of $\eta$, $\eta_{L}=\eta_{R}=1$, 
$\gamma_{L}=\gamma_{R}=10$, 
$\varepsilon^0_{L}=\varepsilon^0_{R}=0$, 
$\mu_{L}=\mu_{R}=0$,
and $\varepsilon_\uparrow=\varepsilon_\downarrow=-5$, $U=15$.
The two spin states are now degenerate and we drop the 
subscript $\alpha$ for the spin DOF.
Since parameters for the left and right leads are the same, 
we can combine them into a single lead with 
$\eta_{tot}=\eta_L+\eta_{R}$,
and the number of basis functions in the HEOM
can be greatly reduced. Our data are calculated using HEOM 
truncated at $N_{\rm{trun}}=6$ to ensure convergence.
In the BSD scheme using the AAA algorithm, the Fermi 
distribution is discretized 
in the range $\mathcal{D}=[-200,200]$, and $tol_A=10^{-3}$, 
which leads to the number of basis functions 
$K_\Gamma+K_f=$ 5, 9, 10, 11 for $k_B T=$ 1, 
0.075, 0.01, 0.001, respectively.

\begin{figure}
\centering
\includegraphics[width=12cm]{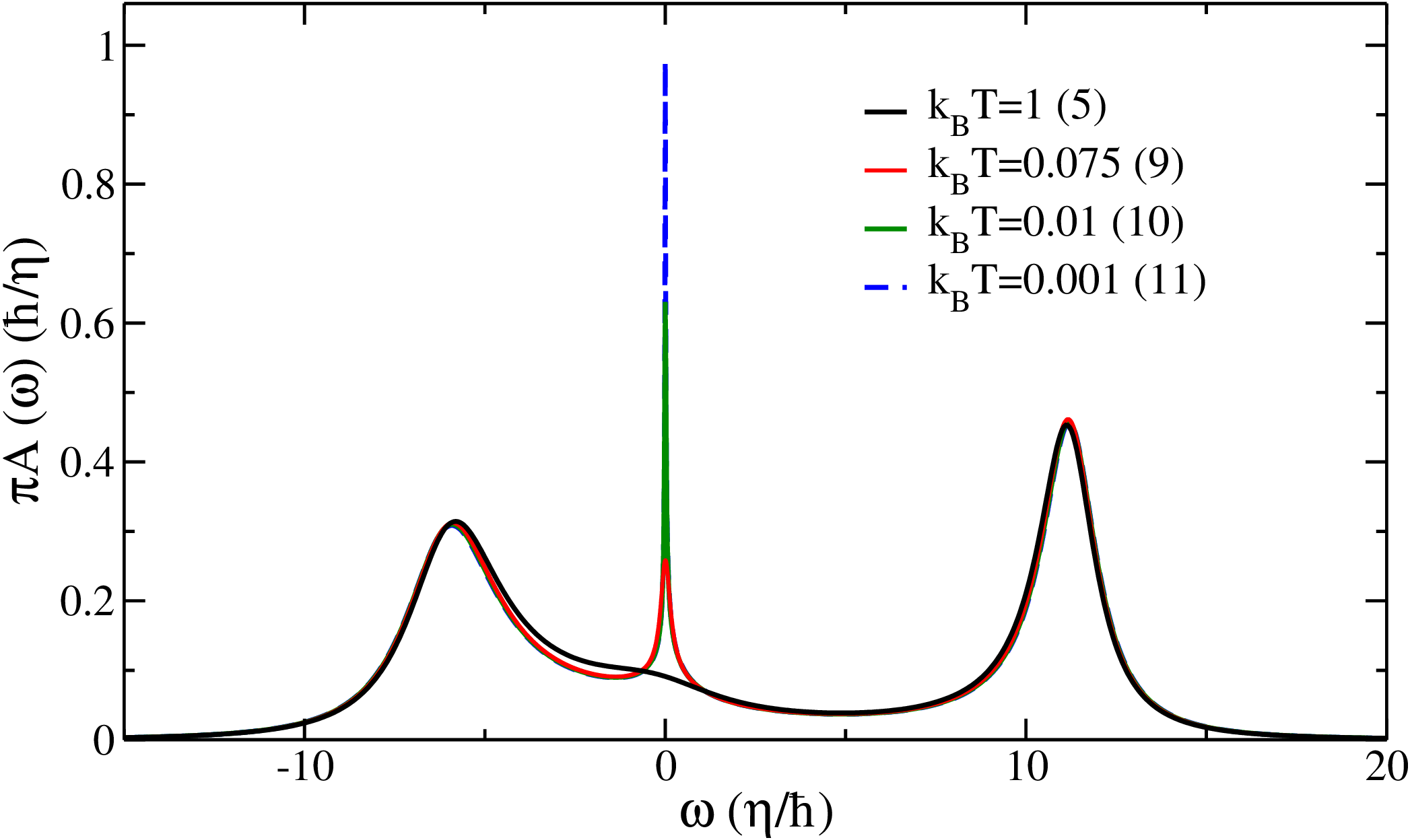}
\vspace{1em}
\caption{
The impurity spectral function $A(\omega)$ 
calculated from $G^R(t)$ in Fig. \ref{Fig:kondo_s1} 
through Eq. (\ref{Eq:Aome})
at different temperatures. 
All the parameters are the same as those in Fig. \ref{Fig:kondo_s1}.}
\label{Fig:kondo_s2}
\vspace{7em}
\end{figure}

The retarded Green's function is calculated after propagating 
the system to equilibrium, which oscillates over short period 
of time and shows exponential decay at longer times. This 
exponential decay is typical for Green's function at longer 
times \cite{barthel09,wolf15,wolf15b}. For comparison, we 
also plot the imaginary part of the reservoir correlation function 
$-{\rm Im} C^+_{L}(t)$ (in units of $\eta^2/\hbar$) in the inset.  
It can be seen that the long-time decay of the Green's function
is essentially determined by the low-frequency effective 
mode of the reservoir correlation function at this temperature, 
which is responsible for the sharp peak of $A(\omega)$ 
near the Fermi energy as the indication of Kondo resonance 
(see Fig. \ref{Fig:kondo_s2}). As the temperature decreases, 
the slope of $-{\rm Im} C^+_{L}(t)$ and $-{\rm Im}G^R(t)$ 
becomes closer, and they decay slowly such that longer simulation 
is required. For $k_B T=$ 0.001, we simulate the Green's function 
to the exponential decay regime 
(about $t=200~\hbar/\eta$ in the inset) 
and then employ the 
“linear prediction” method \cite{white08,barthel09,ganahl15} to 
extrapolate the simulation data to longer times.

Fig.~\ref{Fig:kondo_s2} shows the spectral functions at 
various temperatures. The spectral function 
$A(\omega)$ is obtained from the 
imaginary part of the Fourier transform of $G^R(t)$, 
that is, Eq.(\ref{Eq:Aome}).
For $k_B T=$ 1.0, 0.075, and 0.01, respectively, 
the results are consistent with those in Ref. \cite{zhang20}. 
The spectral function $A(\omega)$ shows two broad peaks 
at impurity level energies $\varepsilon$ and $\varepsilon+U$, 
which do not change significantly with the temperature. 
This is because the broadening is caused by coupling to 
the leads, such that it is on the order of $\eta$ and is 
only slightly affected by temperature.
On the other hand, the peak located at the Fermi energy 
is significantly affected by decreasing the temperature.
A sharp peak at the Fermi energy appears at low temperature,
and for $T=0.001$, the amplitude of this peak is very 
close to 1, which agrees with the theoretical result at 
$T=0$ \cite{langreth66,martinek03}.

\begin{figure}
\centering
\includegraphics[width=12cm]{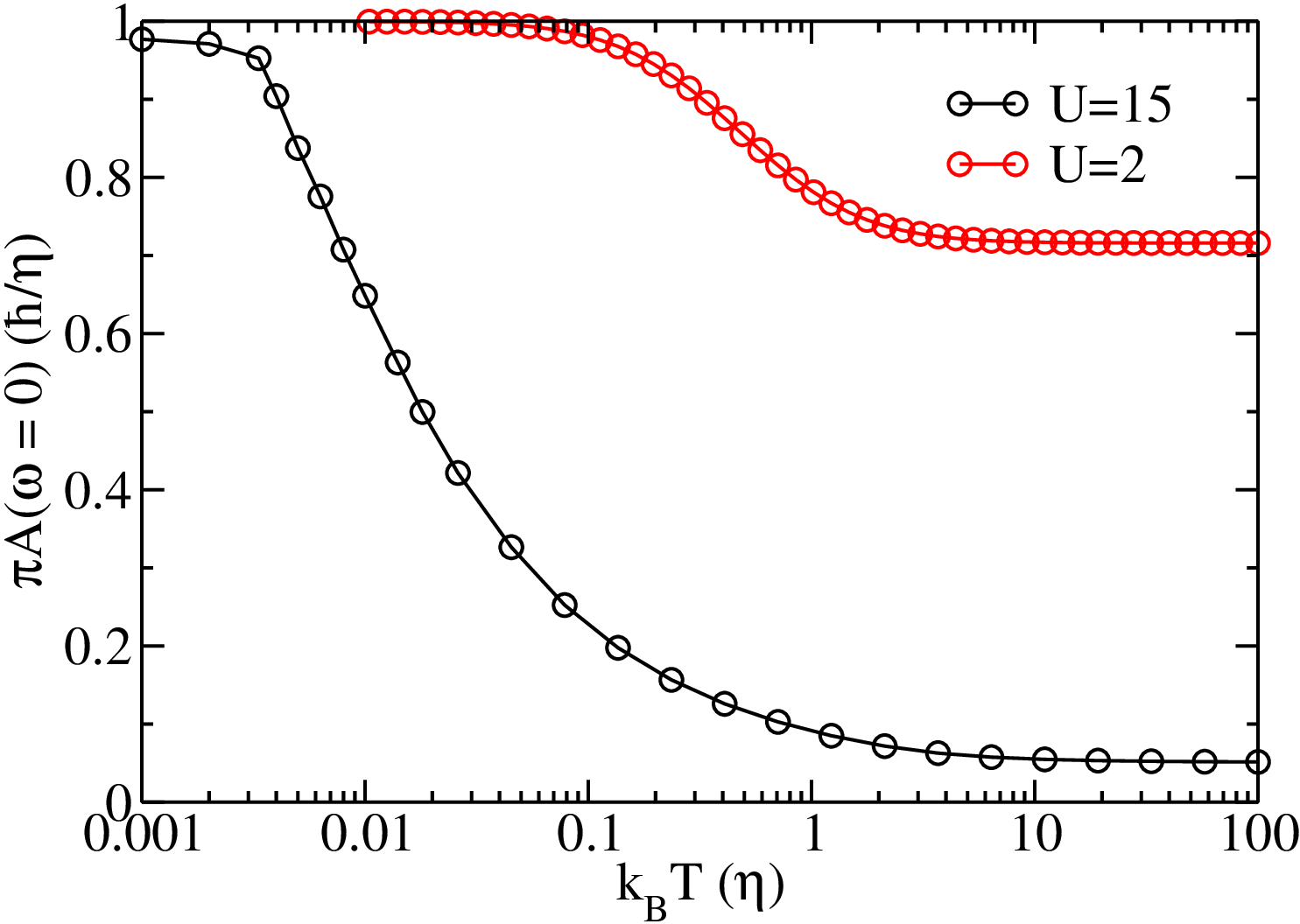}
\vspace{1em}
\caption{
The calculated $A(\omega=0)$ at various temperatures 
with $N_{\rm{trun}}=6$.
The black circles labeled with $U=15$ are obtained using 
the same parameters as those in Fig. \ref{Fig:kondo_s1}.
The red circles labeled with $U=2$ assume a symmetric 
AIM (in units of $\eta$, $\varepsilon_\uparrow
=\varepsilon_\downarrow=-U/2=-1$, with all other parameters 
the same as those in Fig. \ref{Fig:kondo_s1}). The Friedel
sum rule predicts $A(0)=1$ at $T=0$ for the symmetric AIM.}
\label{Fig:FSR}
\end{figure}

We also compare our BSD-HEOM result to the analytic Friedel sum 
rule \cite{langreth66,martinek03} at zero temperature, 
which is given by ${\pi}A(\omega=0)\eta/\hbar=\sin^2(\pi n_\alpha)$, 
with $n_\alpha=\langle \hat{a}_{\alpha}^{\dag} \hat{a}_{\alpha} \rangle$.
Fig.~\ref{Fig:FSR} shows the calculated $A(\omega=0)$ at different 
temperatures, with $N_{\rm{trun}}=6$. The red circle in 
Fig.~\ref{Fig:FSR} corresponds to a symmetric AIM case
with a small Coulomb repulsive energy $U=2$ (in units of $\eta$), 
where the other parameters are the same as those in 
Fig.~\ref{Fig:kondo_s1}. For this symmetric AIM, at equilibrium, 
$n_\alpha=1/2$, the transition temperature (here we define it as 
the temperature at which the Kondo peak changed significantly) 
is about $k_B T = 1$. As the temperature decreases, the amplitude 
of the Kondo peak, $\pi A(\omega=0) \eta/\hbar$ approaches 1, 
which verifies the Friedel sum rule and indicates the accuracy 
of our BSD-based HEOM. 

The black circles in Fig.~\ref{Fig:FSR} show the result for the 
parameters in Fig.~\ref{Fig:kondo_s1} with a larger Coulomb 
repulsive energy $U=15$. In this case, $n_\alpha$ at equilibrium 
for $T=0.001$ is 0.4811, and the Friedel sum rule predicts a Kondo 
peak amplitude at $T=0$ very close to 1. The transition temperature 
now is very low at about $k_B T=0.01$. We can also see that $A(\omega=0)$ 
approaches the analytic result as $T$ goes down. This indicates 
that the calculated data contains the correct long-time behavior 
and the extrapolation scheme works well in these cases.

It is also noted that the results below $k_B T=0.01$ are obtained 
by combining the long-time simulation data with the “linear 
prediction” method \cite{white08,barthel09}. To make the extrapolation 
procedure work, the HEOM needs to capture the correct dynamics 
till the asymptotic region where the extrapolation ansatz works, 
which depends on the long-time convergence of HEOM. 
It has been found previously that, when the truncation tier 
$N_{\rm{trun}}$ is not sufficient, HEOM does not converge and may 
cause overshoots of the Kondo peak \cite{zhang22,ding22}.
In more complex cases or as temperature decreased even further, 
a larger $N_{\rm{trun}}$ might be needed for the convergence 
of HEOM.

\subsection{Low temperature dynamics with 
the tight-binding hybridization function}
To illustrate that the BSD scheme can handle general 
forms of hybridization function, in this section we 
choose a tight-binding hybridization function 
with $\Gamma_{\alpha,l}(\varepsilon)$ given by: 
\begin{subequations}\label{Eq:tightbjome}
\begin{equation}
\Gamma(\varepsilon) =\begin{cases}\frac{\Delta_e^2}{W_e^2}
 \sqrt{4W_e^2-\varepsilon^2} 
&|\varepsilon| <= 2|W_e| \\
0 &|\varepsilon| > 2|W_e|
\end{cases},\\
\end{equation}
\begin{equation}
\Gamma_{\alpha,L}(\varepsilon)=\Gamma(\varepsilon-\mu_L), \quad
\Gamma_{\alpha,R}(\varepsilon)=\Gamma(\varepsilon-\mu_R) \;\; .
\end{equation}
\end{subequations}
This semi-elliptical form of hybridization function has been 
studied previously by Wang {\it et al.} \cite{wang09,
wang13b},  Wolf {\it et al.} \cite{wolf14,
wolf15}, and other groups \cite{karski08,dorda14,
wojtowicz21,kohn21}. Wang {\it et al.} have used the 
ML-MCTDH-SQR method to explore charge transport dynamics 
through single-molecule junctions at zero temperature.
Here, we fix $\Delta_e=0.2~eV$ and $W_e=1~eV$ which are
the same as those in Ref. \cite{wang13b}. 
As in the previous section, we set the two spin states to 
be degenerate.

\begin{figure}
\centering
\includegraphics[width=14cm]{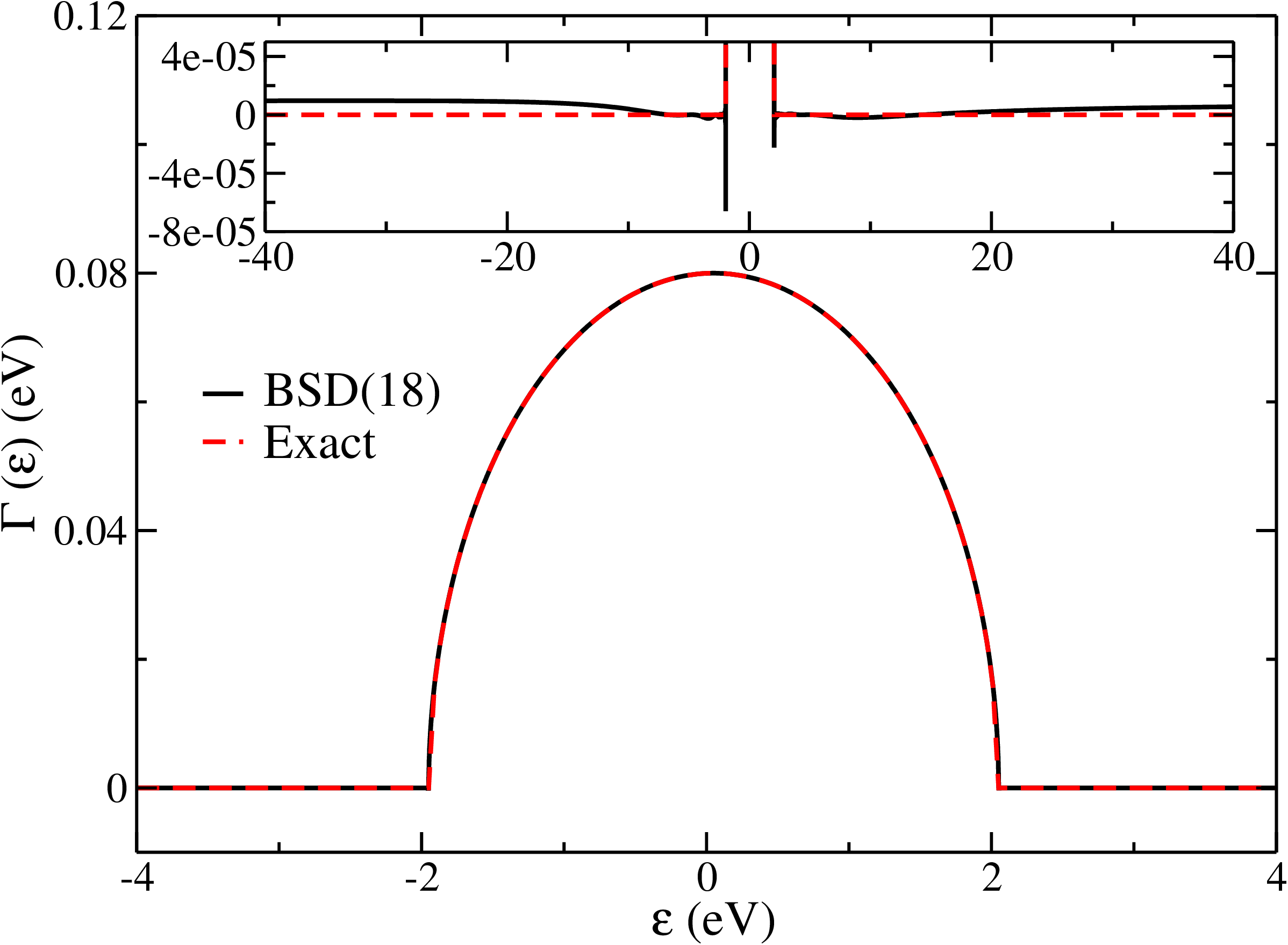}
\caption{Performance of the BSD scheme for 
decomposing the tight-binding hybridization function 
$\Gamma(\varepsilon)$ defined in Eq.\ (\ref{Eq:tightbjome}),
with $\Delta_e=0.2~eV$ and $W_e=1~eV$.
$\Gamma(E)$ is discretized in the domain 
$\mathcal{D}_\Gamma=[-7~eV,7~eV]$, with the minimum 
discretization interval $\delta_\Gamma=0.003~eV$ near 
the band edge. The accuracy control parameter $tol_A=10^{-3}$ 
results in $K_\Gamma=18$. The 
inset shows the error of the 
BSD scheme, for the range close to the band edge and 
beyond the discretization domain $\mathcal{D}$.
}
\label{Fig:hfci_performance}
\end{figure}

Fig.~\ref{Fig:hfci_performance} shows the performance 
of the BSD scheme for decomposing the tight-binding 
hybridization function $\Gamma(E)$ defined in 
Eq.\ (\ref{Eq:tightbjome}).
In the BSD scheme using the AAA algorithm, we set 
$tol_A=10^{-3}$ and discretize the hybridization function 
on the domain $\mathcal{D}=[-7~eV,7~eV]$, with 
the minimum discretization interval $\delta_\Gamma=0.003~eV$.
This results in $K_\Gamma=18$ for the semi-elliptical 
hybridization function. It can be seen
that the approximate hybridization function agrees 
very well with the exact one, 
even beyond the fitting range (see the inset). 
From the inset, we can also see that the approximate 
hybridization function deviates from the exact one 
mainly on the hard edges, and places beyond the fitting domain. 
But the overall errors are smaller than the given accuracy 
control parameter $tol_A$.

\begin{figure}
\centering
\includegraphics[width=14cm]{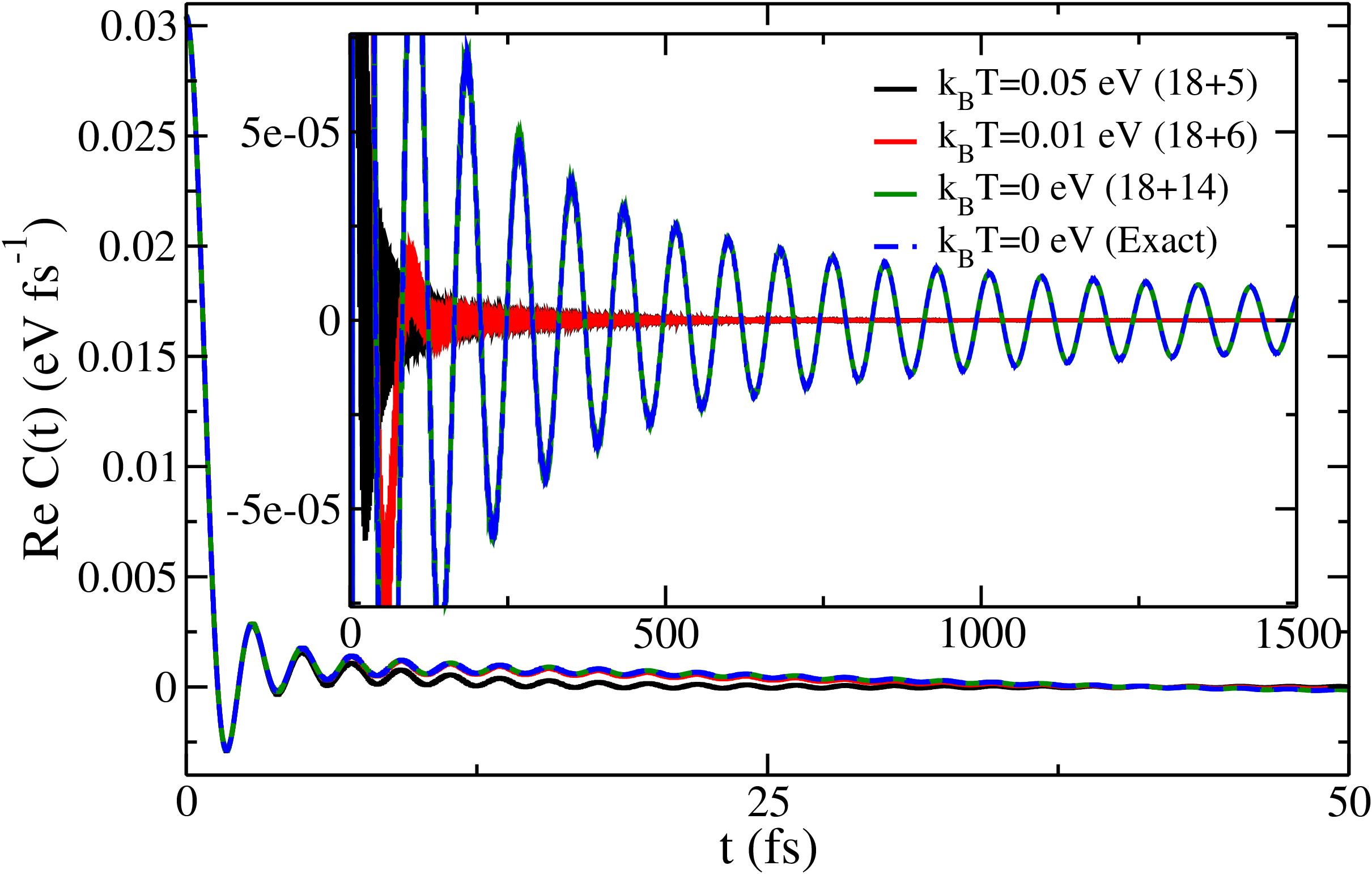}
\caption{
The reservoir correlation function $C_L^+(t)$ at different temperatures, 
for $\Gamma(\varepsilon)$ in Fig.~\ref{Fig:hfci_performance}.
For simplicity, only the real part 
is presented. Here, $\mu_L=0.05~eV$, 
$\Gamma_L(\varepsilon)=\Gamma(\varepsilon-\mu_L)$,
and the Fermi functions are discretized in the domain 
$\mathcal{D}_f=[-3~eV,3~eV]$, with $tol_A=10^{-3}$.
For $T=0$, the minimum discretization 
interval $\delta_F$ is $10^{-4}~eV$. 
The number of basis functions used 
to decompose the Fermi function is $K_f=$~5, 
6, 14 for $k_B T=$~0.05, 0.01, and 0~eV, respectively.
The inset shows the long-time behavior.
}
\label{Fig:hfci_corr}
\end{figure}

Fig.~\ref{Fig:hfci_corr} shows the corresponding reservoir 
correlation function, calculated from $\Gamma(E)$ in 
Fig. \ref{Fig:hfci_performance}, at different temperatures 
$k_B T=$~0.05, 0.01, and 0~eV with the source-drain voltage 
applied symmetrically to the left and right lead $\mu_L=-\mu_R=0.05~eV$. 
For simplicity, only the real part of $C_L^+(t)$ is shown.
The BSD scheme results in $K_f=$~5, 6, and 14 for $k_B T=$~0.05, 
0.01, and 0~eV, respectively.
Compared with the Lorentzian hybridization function, 
the tight-binding hybridization function results in 
oscillating reservoir correlation functions with the 
frequency associated with bandwidth.
The hard edge of the tight-binding hybridization function, 
like the Fermi function at $T=0$ shown in Fig.~\ref{Fig:fermi_T0-1}, 
leads to many poles with small decay constants
${\rm Re} \nu^{\Gamma +}_{L j}$ and weights $d^{\Gamma +}_{L j}$. 
The minimum ${\rm Re} \nu^{\Gamma +}_{L j}$ is found to be 
slightly smaller than $\delta_\Gamma$.

When the minimum ${\rm Re} \nu^{f +}_{L j}$ from the Fermi 
distribution is larger than the minimum ${\rm Re} \nu^{\Gamma +}_{L j}$,
the decay of $C_L^+(t)$ is controlled by temperature at short time, 
and is controlled by the basis functions obtained from 
the tight-binding hybridization function at longer time. 
This is shown in $k_B T=0.05~eV$ and $k_B T=0.01~eV$ curves. 
At short time, the decay of the high-temperature $k_B T=0.05~eV$ 
correlation function is faster than the low-temperature 
one at $k_B T=0.01~eV$. 
But at longer times, the two curves almost coincide as shown 
in the inset of Fig. \ref{Fig:hfci_corr}. When the temperature 
is low enough, the decay of $C_L^+(t)$ is mainly controlled by 
the temperature. Indeed, the approximate $C_L^+(t)$ with $T=0$ 
obtained with $\delta_F=10^{-4}~eV$ shows rather different 
long-time behavior. 

Although as discussed previously in Sec.~\ref{sec:fermi-BSD}, 
the approximate $C_L^+(t)$ corresponds to a very small effective 
temperature as shown in Fig. \ref{Fig:fermi_T0-1},
it can be seen from the inset of Fig.~\ref{Fig:hfci_corr} that, 
the approximate and exact correlation functions agree with each 
other over a very long time range. In this sense,
we can conclude that within this time range, the 
dynamics obtained from the approximate $C_L^+(t)$ should
essentially be the same as the ``true" zero temperature dynamics 
at $T=0$. Moreover, though not shown in the figure, 
in the temperature-controlled decay regime, the decay 
rate of the reservoir correlation function
is the same for different forms of hybridization 
functions, which will result in the same decay rate of the Green's 
function as in Fig.~\ref{Fig:kondo_s1},
such that the Kondo resonance of different types of 
hybridization functions at low temperature should not 
change significantly \cite{chen22}. 

\begin{figure}
\centering
\includegraphics[width=12cm]{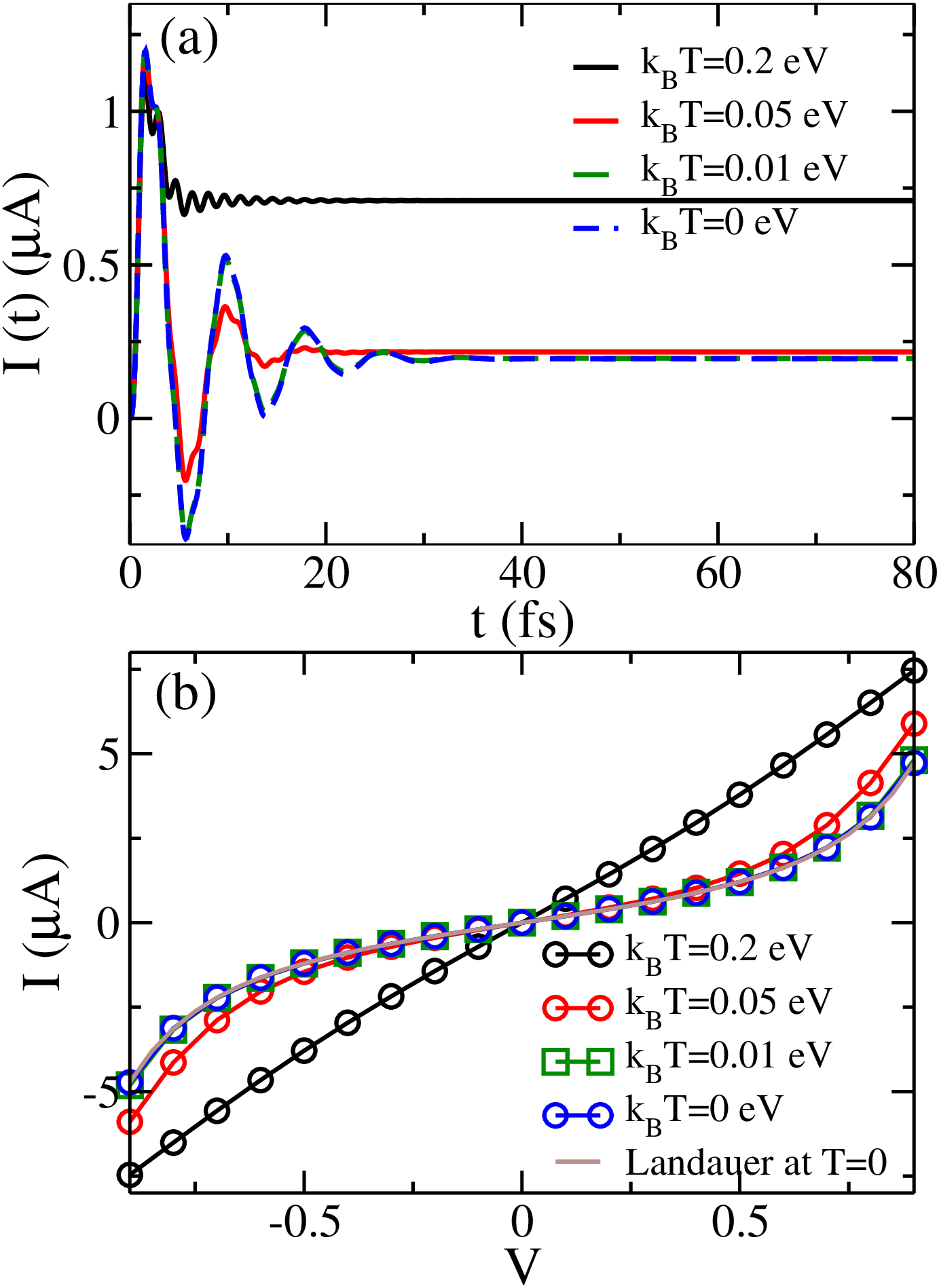}
\caption{(a) The average current $I(t)=[I_L(t)-I_R(t)]/2$ of the 
non-interacting AIM ($U$=0) with the tight-binding 
hybridization function at different temperatures. 
Here, $\Delta_e=0.2~eV$, $W_e=1~eV$, 
$\epsilon_\alpha=-0.5~eV$, $U=0$, and the bias voltage $V=0.1 {~\rm V}$
is symmetrically applied to two leads: $\mu_L=-\mu_R=V/2$. 
The BSD result is the same as in Fig. \ref{Fig:hfci_corr}, 
with $K_\Gamma=18$, $K_f=$~5, 6, 14 for $k_B T=$~0.05, 0.01, $0~eV$, 
respectively. The $k_B T=0.2~eV$ case ($K_f=3$), 
though at an unrealistic high temperature, 
is also shown for comparison.
(b) The average steady state current at different bias voltage, 
the other parameters are same as those in panel (a).}
\label{Fig:hfcicurr_U0}
\end{figure}

We then study the current dynamics of the AIM with the 
tight-binding hybridization function at different temperatures. 
Fig. \ref{Fig:hfcicurr_U0} shows the current dynamics 
for the non-interacting case $U=0$.  In this case, the HEOM 
with $N_{\rm{trun}}=2$ leads to exact results \cite{jin08}.  
The energy of the system state is $\varepsilon_\alpha=-0.5~eV$,
$\mu_L=-\mu_R=0.05~eV$ such that the bias voltage is $0.1~{\rm V}$.
This parameter corresponds to the so called off-resonant 
transport regime \cite{wang13b}. The other parameters for the 
hybridization or Fermi functions are the same as those in 
Fig. \ref{Fig:hfci_corr}. In the simulation, the initial system 
state is doubly occupied. In this case, the current shows 
transient oscillations at short time that decay 
due to the coupling to the leads. 
When the temperature is relatively high, 
we can see another high-frequency oscillation with 
frequency $\omega \approx 3~fs^{-1}$ ($2~eV$) which 
corresponds to the band edge of the semi-elliptical 
hybridization function. 

At low temperatures, the lead states at the band edge do not 
contribute to the transport process.
So as the temperature 
decreases, the high-frequency oscillation is suppressed.
The steady state current is suppressed as the 
temperature decreases, since the conductive window is reduced.
As can be seen from Fig.~\ref{Fig:hfcicurr_U0}(a), 
the $k_B T=0.01~eV$ result is already very close to 
that at $T=0$, indicating that the system is already in 
the low temperature regime. Further lowering the temperature 
will only lead to very minor changes of the dynamics. 
Moreover, our $T=0$ results obtained from the BSD scheme 
agree well with the results in Ref. \cite{wang13b}. 

Fig.~\ref{Fig:hfcicurr_U0}(b) gives the steady state current as 
a function of the bias voltage, with all other parameters fixed.
At high temperature $k_B T=0.2~eV$, 
the $I-V$ characteristics at this bias range in almost linear.
When lowering the temperature, the conductance increases after 
entering the resonant transport regime. 
The $I-V$ curve is almost unchanged for the $k_B T=0.01~eV$ and 
$T=0$ cases obtained from the BSD-based HEOM, 
which further shows that $k_B T=0.01eV$ is already in the 
low temperature regime. It is noted that for the non-interacting 
case $U=0$, the stationary current can be obtained exactly from 
the Landauer-B\"uttiker formula \cite{mahan00,dan22}.
As shown in Fig.~\ref{Fig:hfcicurr_U0}(b), 
the BSD-based HEOM results at $T=0$ agrees well with those 
from the Landauer-B\"uttiker formula, indicating that 
the $T=0$ results are converged. 

We then study the current dynamics of AIM in the presence of 
electron-electron repulsive energy $U$.
This is much more challenging than the non-interacting case 
since the HEOM must be truncated at a higher tier. 
At low temperatures, the needed truncation tier for high 
accuracy is relatively high. Our traditional HEOM code 
based on the filtering approach can not handle the 
computational and memory costs 
to get converged results at $T=0$ ($K_\Gamma+K_f=32$), 
for truncation tier higher than $N_{\rm{trun}}=4$.
In this case, we resort to the MPS-HEOM method \cite{shi18} 
to reduce the computational costs. 

\begin{figure}
\centering
\includegraphics[width=13cm]{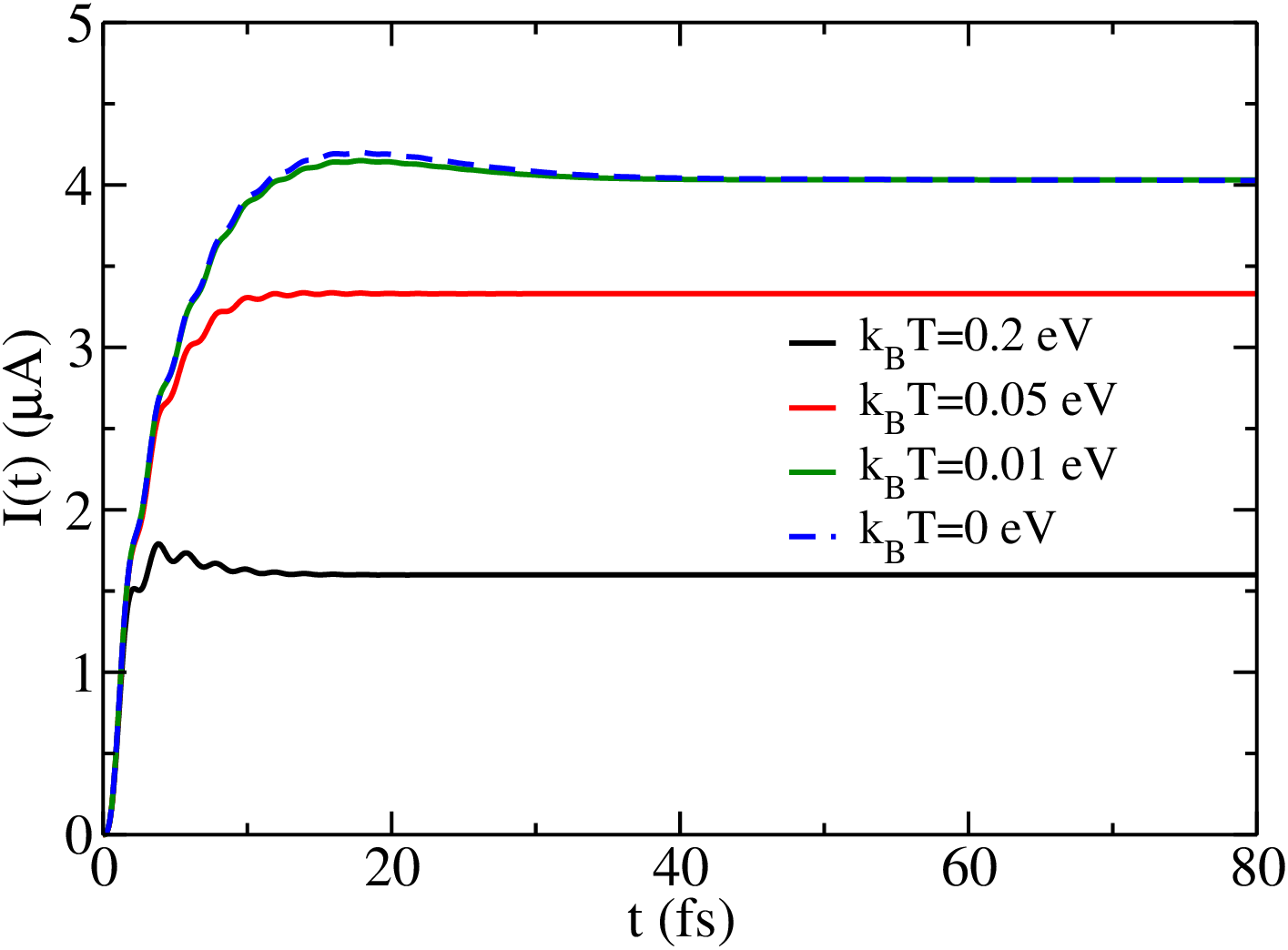}
\caption{
Time evolution of the average transport current of the AIM with the 
tight-binding hybridization function at different temperatures. 
The electron-electron repulsive energy is $U=0.5~eV$, all the other 
parameters are the same as those in Fig. \ref{Fig:hfcicurr_U0}(a).
The $T=0$ result is obtained using MPS-HEOM with the maximum 
bond dimension up to 500.}
\label{Fig:hfcicurr_U0.5}
\end{figure}

Fig. \ref{Fig:hfcicurr_U0.5} shows the current dynamics of the AIM 
with $U=0.5~eV$, with all the other parameters being the same as 
those in Fig.~\ref{Fig:hfcicurr_U0}. Here the $T=0$ result is 
obtained using MPS-HEOM \cite{shi18} with maximum bond dimension 
up to 500. Comparing to Fig. \ref{Fig:hfcicurr_U0}(a),
the $U=0.5~eV$ case has a larger current. This is because that,
the energy change from the single occupied state to the doubly 
occupied state, $\varepsilon+U$, moves close to the Fermi energy 
and provides a channel for resonant transport.
In the $U=0.5~eV$ case, the transport becomes incoherent for 
a short period of time. Moreover, the temperature dependence 
of the current is reversed, that is, decreasing the temperature 
results in an increase in the current. The reason is that,
decreasing the temperature reduces the fluctuations, and 
increases the probability of resonant transport. It can be seen 
that, the high-frequency oscillation of the current 
which is the band edge effect appears even at low temperature. 
While in the $U=0$ case, this effect 
only appears at high temperatures.
The $k_B T=0.01~eV$ case is found to be already in the 
low-temperature regime, where the dynamics are only slightly 
different from the $T=0$ result in the transient regime, 
and the steady-state currents are almost the same. 
Our $T=0$ results agree well with those in Ref. \cite{wang13b}, 
which verifies the validity of the BSD scheme 
for zero temperature dynamics.
The relaxation process of the current to stationary value 
is also slower at low temperatures, due to the long time 
memory effect.

%%%%%%%%%%%%%%
\section{Conclusion and Discussion}\label{sec:conclusion}
In this paper, we have proposed to use the BSD scheme for the
sum-over-poles decomposition of the Fermionic reservoir correlation 
functions. The BSD scheme based on rational function approximation 
of barycentric form can be used to calculate optimized pole structure 
than the analytical Matsubara and PSD poles, thus is superior in 
low-temperature simulations. This decomposition scheme is easy to 
implement with existing software packages. 
By approximating the reservoir correlation 
functions $C^\sigma_{\alpha,l}(t)$ and 
$C^{\bar{\sigma}}_{\alpha,l}(t)$ with the same set of poles, 
the traditional structure of the HEOM is also maintained.
With the BSD scheme, we can apply the HEOM to the AIM in the
low-temperature regime even at $T=0$ for dynamic simulation 
with high efficiency, accuracy, and long-time stability. 
The improvements of BSD over previous HEOM-based methods such 
as PSD \cite{ozaki07,hu10,hu11} and FSD \cite{cui19,zhang20} 
in efficiency and long-time stability, 
as well as the ability to deal with arbitrary band structures, 
offer new opportunities for applications in lower temperatures 
or more realistic systems. 

To demonstrate the performance of the BSD scheme, we first
apply it to approximate the Fermi function, the results 
show that the number of required basis functions grows almost 
linearly as the temperature decreases exponentially, 
compared to the exponential growth of 
the widely used PSD scheme. The low temperature performance 
is further demonstrated by calculating the Kondo 
peak of the impurity spectral function of the AIM 
with the Lorentzian hybridization function.
The $I-V$ characteristics under an $ac$ voltage have also 
been simulated, showing the accuracy and long-time 
stability of the HEOM combined with the BSD scheme.
Furthermore, to illustrate the application to arbitrary 
band structures, we apply the BSD scheme to the AIM with
a tight-binding band structure, where the hybridization 
function has no analytic poles. Combined with the 
MPS-HEOM \cite{shi18},  the current dynamics 
at different temperatures and even at $T=0$ have been explored. 

We have also pointed out that the BSD scheme cannot 
really reach the Fermi function at $T=0$ which is 
discontinuous at the Fermi energy. The approximate function 
for $T=0$ based on BSD can be regarded as the Fermi function 
at a very small temperature related to the smallest
discretizing interval $\delta_F$. In the time domain, the 
simulated results based on BSD reflect the {\it true} $T=0$ 
dynamics within a time range determined by $1/\delta_F$ which 
in many cases can be chosen sufficiently large to approach 
the steady-state regime.

Compared to other widely used methods such as 
the QMC \cite{muhlbacher08,werner09,schiro09, gull11,
antipov16,krivenko19,cohen15,antipov17} 
and time-dependent NRG (TD-NRG) \cite{anders05,anders06,bulla08,
nghiem18}, the long-time dynamics of HEOM do not 
suffer from discretization problems, and its computation 
cost grows linearly as time increases. 
The BSD scheme in this work also significantly increases the 
capability of HEOM in the low-temperature regime and for 
more complex band structures. Thus, we believe that BSD-based 
HEOM could become the method of choice for certain types of 
applications, especially in long-time simulations that may
be difficult for the QMC and TD-NRG methods.

Recently, some of us have developed the generalized master 
equation (GME) method to calculate the exact memory 
kernel from physical quantities such 
as population and current \cite{dan22}.
The memory kernels, which usually decay within a short 
period of time, can be used to produce reliable long-time 
dynamics \cite{cohen11,cohen13b}. 
Combining this method with the BSD scheme may allow us 
to calculate efficiently asymptotic long-time dynamics and 
stationary state properties 
at $T=0$ for even broader classes of hybridization functions.

With the development of the HEOM method and new algorithms 
including the MPS-HEOM \cite{shi18} with efficient time 
evolution methods \cite{yang20,borrelli21},  
Tree-Tensor HEOM \cite{yan21},  and the hierarchical 
Schr\"odinger equations of motion (HSEOM) \cite{nakamura18},  
the BSD scheme provides a new method for the simulation of 
realistic systems at low temperatures. 
Moreover, the methodology can be applied to other approaches 
based on an expansion of reservoir correlation functions 
or Green's functions, 
such as the hierarchy of pure states (HOPS) \cite{suess14}, 
the non-equilibrium Green's function (NEGF) \cite{beach00,croy09,gu20} 
methods, and the nonequilibrium dynamical mean-field theory 
(DMFT) \cite{arrigoni13}.

\acknowledgements
X.D. and Q.S. thank the financial support from NSFC (Grant No. 21933011)
and the K. C. Wong Education Foundation.
M.X., J.T.S., and J.A. gratefully acknowledge financial support from 
the DFG via AN336/12-1 (FOR2724) and the BMBF through QCOMP within the 
Cluster4Future QSens.

\bibliography{./quantum}

%merlin.mbs apsrev4-1.bst 2010-07-25 4.21a (PWD, AO, DPC) hacked
%Control: key (0)
%Control: author (8) initials jnrlst
%Control: editor formatted (1) identically to author
%Control: production of article title (-1) disabled
%Control: page (0) single
%Control: year (1) truncated
%Control: production of eprint (0) enabled
\begin{thebibliography}{125}%
\makeatletter
\providecommand \@ifxundefined [1]{%
 \@ifx{#1\undefined}
}%
\providecommand \@ifnum [1]{%
 \ifnum #1\expandafter \@firstoftwo
 \else \expandafter \@secondoftwo
 \fi
}%
\providecommand \@ifx [1]{%
 \ifx #1\expandafter \@firstoftwo
 \else \expandafter \@secondoftwo
 \fi
}%
\providecommand \natexlab [1]{#1}%
\providecommand \enquote  [1]{``#1''}%
\providecommand \bibnamefont  [1]{#1}%
\providecommand \bibfnamefont [1]{#1}%
\providecommand \citenamefont [1]{#1}%
\providecommand \href@noop [0]{\@secondoftwo}%
\providecommand \href [0]{\begingroup \@sanitize@url \@href}%
\providecommand \@href[1]{\@@startlink{#1}\@@href}%
\providecommand \@@href[1]{\endgroup#1\@@endlink}%
\providecommand \@sanitize@url [0]{\catcode `\\12\catcode `\$12\catcode
  `\&12\catcode `\#12\catcode `\^12\catcode `\_12\catcode `\%12\relax}%
\providecommand \@@startlink[1]{}%
\providecommand \@@endlink[0]{}%
\providecommand \url  [0]{\begingroup\@sanitize@url \@url }%
\providecommand \@url [1]{\endgroup\@href {#1}{\urlprefix }}%
\providecommand \urlprefix  [0]{URL }%
\providecommand \Eprint [0]{\href }%
\providecommand \doibase [0]{http://dx.doi.org/}%
\providecommand \selectlanguage [0]{\@gobble}%
\providecommand \bibinfo  [0]{\@secondoftwo}%
\providecommand \bibfield  [0]{\@secondoftwo}%
\providecommand \translation [1]{[#1]}%
\providecommand \BibitemOpen [0]{}%
\providecommand \bibitemStop [0]{}%
\providecommand \bibitemNoStop [0]{.\EOS\space}%
\providecommand \EOS [0]{\spacefactor3000\relax}%
\providecommand \BibitemShut  [1]{\csname bibitem#1\endcsname}%
\let\auto@bib@innerbib\@empty
%</preamble>
\bibitem [{\citenamefont {Galperin}\ \emph {et~al.}(2008)\citenamefont
  {Galperin}, \citenamefont {Ratner}, \citenamefont {Nitzan},\ and\
  \citenamefont {Troisi}}]{galperin08}%
  \BibitemOpen
  \bibfield  {author} {\bibinfo {author} {\bibfnamefont {M.}~\bibnamefont
  {Galperin}}, \bibinfo {author} {\bibfnamefont {M.~A.}\ \bibnamefont
  {Ratner}}, \bibinfo {author} {\bibfnamefont {A.}~\bibnamefont {Nitzan}}, \
  and\ \bibinfo {author} {\bibfnamefont {A.}~\bibnamefont {Troisi}},\
  }\href@noop {} {\bibfield  {journal} {\bibinfo  {journal} {Science}\ }\textbf
  {\bibinfo {volume} {319}},\ \bibinfo {pages} {1056} (\bibinfo {year}
  {2008})}\BibitemShut {NoStop}%
\bibitem [{\citenamefont {Zimbovskaya}(2013)}]{zimbovskaya13}%
  \BibitemOpen
  \bibfield  {author} {\bibinfo {author} {\bibfnamefont {N.~A.}\ \bibnamefont
  {Zimbovskaya}},\ }\href {\doibase 10.1007/978-1-4614-8011-2} {\emph {\bibinfo
  {title} {Transport Properties of Molecular Junctions}}},\ Vol.\ \bibinfo
  {volume} {254}\ (\bibinfo  {publisher} {Springer},\ \bibinfo {address} {New
  York},\ \bibinfo {year} {2013})\BibitemShut {NoStop}%
\bibitem [{\citenamefont {Ratner}(2013)}]{ratner13}%
  \BibitemOpen
  \bibfield  {author} {\bibinfo {author} {\bibfnamefont {M.}~\bibnamefont
  {Ratner}},\ }\href@noop {} {\bibfield  {journal} {\bibinfo  {journal} {Nature
  Nanotech.}\ }\textbf {\bibinfo {volume} {8}},\ \bibinfo {pages} {378}
  (\bibinfo {year} {2013})}\BibitemShut {NoStop}%
\bibitem [{\citenamefont {Xiang}\ \emph {et~al.}(2016)\citenamefont {Xiang},
  \citenamefont {Wang}, \citenamefont {Jia}, \citenamefont {Lee},\ and\
  \citenamefont {Guo}}]{xiang16}%
  \BibitemOpen
  \bibfield  {author} {\bibinfo {author} {\bibfnamefont {D.}~\bibnamefont
  {Xiang}}, \bibinfo {author} {\bibfnamefont {X.}~\bibnamefont {Wang}},
  \bibinfo {author} {\bibfnamefont {C.}~\bibnamefont {Jia}}, \bibinfo {author}
  {\bibfnamefont {T.}~\bibnamefont {Lee}}, \ and\ \bibinfo {author}
  {\bibfnamefont {X.}~\bibnamefont {Guo}},\ }\href@noop {} {\bibfield
  {journal} {\bibinfo  {journal} {Chem. Rev.}\ }\textbf {\bibinfo {volume}
  {116}},\ \bibinfo {pages} {4318} (\bibinfo {year} {2016})}\BibitemShut
  {NoStop}%
\bibitem [{\citenamefont {Thoss}\ and\ \citenamefont {Evers}(2018)}]{thoss18}%
  \BibitemOpen
  \bibfield  {author} {\bibinfo {author} {\bibfnamefont {M.}~\bibnamefont
  {Thoss}}\ and\ \bibinfo {author} {\bibfnamefont {F.}~\bibnamefont {Evers}},\
  }\href@noop {} {\bibfield  {journal} {\bibinfo  {journal} {J.~Chem.~Phys.}\
  }\textbf {\bibinfo {volume} {148}},\ \bibinfo {pages} {030901} (\bibinfo
  {year} {2018})}\BibitemShut {NoStop}%
\bibitem [{\citenamefont {Chen}\ \emph {et~al.}(1999)\citenamefont {Chen},
  \citenamefont {Reed}, \citenamefont {Rawlett},\ and\ \citenamefont
  {Tour}}]{chen99}%
  \BibitemOpen
  \bibfield  {author} {\bibinfo {author} {\bibfnamefont {J.}~\bibnamefont
  {Chen}}, \bibinfo {author} {\bibfnamefont {M.}~\bibnamefont {Reed}}, \bibinfo
  {author} {\bibfnamefont {A.}~\bibnamefont {Rawlett}}, \ and\ \bibinfo
  {author} {\bibfnamefont {J.}~\bibnamefont {Tour}},\ }\href@noop {} {\bibfield
   {journal} {\bibinfo  {journal} {Science}\ }\textbf {\bibinfo {volume}
  {286}},\ \bibinfo {pages} {1550} (\bibinfo {year} {1999})}\BibitemShut
  {NoStop}%
\bibitem [{\citenamefont {Park}\ \emph {et~al.}(2000)\citenamefont {Park},
  \citenamefont {Park}, \citenamefont {Lim}, \citenamefont {Anderson},
  \citenamefont {Alivisatos},\ and\ \citenamefont {McEuen}}]{park00}%
  \BibitemOpen
  \bibfield  {author} {\bibinfo {author} {\bibfnamefont {H.}~\bibnamefont
  {Park}}, \bibinfo {author} {\bibfnamefont {J.}~\bibnamefont {Park}}, \bibinfo
  {author} {\bibfnamefont {A.~K.}\ \bibnamefont {Lim}}, \bibinfo {author}
  {\bibfnamefont {E.~H.}\ \bibnamefont {Anderson}}, \bibinfo {author}
  {\bibfnamefont {A.~P.}\ \bibnamefont {Alivisatos}}, \ and\ \bibinfo {author}
  {\bibfnamefont {P.~L.}\ \bibnamefont {McEuen}},\ }\href@noop {} {\bibfield
  {journal} {\bibinfo  {journal} {Nature}\ }\textbf {\bibinfo {volume} {407}},\
  \bibinfo {pages} {57} (\bibinfo {year} {2000})}\BibitemShut {NoStop}%
\bibitem [{\citenamefont {Park}\ \emph {et~al.}(2002)\citenamefont {Park},
  \citenamefont {Pasupathy}, \citenamefont {Goldsmith}, \citenamefont {Chang},
  \citenamefont {Yaish}, \citenamefont {Petta}, \citenamefont {Rinkoski},
  \citenamefont {Sethna}, \citenamefont {Abru{\~n}a}, \citenamefont {McEuen}
  \emph {et~al.}}]{park02}%
  \BibitemOpen
  \bibfield  {author} {\bibinfo {author} {\bibfnamefont {J.}~\bibnamefont
  {Park}}, \bibinfo {author} {\bibfnamefont {A.~N.}\ \bibnamefont {Pasupathy}},
  \bibinfo {author} {\bibfnamefont {J.~I.}\ \bibnamefont {Goldsmith}}, \bibinfo
  {author} {\bibfnamefont {C.}~\bibnamefont {Chang}}, \bibinfo {author}
  {\bibfnamefont {Y.}~\bibnamefont {Yaish}}, \bibinfo {author} {\bibfnamefont
  {J.~R.}\ \bibnamefont {Petta}}, \bibinfo {author} {\bibfnamefont
  {M.}~\bibnamefont {Rinkoski}}, \bibinfo {author} {\bibfnamefont {J.~P.}\
  \bibnamefont {Sethna}}, \bibinfo {author} {\bibfnamefont {H.~D.}\
  \bibnamefont {Abru{\~n}a}}, \bibinfo {author} {\bibfnamefont {P.~L.}\
  \bibnamefont {McEuen}},  \emph {et~al.},\ }\href@noop {} {\bibfield
  {journal} {\bibinfo  {journal} {Nature}\ }\textbf {\bibinfo {volume} {417}},\
  \bibinfo {pages} {722} (\bibinfo {year} {2002})}\BibitemShut {NoStop}%
\bibitem [{\citenamefont {Xu}\ and\ \citenamefont {Tao}(2003)}]{xu03}%
  \BibitemOpen
  \bibfield  {author} {\bibinfo {author} {\bibfnamefont {B.}~\bibnamefont
  {Xu}}\ and\ \bibinfo {author} {\bibfnamefont {N.~J.}\ \bibnamefont {Tao}},\
  }\href@noop {} {\bibfield  {journal} {\bibinfo  {journal} {Science}\ }\textbf
  {\bibinfo {volume} {301}},\ \bibinfo {pages} {1221} (\bibinfo {year}
  {2003})}\BibitemShut {NoStop}%
\bibitem [{\citenamefont {Qiu}\ \emph {et~al.}(2004)\citenamefont {Qiu},
  \citenamefont {Nazin},\ and\ \citenamefont {Ho}}]{qiu04}%
  \BibitemOpen
  \bibfield  {author} {\bibinfo {author} {\bibfnamefont {X.}~\bibnamefont
  {Qiu}}, \bibinfo {author} {\bibfnamefont {G.~V.}\ \bibnamefont {Nazin}}, \
  and\ \bibinfo {author} {\bibfnamefont {W.}~\bibnamefont {Ho}},\ }\href@noop
  {} {\bibfield  {journal} {\bibinfo  {journal} {Phys.~Rev.~Lett.}\ }\textbf
  {\bibinfo {volume} {92}},\ \bibinfo {pages} {206102} (\bibinfo {year}
  {2004})}\BibitemShut {NoStop}%
\bibitem [{\citenamefont {Heersche}\ \emph {et~al.}(2006)\citenamefont
  {Heersche}, \citenamefont {De~Groot}, \citenamefont {Folk}, \citenamefont
  {Van Der~Zant}, \citenamefont {Romeike}, \citenamefont {Wegewijs},
  \citenamefont {Zobbi}, \citenamefont {Barreca}, \citenamefont {Tondello},\
  and\ \citenamefont {Cornia}}]{heersche06}%
  \BibitemOpen
  \bibfield  {author} {\bibinfo {author} {\bibfnamefont {H.~B.}\ \bibnamefont
  {Heersche}}, \bibinfo {author} {\bibfnamefont {Z.}~\bibnamefont {De~Groot}},
  \bibinfo {author} {\bibfnamefont {J.~A.}\ \bibnamefont {Folk}}, \bibinfo
  {author} {\bibfnamefont {H.~S.~J.}\ \bibnamefont {Van Der~Zant}}, \bibinfo
  {author} {\bibfnamefont {C.}~\bibnamefont {Romeike}}, \bibinfo {author}
  {\bibfnamefont {M.~R.}\ \bibnamefont {Wegewijs}}, \bibinfo {author}
  {\bibfnamefont {L.}~\bibnamefont {Zobbi}}, \bibinfo {author} {\bibfnamefont
  {D.}~\bibnamefont {Barreca}}, \bibinfo {author} {\bibfnamefont
  {E.}~\bibnamefont {Tondello}}, \ and\ \bibinfo {author} {\bibfnamefont
  {A.}~\bibnamefont {Cornia}},\ }\href@noop {} {\bibfield  {journal} {\bibinfo
  {journal} {Phys.~Rev.~Lett.}\ }\textbf {\bibinfo {volume} {96}},\ \bibinfo
  {pages} {206801} (\bibinfo {year} {2006})}\BibitemShut {NoStop}%
\bibitem [{\citenamefont {Liang}\ \emph {et~al.}(2002)\citenamefont {Liang},
  \citenamefont {Shores}, \citenamefont {Bockrath}, \citenamefont {Long},\ and\
  \citenamefont {Park}}]{liang02}%
  \BibitemOpen
  \bibfield  {author} {\bibinfo {author} {\bibfnamefont {W.}~\bibnamefont
  {Liang}}, \bibinfo {author} {\bibfnamefont {M.~P.}\ \bibnamefont {Shores}},
  \bibinfo {author} {\bibfnamefont {M.}~\bibnamefont {Bockrath}}, \bibinfo
  {author} {\bibfnamefont {J.~R.}\ \bibnamefont {Long}}, \ and\ \bibinfo
  {author} {\bibfnamefont {H.}~\bibnamefont {Park}},\ }\href@noop {} {\bibfield
   {journal} {\bibinfo  {journal} {Nature}\ }\textbf {\bibinfo {volume}
  {417}},\ \bibinfo {pages} {725} (\bibinfo {year} {2002})}\BibitemShut
  {NoStop}%
\bibitem [{\citenamefont {Parks}\ \emph {et~al.}(2010)\citenamefont {Parks},
  \citenamefont {Champagne}, \citenamefont {Costi}, \citenamefont {Shum},
  \citenamefont {Pasupathy}, \citenamefont {Neuscamman}, \citenamefont
  {Flores-Torres}, \citenamefont {Cornaglia}, \citenamefont {Aligia},
  \citenamefont {Balseiro} \emph {et~al.}}]{parks10}%
  \BibitemOpen
  \bibfield  {author} {\bibinfo {author} {\bibfnamefont {J.~J.}\ \bibnamefont
  {Parks}}, \bibinfo {author} {\bibfnamefont {A.~R.}\ \bibnamefont
  {Champagne}}, \bibinfo {author} {\bibfnamefont {T.~A.}\ \bibnamefont
  {Costi}}, \bibinfo {author} {\bibfnamefont {W.~W.}\ \bibnamefont {Shum}},
  \bibinfo {author} {\bibfnamefont {A.~N.}\ \bibnamefont {Pasupathy}}, \bibinfo
  {author} {\bibfnamefont {E.}~\bibnamefont {Neuscamman}}, \bibinfo {author}
  {\bibfnamefont {S.}~\bibnamefont {Flores-Torres}}, \bibinfo {author}
  {\bibfnamefont {P.~S.}\ \bibnamefont {Cornaglia}}, \bibinfo {author}
  {\bibfnamefont {A.~A.}\ \bibnamefont {Aligia}}, \bibinfo {author}
  {\bibfnamefont {C.~A.}\ \bibnamefont {Balseiro}},  \emph {et~al.},\
  }\href@noop {} {\bibfield  {journal} {\bibinfo  {journal} {Science}\ }\textbf
  {\bibinfo {volume} {328}},\ \bibinfo {pages} {1370} (\bibinfo {year}
  {2010})}\BibitemShut {NoStop}%
\bibitem [{\citenamefont {Osorio}\ \emph {et~al.}(2010)\citenamefont {Osorio},
  \citenamefont {Ruben}, \citenamefont {Seldenthuis}, \citenamefont {Lehn},\
  and\ \citenamefont {van~der Zant}}]{osorio10}%
  \BibitemOpen
  \bibfield  {author} {\bibinfo {author} {\bibfnamefont {E.~A.}\ \bibnamefont
  {Osorio}}, \bibinfo {author} {\bibfnamefont {M.}~\bibnamefont {Ruben}},
  \bibinfo {author} {\bibfnamefont {J.~S.}\ \bibnamefont {Seldenthuis}},
  \bibinfo {author} {\bibfnamefont {J.~M.}\ \bibnamefont {Lehn}}, \ and\
  \bibinfo {author} {\bibfnamefont {H.~S.}\ \bibnamefont {van~der Zant}},\
  }\href@noop {} {\bibfield  {journal} {\bibinfo  {journal} {Small}\ }\textbf
  {\bibinfo {volume} {6}},\ \bibinfo {pages} {174} (\bibinfo {year}
  {2010})}\BibitemShut {NoStop}%
\bibitem [{\citenamefont {Gaudioso}\ \emph {et~al.}(2000)\citenamefont
  {Gaudioso}, \citenamefont {Lauhon},\ and\ \citenamefont {Ho}}]{gaudioso00}%
  \BibitemOpen
  \bibfield  {author} {\bibinfo {author} {\bibfnamefont {J.}~\bibnamefont
  {Gaudioso}}, \bibinfo {author} {\bibfnamefont {L.~J.}\ \bibnamefont
  {Lauhon}}, \ and\ \bibinfo {author} {\bibfnamefont {W.}~\bibnamefont {Ho}},\
  }\href@noop {} {\bibfield  {journal} {\bibinfo  {journal} {Phys.~Rev.~Lett.}\
  }\textbf {\bibinfo {volume} {85}},\ \bibinfo {pages} {1918} (\bibinfo {year}
  {2000})}\BibitemShut {NoStop}%
\bibitem [{\citenamefont {Xu}\ and\ \citenamefont {Dubi}(2015)}]{xu15}%
  \BibitemOpen
  \bibfield  {author} {\bibinfo {author} {\bibfnamefont {B.}~\bibnamefont
  {Xu}}\ and\ \bibinfo {author} {\bibfnamefont {Y.}~\bibnamefont {Dubi}},\
  }\href@noop {} {\bibfield  {journal} {\bibinfo  {journal} {J. Phys.: Condens.
  Matter}\ }\textbf {\bibinfo {volume} {27}},\ \bibinfo {pages} {263202}
  (\bibinfo {year} {2015})}\BibitemShut {NoStop}%
\bibitem [{\citenamefont {Blum}\ \emph {et~al.}(2005)\citenamefont {Blum},
  \citenamefont {Kushmerick}, \citenamefont {Long}, \citenamefont {Patterson},
  \citenamefont {Yang}, \citenamefont {Henderson}, \citenamefont {Yao},
  \citenamefont {Tour}, \citenamefont {Shashidhar},\ and\ \citenamefont
  {Ratna}}]{blum05}%
  \BibitemOpen
  \bibfield  {author} {\bibinfo {author} {\bibfnamefont {A.~S.}\ \bibnamefont
  {Blum}}, \bibinfo {author} {\bibfnamefont {J.~G.}\ \bibnamefont
  {Kushmerick}}, \bibinfo {author} {\bibfnamefont {D.~P.}\ \bibnamefont
  {Long}}, \bibinfo {author} {\bibfnamefont {C.~H.}\ \bibnamefont {Patterson}},
  \bibinfo {author} {\bibfnamefont {J.~C.}\ \bibnamefont {Yang}}, \bibinfo
  {author} {\bibfnamefont {J.~C.}\ \bibnamefont {Henderson}}, \bibinfo {author}
  {\bibfnamefont {Y.}~\bibnamefont {Yao}}, \bibinfo {author} {\bibfnamefont
  {J.~M.}\ \bibnamefont {Tour}}, \bibinfo {author} {\bibfnamefont
  {R.}~\bibnamefont {Shashidhar}}, \ and\ \bibinfo {author} {\bibfnamefont
  {B.~R.}\ \bibnamefont {Ratna}},\ }\href@noop {} {\bibfield  {journal}
  {\bibinfo  {journal} {Nat. Mater.}\ }\textbf {\bibinfo {volume} {4}},\
  \bibinfo {pages} {167} (\bibinfo {year} {2005})}\BibitemShut {NoStop}%
\bibitem [{\citenamefont {Choi}\ \emph {et~al.}(2006)\citenamefont {Choi},
  \citenamefont {Kahng}, \citenamefont {Kim}, \citenamefont {Kim},
  \citenamefont {Kim}, \citenamefont {Song}, \citenamefont {Ihm},\ and\
  \citenamefont {Kuk}}]{choi06}%
  \BibitemOpen
  \bibfield  {author} {\bibinfo {author} {\bibfnamefont {B.-Y.}\ \bibnamefont
  {Choi}}, \bibinfo {author} {\bibfnamefont {S.-J.}\ \bibnamefont {Kahng}},
  \bibinfo {author} {\bibfnamefont {S.}~\bibnamefont {Kim}}, \bibinfo {author}
  {\bibfnamefont {H.}~\bibnamefont {Kim}}, \bibinfo {author} {\bibfnamefont
  {H.~W.}\ \bibnamefont {Kim}}, \bibinfo {author} {\bibfnamefont {Y.~J.}\
  \bibnamefont {Song}}, \bibinfo {author} {\bibfnamefont {J.}~\bibnamefont
  {Ihm}}, \ and\ \bibinfo {author} {\bibfnamefont {Y.}~\bibnamefont {Kuk}},\
  }\href@noop {} {\bibfield  {journal} {\bibinfo  {journal} {Phys.~Rev.~Lett.}\
  }\textbf {\bibinfo {volume} {96}},\ \bibinfo {pages} {156106} (\bibinfo
  {year} {2006})}\BibitemShut {NoStop}%
\bibitem [{\citenamefont {Ballmann}\ \emph {et~al.}(2012)\citenamefont
  {Ballmann}, \citenamefont {H{\"a}rtle}, \citenamefont {Coto}, \citenamefont
  {Elbing}, \citenamefont {Mayor}, \citenamefont {Bryce}, \citenamefont
  {Thoss},\ and\ \citenamefont {Weber}}]{ballmann12}%
  \BibitemOpen
  \bibfield  {author} {\bibinfo {author} {\bibfnamefont {S.}~\bibnamefont
  {Ballmann}}, \bibinfo {author} {\bibfnamefont {R.}~\bibnamefont
  {H{\"a}rtle}}, \bibinfo {author} {\bibfnamefont {P.~B.}\ \bibnamefont
  {Coto}}, \bibinfo {author} {\bibfnamefont {M.}~\bibnamefont {Elbing}},
  \bibinfo {author} {\bibfnamefont {M.}~\bibnamefont {Mayor}}, \bibinfo
  {author} {\bibfnamefont {M.~R.}\ \bibnamefont {Bryce}}, \bibinfo {author}
  {\bibfnamefont {M.}~\bibnamefont {Thoss}}, \ and\ \bibinfo {author}
  {\bibfnamefont {H.~B.}\ \bibnamefont {Weber}},\ }\href@noop {} {\bibfield
  {journal} {\bibinfo  {journal} {Phys.~Rev.~Lett.}\ }\textbf {\bibinfo
  {volume} {109}},\ \bibinfo {pages} {056801} (\bibinfo {year}
  {2012})}\BibitemShut {NoStop}%
\bibitem [{\citenamefont {Mitra}\ \emph {et~al.}(2004)\citenamefont {Mitra},
  \citenamefont {Aleiner},\ and\ \citenamefont {Millis}}]{mitra04}%
  \BibitemOpen
  \bibfield  {author} {\bibinfo {author} {\bibfnamefont {A.}~\bibnamefont
  {Mitra}}, \bibinfo {author} {\bibfnamefont {I.}~\bibnamefont {Aleiner}}, \
  and\ \bibinfo {author} {\bibfnamefont {A.}~\bibnamefont {Millis}},\
  }\href@noop {} {\bibfield  {journal} {\bibinfo  {journal} {Phys.~Rev.~B}\
  }\textbf {\bibinfo {volume} {69}},\ \bibinfo {pages} {245302} (\bibinfo
  {year} {2004})}\BibitemShut {NoStop}%
\bibitem [{\citenamefont {Donarini}\ \emph {et~al.}(2006)\citenamefont
  {Donarini}, \citenamefont {Grifoni},\ and\ \citenamefont
  {Richter}}]{donarini06}%
  \BibitemOpen
  \bibfield  {author} {\bibinfo {author} {\bibfnamefont {A.}~\bibnamefont
  {Donarini}}, \bibinfo {author} {\bibfnamefont {M.}~\bibnamefont {Grifoni}}, \
  and\ \bibinfo {author} {\bibfnamefont {K.}~\bibnamefont {Richter}},\
  }\href@noop {} {\bibfield  {journal} {\bibinfo  {journal} {Phys.~Rev.~Lett.}\
  }\textbf {\bibinfo {volume} {97}},\ \bibinfo {pages} {166801} (\bibinfo
  {year} {2006})}\BibitemShut {NoStop}%
\bibitem [{\citenamefont {Timm}(2008)}]{timm08}%
  \BibitemOpen
  \bibfield  {author} {\bibinfo {author} {\bibfnamefont {C.}~\bibnamefont
  {Timm}},\ }\href@noop {} {\bibfield  {journal} {\bibinfo  {journal}
  {Phys.~Rev.~B}\ }\textbf {\bibinfo {volume} {77}},\ \bibinfo {pages} {195416}
  (\bibinfo {year} {2008})}\BibitemShut {NoStop}%
\bibitem [{\citenamefont {Leijnse}\ and\ \citenamefont
  {Wegewijs}(2008)}]{leijnse08}%
  \BibitemOpen
  \bibfield  {author} {\bibinfo {author} {\bibfnamefont {M.}~\bibnamefont
  {Leijnse}}\ and\ \bibinfo {author} {\bibfnamefont {M.~R.}\ \bibnamefont
  {Wegewijs}},\ }\href@noop {} {\bibfield  {journal} {\bibinfo  {journal}
  {Phys.~Rev.~B}\ }\textbf {\bibinfo {volume} {78}},\ \bibinfo {pages} {235424}
  (\bibinfo {year} {2008})}\BibitemShut {NoStop}%
\bibitem [{\citenamefont {Esposito}\ and\ \citenamefont
  {Galperin}(2009)}]{esposito09b}%
  \BibitemOpen
  \bibfield  {author} {\bibinfo {author} {\bibfnamefont {M.}~\bibnamefont
  {Esposito}}\ and\ \bibinfo {author} {\bibfnamefont {M.}~\bibnamefont
  {Galperin}},\ }\href@noop {} {\bibfield  {journal} {\bibinfo  {journal}
  {Phys.~Rev.~B}\ }\textbf {\bibinfo {volume} {79}},\ \bibinfo {pages} {205303}
  (\bibinfo {year} {2009})}\BibitemShut {NoStop}%
\bibitem [{\citenamefont {H{\"a}rtle}\ and\ \citenamefont
  {Thoss}(2011)}]{hartle11}%
  \BibitemOpen
  \bibfield  {author} {\bibinfo {author} {\bibfnamefont {R.}~\bibnamefont
  {H{\"a}rtle}}\ and\ \bibinfo {author} {\bibfnamefont {M.}~\bibnamefont
  {Thoss}},\ }\href@noop {} {\bibfield  {journal} {\bibinfo  {journal}
  {Phys.~Rev.~B}\ }\textbf {\bibinfo {volume} {83}},\ \bibinfo {pages} {115414}
  (\bibinfo {year} {2011})}\BibitemShut {NoStop}%
\bibitem [{\citenamefont {Levy}\ \emph {et~al.}(2019)\citenamefont {Levy},
  \citenamefont {Kidon}, \citenamefont {Batge}, \citenamefont {Okamoto},
  \citenamefont {Thoss}, \citenamefont {Limmer},\ and\ \citenamefont
  {Rabani}}]{levy19}%
  \BibitemOpen
  \bibfield  {author} {\bibinfo {author} {\bibfnamefont {A.}~\bibnamefont
  {Levy}}, \bibinfo {author} {\bibfnamefont {L.}~\bibnamefont {Kidon}},
  \bibinfo {author} {\bibfnamefont {J.}~\bibnamefont {Batge}}, \bibinfo
  {author} {\bibfnamefont {J.}~\bibnamefont {Okamoto}}, \bibinfo {author}
  {\bibfnamefont {M.}~\bibnamefont {Thoss}}, \bibinfo {author} {\bibfnamefont
  {D.~T.}\ \bibnamefont {Limmer}}, \ and\ \bibinfo {author} {\bibfnamefont
  {E.}~\bibnamefont {Rabani}},\ }\href@noop {} {\bibfield  {journal} {\bibinfo
  {journal} {J.~Phys.~Chem.~C}\ }\textbf {\bibinfo {volume} {123}},\ \bibinfo
  {pages} {13538} (\bibinfo {year} {2019})}\BibitemShut {NoStop}%
\bibitem [{\citenamefont {Dan}\ \emph {et~al.}(2022)\citenamefont {Dan},
  \citenamefont {Xu}, \citenamefont {Yan},\ and\ \citenamefont {Shi}}]{dan22}%
  \BibitemOpen
  \bibfield  {author} {\bibinfo {author} {\bibfnamefont {X.}~\bibnamefont
  {Dan}}, \bibinfo {author} {\bibfnamefont {M.}~\bibnamefont {Xu}}, \bibinfo
  {author} {\bibfnamefont {Y.}~\bibnamefont {Yan}}, \ and\ \bibinfo {author}
  {\bibfnamefont {Q.}~\bibnamefont {Shi}},\ }\href@noop {} {\bibfield
  {journal} {\bibinfo  {journal} {J.~Chem.~Phys.}\ }\textbf {\bibinfo {volume}
  {156}},\ \bibinfo {pages} {134114} (\bibinfo {year} {2022})}\BibitemShut
  {NoStop}%
\bibitem [{\citenamefont {Anders}\ and\ \citenamefont
  {Schiller}(2005)}]{anders05}%
  \BibitemOpen
  \bibfield  {author} {\bibinfo {author} {\bibfnamefont {F.~B.}\ \bibnamefont
  {Anders}}\ and\ \bibinfo {author} {\bibfnamefont {A.}~\bibnamefont
  {Schiller}},\ }\href {\doibase 10.1103/PhysRevLett.95.196801} {\bibfield
  {journal} {\bibinfo  {journal} {Phys. Rev. Lett.}\ }\textbf {\bibinfo
  {volume} {95}},\ \bibinfo {pages} {196801} (\bibinfo {year}
  {2005})}\BibitemShut {NoStop}%
\bibitem [{\citenamefont {Anders}\ and\ \citenamefont
  {Schiller}(2006)}]{anders06}%
  \BibitemOpen
  \bibfield  {author} {\bibinfo {author} {\bibfnamefont {F.~B.}\ \bibnamefont
  {Anders}}\ and\ \bibinfo {author} {\bibfnamefont {A.}~\bibnamefont
  {Schiller}},\ }\href@noop {} {\bibfield  {journal} {\bibinfo  {journal}
  {Phys.~Rev.~B}\ }\textbf {\bibinfo {volume} {74}},\ \bibinfo {pages} {245113}
  (\bibinfo {year} {2006})}\BibitemShut {NoStop}%
\bibitem [{\citenamefont {Bulla}\ \emph {et~al.}(2008)\citenamefont {Bulla},
  \citenamefont {Costi},\ and\ \citenamefont {Pruschke}}]{bulla08}%
  \BibitemOpen
  \bibfield  {author} {\bibinfo {author} {\bibfnamefont {R.}~\bibnamefont
  {Bulla}}, \bibinfo {author} {\bibfnamefont {T.~A.}\ \bibnamefont {Costi}}, \
  and\ \bibinfo {author} {\bibfnamefont {T.}~\bibnamefont {Pruschke}},\
  }\href@noop {} {\bibfield  {journal} {\bibinfo  {journal} {Rev.~Mod.~Phys.}\
  }\textbf {\bibinfo {volume} {80}},\ \bibinfo {pages} {395} (\bibinfo {year}
  {2008})}\BibitemShut {NoStop}%
\bibitem [{\citenamefont {Nghiem}\ and\ \citenamefont
  {Costi}(2018)}]{nghiem18}%
  \BibitemOpen
  \bibfield  {author} {\bibinfo {author} {\bibfnamefont {H.~T.~M.}\
  \bibnamefont {Nghiem}}\ and\ \bibinfo {author} {\bibfnamefont {T.~A.}\
  \bibnamefont {Costi}},\ }\href {\doibase 10.1103/PhysRevB.98.155107}
  {\bibfield  {journal} {\bibinfo  {journal} {Phys. Rev. B}\ }\textbf {\bibinfo
  {volume} {98}},\ \bibinfo {pages} {155107} (\bibinfo {year}
  {2018})}\BibitemShut {NoStop}%
\bibitem [{\citenamefont {de~Souza~Melo}\ \emph {et~al.}(2019)\citenamefont
  {de~Souza~Melo}, \citenamefont {da~Silva}, \citenamefont {Rocha},\ and\
  \citenamefont {Lewenkopf}}]{de19}%
  \BibitemOpen
  \bibfield  {author} {\bibinfo {author} {\bibfnamefont {B.~M.}\ \bibnamefont
  {de~Souza~Melo}}, \bibinfo {author} {\bibfnamefont {L.~G. G. V.~D.}\
  \bibnamefont {da~Silva}}, \bibinfo {author} {\bibfnamefont {A.~R.}\
  \bibnamefont {Rocha}}, \ and\ \bibinfo {author} {\bibfnamefont
  {C.}~\bibnamefont {Lewenkopf}},\ }\href@noop {} {\bibfield  {journal}
  {\bibinfo  {journal} {J. Phys.: Condens. Matter}\ }\textbf {\bibinfo {volume}
  {32}},\ \bibinfo {pages} {095602} (\bibinfo {year} {2019})}\BibitemShut
  {NoStop}%
\bibitem [{\citenamefont {Schollw\"ock}(2005)}]{schollwock05}%
  \BibitemOpen
  \bibfield  {author} {\bibinfo {author} {\bibfnamefont {U.}~\bibnamefont
  {Schollw\"ock}},\ }\href {\doibase 10.1103/RevModPhys.77.259} {\bibfield
  {journal} {\bibinfo  {journal} {Rev. Mod. Phys.}\ }\textbf {\bibinfo {volume}
  {77}},\ \bibinfo {pages} {259} (\bibinfo {year} {2005})}\BibitemShut
  {NoStop}%
\bibitem [{\citenamefont {Heidrich-Meisner}\ \emph {et~al.}(2009)\citenamefont
  {Heidrich-Meisner}, \citenamefont {Feiguin},\ and\ \citenamefont
  {Dagotto}}]{heidrich09}%
  \BibitemOpen
  \bibfield  {author} {\bibinfo {author} {\bibfnamefont {F.}~\bibnamefont
  {Heidrich-Meisner}}, \bibinfo {author} {\bibfnamefont {A.~E.}\ \bibnamefont
  {Feiguin}}, \ and\ \bibinfo {author} {\bibfnamefont {E.}~\bibnamefont
  {Dagotto}},\ }\href@noop {} {\bibfield  {journal} {\bibinfo  {journal}
  {Phys.~Rev.~B}\ }\textbf {\bibinfo {volume} {79}},\ \bibinfo {pages} {235336}
  (\bibinfo {year} {2009})}\BibitemShut {NoStop}%
\bibitem [{\citenamefont {He}\ and\ \citenamefont {Millis}(2019)}]{he19}%
  \BibitemOpen
  \bibfield  {author} {\bibinfo {author} {\bibfnamefont {Z.}~\bibnamefont
  {He}}\ and\ \bibinfo {author} {\bibfnamefont {A.~J.}\ \bibnamefont
  {Millis}},\ }\href@noop {} {\bibfield  {journal} {\bibinfo  {journal}
  {Phys.~Rev.~B}\ }\textbf {\bibinfo {volume} {99}},\ \bibinfo {pages} {205138}
  (\bibinfo {year} {2019})}\BibitemShut {NoStop}%
\bibitem [{\citenamefont {Kohn}\ and\ \citenamefont {Santoro}(2021)}]{kohn21}%
  \BibitemOpen
  \bibfield  {author} {\bibinfo {author} {\bibfnamefont {L.}~\bibnamefont
  {Kohn}}\ and\ \bibinfo {author} {\bibfnamefont {G.~E.}\ \bibnamefont
  {Santoro}},\ }\href@noop {} {\bibfield  {journal} {\bibinfo  {journal}
  {Phys.~Rev.~B}\ }\textbf {\bibinfo {volume} {104}},\ \bibinfo {pages}
  {014303} (\bibinfo {year} {2021})}\BibitemShut {NoStop}%
\bibitem [{\citenamefont {Wang}\ and\ \citenamefont {Thoss}(2009)}]{wang09}%
  \BibitemOpen
  \bibfield  {author} {\bibinfo {author} {\bibfnamefont {H.}~\bibnamefont
  {Wang}}\ and\ \bibinfo {author} {\bibfnamefont {M.}~\bibnamefont {Thoss}},\
  }\href@noop {} {\bibfield  {journal} {\bibinfo  {journal} {J.~Chem.~Phys.}\
  }\textbf {\bibinfo {volume} {131}},\ \bibinfo {pages} {024114} (\bibinfo
  {year} {2009})}\BibitemShut {NoStop}%
\bibitem [{\citenamefont {Wang}\ \emph {et~al.}(2011)\citenamefont {Wang},
  \citenamefont {Pshenichnyuk}, \citenamefont {Haertle},\ and\ \citenamefont
  {Thoss}}]{wang11}%
  \BibitemOpen
  \bibfield  {author} {\bibinfo {author} {\bibfnamefont {H.}~\bibnamefont
  {Wang}}, \bibinfo {author} {\bibfnamefont {I.}~\bibnamefont {Pshenichnyuk}},
  \bibinfo {author} {\bibfnamefont {R.}~\bibnamefont {Haertle}}, \ and\
  \bibinfo {author} {\bibfnamefont {M.}~\bibnamefont {Thoss}},\ }\href@noop {}
  {\bibfield  {journal} {\bibinfo  {journal} {J.~Chem.~Phys.}\ }\textbf
  {\bibinfo {volume} {135}},\ \bibinfo {pages} {244506} (\bibinfo {year}
  {2011})}\BibitemShut {NoStop}%
\bibitem [{\citenamefont {Wang}\ and\ \citenamefont {Thoss}(2018)}]{wang18}%
  \BibitemOpen
  \bibfield  {author} {\bibinfo {author} {\bibfnamefont {H.}~\bibnamefont
  {Wang}}\ and\ \bibinfo {author} {\bibfnamefont {M.}~\bibnamefont {Thoss}},\
  }\href@noop {} {\bibfield  {journal} {\bibinfo  {journal} {Chem.~Phys.}\
  }\textbf {\bibinfo {volume} {509}},\ \bibinfo {pages} {13} (\bibinfo {year}
  {2018})}\BibitemShut {NoStop}%
\bibitem [{\citenamefont {Nishimoto}\ and\ \citenamefont
  {Jeckelmann}(2004)}]{nishimoto04}%
  \BibitemOpen
  \bibfield  {author} {\bibinfo {author} {\bibfnamefont {S.}~\bibnamefont
  {Nishimoto}}\ and\ \bibinfo {author} {\bibfnamefont {E.}~\bibnamefont
  {Jeckelmann}},\ }\href@noop {} {\bibfield  {journal} {\bibinfo  {journal} {J.
  Phys.: Condens. Matter}\ }\textbf {\bibinfo {volume} {16}},\ \bibinfo {pages}
  {613} (\bibinfo {year} {2004})}\BibitemShut {NoStop}%
\bibitem [{\citenamefont {\ifmmode~\check{Z}\else \v{Z}\fi{}itko}\ and\
  \citenamefont {Pruschke}(2009)}]{zitko09}%
  \BibitemOpen
  \bibfield  {author} {\bibinfo {author} {\bibfnamefont {R.}~\bibnamefont
  {\ifmmode~\check{Z}\else \v{Z}\fi{}itko}}\ and\ \bibinfo {author}
  {\bibfnamefont {T.}~\bibnamefont {Pruschke}},\ }\href {\doibase
  10.1103/PhysRevB.79.085106} {\bibfield  {journal} {\bibinfo  {journal} {Phys.
  Rev. B}\ }\textbf {\bibinfo {volume} {79}},\ \bibinfo {pages} {085106}
  (\bibinfo {year} {2009})}\BibitemShut {NoStop}%
\bibitem [{\citenamefont {Weiss}\ \emph {et~al.}(2008)\citenamefont {Weiss},
  \citenamefont {Eckel}, \citenamefont {Thorwart},\ and\ \citenamefont
  {Egger}}]{weiss08b}%
  \BibitemOpen
  \bibfield  {author} {\bibinfo {author} {\bibfnamefont {S.}~\bibnamefont
  {Weiss}}, \bibinfo {author} {\bibfnamefont {J.}~\bibnamefont {Eckel}},
  \bibinfo {author} {\bibfnamefont {M.}~\bibnamefont {Thorwart}}, \ and\
  \bibinfo {author} {\bibfnamefont {R.}~\bibnamefont {Egger}},\ }\href@noop {}
  {\bibfield  {journal} {\bibinfo  {journal} {Phys.~Rev.~B}\ }\textbf {\bibinfo
  {volume} {77}},\ \bibinfo {pages} {195316} (\bibinfo {year}
  {2008})}\BibitemShut {NoStop}%
\bibitem [{\citenamefont {Segal}\ \emph {et~al.}(2010)\citenamefont {Segal},
  \citenamefont {Millis},\ and\ \citenamefont {Reichman}}]{segal10}%
  \BibitemOpen
  \bibfield  {author} {\bibinfo {author} {\bibfnamefont {D.}~\bibnamefont
  {Segal}}, \bibinfo {author} {\bibfnamefont {A.~J.}\ \bibnamefont {Millis}}, \
  and\ \bibinfo {author} {\bibfnamefont {D.~R.}\ \bibnamefont {Reichman}},\
  }\href@noop {} {\bibfield  {journal} {\bibinfo  {journal} {Phys.~Rev.~B}\
  }\textbf {\bibinfo {volume} {82}},\ \bibinfo {pages} {205323} (\bibinfo
  {year} {2010})}\BibitemShut {NoStop}%
\bibitem [{\citenamefont {Weiss}\ \emph {et~al.}(2013)\citenamefont {Weiss},
  \citenamefont {H{\"u}tzen}, \citenamefont {Becker}, \citenamefont {Eckel},
  \citenamefont {Egger},\ and\ \citenamefont {Thorwart}}]{weiss13}%
  \BibitemOpen
  \bibfield  {author} {\bibinfo {author} {\bibfnamefont {S.}~\bibnamefont
  {Weiss}}, \bibinfo {author} {\bibfnamefont {R.}~\bibnamefont {H{\"u}tzen}},
  \bibinfo {author} {\bibfnamefont {D.}~\bibnamefont {Becker}}, \bibinfo
  {author} {\bibfnamefont {J.}~\bibnamefont {Eckel}}, \bibinfo {author}
  {\bibfnamefont {R.}~\bibnamefont {Egger}}, \ and\ \bibinfo {author}
  {\bibfnamefont {M.}~\bibnamefont {Thorwart}},\ }\href@noop {} {\bibfield
  {journal} {\bibinfo  {journal} {Phys. Status Solidi B}\ }\textbf {\bibinfo
  {volume} {250}},\ \bibinfo {pages} {2298} (\bibinfo {year}
  {2013})}\BibitemShut {NoStop}%
\bibitem [{\citenamefont {M{\"u}hlbacher}\ and\ \citenamefont
  {Rabani}(2008)}]{muhlbacher08}%
  \BibitemOpen
  \bibfield  {author} {\bibinfo {author} {\bibfnamefont {L.}~\bibnamefont
  {M{\"u}hlbacher}}\ and\ \bibinfo {author} {\bibfnamefont {E.}~\bibnamefont
  {Rabani}},\ }\href@noop {} {\bibfield  {journal} {\bibinfo  {journal}
  {Phys.~Rev.~Lett.}\ }\textbf {\bibinfo {volume} {100}},\ \bibinfo {pages}
  {176403} (\bibinfo {year} {2008})}\BibitemShut {NoStop}%
\bibitem [{\citenamefont {Werner}\ \emph {et~al.}(2009)\citenamefont {Werner},
  \citenamefont {Oka},\ and\ \citenamefont {Millis}}]{werner09}%
  \BibitemOpen
  \bibfield  {author} {\bibinfo {author} {\bibfnamefont {P.}~\bibnamefont
  {Werner}}, \bibinfo {author} {\bibfnamefont {T.}~\bibnamefont {Oka}}, \ and\
  \bibinfo {author} {\bibfnamefont {A.~J.}\ \bibnamefont {Millis}},\
  }\href@noop {} {\bibfield  {journal} {\bibinfo  {journal} {Phys.~Rev.~B}\
  }\textbf {\bibinfo {volume} {79}},\ \bibinfo {pages} {035320} (\bibinfo
  {year} {2009})}\BibitemShut {NoStop}%
\bibitem [{\citenamefont {Schir{\'o}}\ and\ \citenamefont
  {Fabrizio}(2009)}]{schiro09}%
  \BibitemOpen
  \bibfield  {author} {\bibinfo {author} {\bibfnamefont {M.}~\bibnamefont
  {Schir{\'o}}}\ and\ \bibinfo {author} {\bibfnamefont {M.}~\bibnamefont
  {Fabrizio}},\ }\href@noop {} {\bibfield  {journal} {\bibinfo  {journal}
  {Phys.~Rev.~B}\ }\textbf {\bibinfo {volume} {79}},\ \bibinfo {pages} {153302}
  (\bibinfo {year} {2009})}\BibitemShut {NoStop}%
\bibitem [{\citenamefont {Gull}\ \emph {et~al.}(2011)\citenamefont {Gull},
  \citenamefont {Millis}, \citenamefont {Lichtenstein}, \citenamefont
  {Rubtsov}, \citenamefont {Troyer},\ and\ \citenamefont {Werner}}]{gull11}%
  \BibitemOpen
  \bibfield  {author} {\bibinfo {author} {\bibfnamefont {E.}~\bibnamefont
  {Gull}}, \bibinfo {author} {\bibfnamefont {A.~J.}\ \bibnamefont {Millis}},
  \bibinfo {author} {\bibfnamefont {A.~I.}\ \bibnamefont {Lichtenstein}},
  \bibinfo {author} {\bibfnamefont {A.~N.}\ \bibnamefont {Rubtsov}}, \bibinfo
  {author} {\bibfnamefont {M.}~\bibnamefont {Troyer}}, \ and\ \bibinfo {author}
  {\bibfnamefont {P.}~\bibnamefont {Werner}},\ }\href@noop {} {\bibfield
  {journal} {\bibinfo  {journal} {Rev.~Mod.~Phys.}\ }\textbf {\bibinfo {volume}
  {83}},\ \bibinfo {pages} {349} (\bibinfo {year} {2011})}\BibitemShut
  {NoStop}%
\bibitem [{\citenamefont {Antipov}\ \emph {et~al.}(2016)\citenamefont
  {Antipov}, \citenamefont {Dong},\ and\ \citenamefont {Gull}}]{antipov16}%
  \BibitemOpen
  \bibfield  {author} {\bibinfo {author} {\bibfnamefont {A.~E.}\ \bibnamefont
  {Antipov}}, \bibinfo {author} {\bibfnamefont {Q.}~\bibnamefont {Dong}}, \
  and\ \bibinfo {author} {\bibfnamefont {E.}~\bibnamefont {Gull}},\ }\href@noop
  {} {\bibfield  {journal} {\bibinfo  {journal} {Phys.~Rev.~Lett.}\ }\textbf
  {\bibinfo {volume} {116}},\ \bibinfo {pages} {036801} (\bibinfo {year}
  {2016})}\BibitemShut {NoStop}%
\bibitem [{\citenamefont {Krivenko}\ \emph {et~al.}(2019)\citenamefont
  {Krivenko}, \citenamefont {Kleinhenz}, \citenamefont {Cohen},\ and\
  \citenamefont {Gull}}]{krivenko19}%
  \BibitemOpen
  \bibfield  {author} {\bibinfo {author} {\bibfnamefont {I.}~\bibnamefont
  {Krivenko}}, \bibinfo {author} {\bibfnamefont {J.}~\bibnamefont {Kleinhenz}},
  \bibinfo {author} {\bibfnamefont {G.}~\bibnamefont {Cohen}}, \ and\ \bibinfo
  {author} {\bibfnamefont {E.}~\bibnamefont {Gull}},\ }\href@noop {} {\bibfield
   {journal} {\bibinfo  {journal} {Phys.~Rev.~B}\ }\textbf {\bibinfo {volume}
  {100}},\ \bibinfo {pages} {201104} (\bibinfo {year} {2019})}\BibitemShut
  {NoStop}%
\bibitem [{\citenamefont {Cohen}\ \emph {et~al.}(2015)\citenamefont {Cohen},
  \citenamefont {Gull}, \citenamefont {Reichman},\ and\ \citenamefont
  {Millis}}]{cohen15}%
  \BibitemOpen
  \bibfield  {author} {\bibinfo {author} {\bibfnamefont {G.}~\bibnamefont
  {Cohen}}, \bibinfo {author} {\bibfnamefont {E.}~\bibnamefont {Gull}},
  \bibinfo {author} {\bibfnamefont {D.~R.}\ \bibnamefont {Reichman}}, \ and\
  \bibinfo {author} {\bibfnamefont {A.~J.}\ \bibnamefont {Millis}},\
  }\href@noop {} {\bibfield  {journal} {\bibinfo  {journal} {Phys.~Rev.~Lett.}\
  }\textbf {\bibinfo {volume} {115}},\ \bibinfo {pages} {266802} (\bibinfo
  {year} {2015})}\BibitemShut {NoStop}%
\bibitem [{\citenamefont {Antipov}\ \emph {et~al.}(2017)\citenamefont
  {Antipov}, \citenamefont {Dong}, \citenamefont {Kleinhenz}, \citenamefont
  {Cohen},\ and\ \citenamefont {Gull}}]{antipov17}%
  \BibitemOpen
  \bibfield  {author} {\bibinfo {author} {\bibfnamefont {A.~E.}\ \bibnamefont
  {Antipov}}, \bibinfo {author} {\bibfnamefont {Q.}~\bibnamefont {Dong}},
  \bibinfo {author} {\bibfnamefont {J.}~\bibnamefont {Kleinhenz}}, \bibinfo
  {author} {\bibfnamefont {G.}~\bibnamefont {Cohen}}, \ and\ \bibinfo {author}
  {\bibfnamefont {E.}~\bibnamefont {Gull}},\ }\href {\doibase
  10.1103/PhysRevB.95.085144} {\bibfield  {journal} {\bibinfo  {journal} {Phys.
  Rev. B}\ }\textbf {\bibinfo {volume} {95}},\ \bibinfo {pages} {085144}
  (\bibinfo {year} {2017})}\BibitemShut {NoStop}%
\bibitem [{\citenamefont {Jin}\ \emph {et~al.}(2008)\citenamefont {Jin},
  \citenamefont {Zheng},\ and\ \citenamefont {Yan}}]{jin08}%
  \BibitemOpen
  \bibfield  {author} {\bibinfo {author} {\bibfnamefont {J.-S.}\ \bibnamefont
  {Jin}}, \bibinfo {author} {\bibfnamefont {X.}~\bibnamefont {Zheng}}, \ and\
  \bibinfo {author} {\bibfnamefont {Y.-J.}\ \bibnamefont {Yan}},\ }\href@noop
  {} {\bibfield  {journal} {\bibinfo  {journal} {J. Chem. Phys.}\ }\textbf
  {\bibinfo {volume} {128}},\ \bibinfo {pages} {234703} (\bibinfo {year}
  {2008})}\BibitemShut {NoStop}%
\bibitem [{\citenamefont {Li}\ \emph {et~al.}(2012)\citenamefont {Li},
  \citenamefont {Tong}, \citenamefont {Zheng}, \citenamefont {Hou},
  \citenamefont {Wei}, \citenamefont {Hu},\ and\ \citenamefont {Yan}}]{li12}%
  \BibitemOpen
  \bibfield  {author} {\bibinfo {author} {\bibfnamefont {Z.-H.}\ \bibnamefont
  {Li}}, \bibinfo {author} {\bibfnamefont {N.-H.}\ \bibnamefont {Tong}},
  \bibinfo {author} {\bibfnamefont {X.}~\bibnamefont {Zheng}}, \bibinfo
  {author} {\bibfnamefont {D.}~\bibnamefont {Hou}}, \bibinfo {author}
  {\bibfnamefont {J.-H.}\ \bibnamefont {Wei}}, \bibinfo {author} {\bibfnamefont
  {J.}~\bibnamefont {Hu}}, \ and\ \bibinfo {author} {\bibfnamefont {Y.-J.}\
  \bibnamefont {Yan}},\ }\href@noop {} {\bibfield  {journal} {\bibinfo
  {journal} {Phys.~Rev.~Lett.}\ }\textbf {\bibinfo {volume} {109}},\ \bibinfo
  {pages} {266403} (\bibinfo {year} {2012})}\BibitemShut {NoStop}%
\bibitem [{\citenamefont {H\"{a}rtle}\ \emph {et~al.}(2013)\citenamefont
  {H\"{a}rtle}, \citenamefont {Cohen}, \citenamefont {Reichman},\ and\
  \citenamefont {Millis}}]{hartle13}%
  \BibitemOpen
  \bibfield  {author} {\bibinfo {author} {\bibfnamefont {R.}~\bibnamefont
  {H\"{a}rtle}}, \bibinfo {author} {\bibfnamefont {G.}~\bibnamefont {Cohen}},
  \bibinfo {author} {\bibfnamefont {D.~R.}\ \bibnamefont {Reichman}}, \ and\
  \bibinfo {author} {\bibfnamefont {A.~J.}\ \bibnamefont {Millis}},\
  }\href@noop {} {\bibfield  {journal} {\bibinfo  {journal} {Phys.~Rev.~B}\
  }\textbf {\bibinfo {volume} {88}},\ \bibinfo {pages} {235426} (\bibinfo
  {year} {2013})}\BibitemShut {NoStop}%
\bibitem [{\citenamefont {H{\"a}rtle}\ \emph {et~al.}(2015)\citenamefont
  {H{\"a}rtle}, \citenamefont {Cohen}, \citenamefont {Reichman},\ and\
  \citenamefont {Millis}}]{hartle15b}%
  \BibitemOpen
  \bibfield  {author} {\bibinfo {author} {\bibfnamefont {R.}~\bibnamefont
  {H{\"a}rtle}}, \bibinfo {author} {\bibfnamefont {G.}~\bibnamefont {Cohen}},
  \bibinfo {author} {\bibfnamefont {D.}~\bibnamefont {Reichman}}, \ and\
  \bibinfo {author} {\bibfnamefont {A.}~\bibnamefont {Millis}},\ }\href@noop {}
  {\bibfield  {journal} {\bibinfo  {journal} {Phys.~Rev.~B}\ }\textbf {\bibinfo
  {volume} {92}},\ \bibinfo {pages} {085430} (\bibinfo {year}
  {2015})}\BibitemShut {NoStop}%
\bibitem [{\citenamefont {Shi}\ \emph {et~al.}(2018)\citenamefont {Shi},
  \citenamefont {Xu}, \citenamefont {Yan},\ and\ \citenamefont {Xu}}]{shi18}%
  \BibitemOpen
  \bibfield  {author} {\bibinfo {author} {\bibfnamefont {Q.}~\bibnamefont
  {Shi}}, \bibinfo {author} {\bibfnamefont {Y.}~\bibnamefont {Xu}}, \bibinfo
  {author} {\bibfnamefont {Y.}~\bibnamefont {Yan}}, \ and\ \bibinfo {author}
  {\bibfnamefont {M.}~\bibnamefont {Xu}},\ }\href@noop {} {\bibfield  {journal}
  {\bibinfo  {journal} {J.~Chem.~Phys.}\ }\textbf {\bibinfo {volume} {148}},\
  \bibinfo {pages} {174102} (\bibinfo {year} {2018})}\BibitemShut {NoStop}%
\bibitem [{\citenamefont {Erpenbeck}\ \emph {et~al.}(2019)\citenamefont
  {Erpenbeck}, \citenamefont {G{\"o}tzend{\"o}rfer}, \citenamefont
  {Schinabeck},\ and\ \citenamefont {Thoss}}]{erpenbeck19}%
  \BibitemOpen
  \bibfield  {author} {\bibinfo {author} {\bibfnamefont {A.}~\bibnamefont
  {Erpenbeck}}, \bibinfo {author} {\bibfnamefont {L.}~\bibnamefont
  {G{\"o}tzend{\"o}rfer}}, \bibinfo {author} {\bibfnamefont {C.}~\bibnamefont
  {Schinabeck}}, \ and\ \bibinfo {author} {\bibfnamefont {M.}~\bibnamefont
  {Thoss}},\ }\href@noop {} {\bibfield  {journal} {\bibinfo  {journal} {Eur.
  Phys. J.-Spec. Top.}\ }\textbf {\bibinfo {volume} {227}},\ \bibinfo {pages}
  {1981} (\bibinfo {year} {2019})}\BibitemShut {NoStop}%
\bibitem [{\citenamefont {Zhang}\ \emph {et~al.}(2020)\citenamefont {Zhang},
  \citenamefont {Cui}, \citenamefont {Gong}, \citenamefont {Xu}, \citenamefont
  {Zheng},\ and\ \citenamefont {Yan}}]{zhang20}%
  \BibitemOpen
  \bibfield  {author} {\bibinfo {author} {\bibfnamefont {H.-D.}\ \bibnamefont
  {Zhang}}, \bibinfo {author} {\bibfnamefont {L.}~\bibnamefont {Cui}}, \bibinfo
  {author} {\bibfnamefont {H.}~\bibnamefont {Gong}}, \bibinfo {author}
  {\bibfnamefont {R.-X.}\ \bibnamefont {Xu}}, \bibinfo {author} {\bibfnamefont
  {X.}~\bibnamefont {Zheng}}, \ and\ \bibinfo {author} {\bibfnamefont
  {Y.}~\bibnamefont {Yan}},\ }\href@noop {} {\bibfield  {journal} {\bibinfo
  {journal} {J.~Chem.~Phys.}\ }\textbf {\bibinfo {volume} {152}},\ \bibinfo
  {pages} {064107} (\bibinfo {year} {2020})}\BibitemShut {NoStop}%
\bibitem [{\citenamefont {Tanimura}\ and\ \citenamefont
  {Kubo}(1989)}]{tanimura89}%
  \BibitemOpen
  \bibfield  {author} {\bibinfo {author} {\bibfnamefont {Y.}~\bibnamefont
  {Tanimura}}\ and\ \bibinfo {author} {\bibfnamefont {R.}~\bibnamefont
  {Kubo}},\ }\href@noop {} {\bibfield  {journal} {\bibinfo  {journal} {J. Phys.
  Soc. Jpn.}\ }\textbf {\bibinfo {volume} {58}},\ \bibinfo {pages} {101}
  (\bibinfo {year} {1989})}\BibitemShut {NoStop}%
\bibitem [{\citenamefont {Tanimura}(2006)}]{tanimura06}%
  \BibitemOpen
  \bibfield  {author} {\bibinfo {author} {\bibfnamefont {Y.}~\bibnamefont
  {Tanimura}},\ }\href@noop {} {\bibfield  {journal} {\bibinfo  {journal} {J.
  Phys. Soc. Jpn.}\ }\textbf {\bibinfo {volume} {75}},\ \bibinfo {pages}
  {082001} (\bibinfo {year} {2006})}\BibitemShut {NoStop}%
\bibitem [{\citenamefont {Jin}\ \emph {et~al.}(2007)\citenamefont {Jin},
  \citenamefont {Welack}, \citenamefont {Luo}, \citenamefont {Li},
  \citenamefont {Cui}, \citenamefont {Xu},\ and\ \citenamefont {Yan}}]{jin07}%
  \BibitemOpen
  \bibfield  {author} {\bibinfo {author} {\bibfnamefont {J.-S.}\ \bibnamefont
  {Jin}}, \bibinfo {author} {\bibfnamefont {S.}~\bibnamefont {Welack}},
  \bibinfo {author} {\bibfnamefont {J.-Y.}\ \bibnamefont {Luo}}, \bibinfo
  {author} {\bibfnamefont {X.-Q.}\ \bibnamefont {Li}}, \bibinfo {author}
  {\bibfnamefont {P.}~\bibnamefont {Cui}}, \bibinfo {author} {\bibfnamefont
  {R.-X.}\ \bibnamefont {Xu}}, \ and\ \bibinfo {author} {\bibfnamefont {Y.-J.}\
  \bibnamefont {Yan}},\ }\href@noop {} {\bibfield  {journal} {\bibinfo
  {journal} {J. Chem. Phys.}\ }\textbf {\bibinfo {volume} {126}},\ \bibinfo
  {pages} {134113} (\bibinfo {year} {2007})}\BibitemShut {NoStop}%
\bibitem [{\citenamefont {Zheng}\ \emph {et~al.}(2013)\citenamefont {Zheng},
  \citenamefont {Yan},\ and\ \citenamefont {{Di~Ventra}}}]{zheng13}%
  \BibitemOpen
  \bibfield  {author} {\bibinfo {author} {\bibfnamefont {X.}~\bibnamefont
  {Zheng}}, \bibinfo {author} {\bibfnamefont {Y.}~\bibnamefont {Yan}}, \ and\
  \bibinfo {author} {\bibfnamefont {M.}~\bibnamefont {{Di~Ventra}}},\
  }\href@noop {} {\bibfield  {journal} {\bibinfo  {journal} {Phys.~Rev.~Lett.}\
  }\textbf {\bibinfo {volume} {111}},\ \bibinfo {pages} {086601} (\bibinfo
  {year} {2013})}\BibitemShut {NoStop}%
\bibitem [{\citenamefont {Wang}\ \emph
  {et~al.}(2016{\natexlab{a}})\citenamefont {Wang}, \citenamefont {Hou},
  \citenamefont {Zheng},\ and\ \citenamefont {Yan}}]{wang16}%
  \BibitemOpen
  \bibfield  {author} {\bibinfo {author} {\bibfnamefont {X.}~\bibnamefont
  {Wang}}, \bibinfo {author} {\bibfnamefont {D.}~\bibnamefont {Hou}}, \bibinfo
  {author} {\bibfnamefont {X.}~\bibnamefont {Zheng}}, \ and\ \bibinfo {author}
  {\bibfnamefont {Y.}~\bibnamefont {Yan}},\ }\href {\doibase 10.1063/1.4939843}
  {\bibfield  {journal} {\bibinfo  {journal} {J.~Chem.~Phys.}\ }\textbf
  {\bibinfo {volume} {144}},\ \bibinfo {pages} {034101} (\bibinfo {year}
  {2016}{\natexlab{a}})}\BibitemShut {NoStop}%
\bibitem [{\citenamefont {Wang}\ \emph
  {et~al.}(2016{\natexlab{b}})\citenamefont {Wang}, \citenamefont {Zheng},\
  and\ \citenamefont {Yang}}]{wang16c}%
  \BibitemOpen
  \bibfield  {author} {\bibinfo {author} {\bibfnamefont {Y.}~\bibnamefont
  {Wang}}, \bibinfo {author} {\bibfnamefont {X.}~\bibnamefont {Zheng}}, \ and\
  \bibinfo {author} {\bibfnamefont {J.}~\bibnamefont {Yang}},\ }\href@noop {}
  {\bibfield  {journal} {\bibinfo  {journal} {J.~Chem.~Phys.}\ }\textbf
  {\bibinfo {volume} {145}},\ \bibinfo {pages} {154301} (\bibinfo {year}
  {2016}{\natexlab{b}})}\BibitemShut {NoStop}%
\bibitem [{\citenamefont {Schinabeck}\ \emph {et~al.}(2016)\citenamefont
  {Schinabeck}, \citenamefont {Erpenbeck}, \citenamefont {H{\"a}rtle},\ and\
  \citenamefont {Thoss}}]{schinabeck16}%
  \BibitemOpen
  \bibfield  {author} {\bibinfo {author} {\bibfnamefont {C.}~\bibnamefont
  {Schinabeck}}, \bibinfo {author} {\bibfnamefont {A.}~\bibnamefont
  {Erpenbeck}}, \bibinfo {author} {\bibfnamefont {R.}~\bibnamefont
  {H{\"a}rtle}}, \ and\ \bibinfo {author} {\bibfnamefont {M.}~\bibnamefont
  {Thoss}},\ }\href@noop {} {\bibfield  {journal} {\bibinfo  {journal}
  {Phys.~Rev.~B}\ }\textbf {\bibinfo {volume} {94}},\ \bibinfo {pages} {201407}
  (\bibinfo {year} {2016})}\BibitemShut {NoStop}%
\bibitem [{\citenamefont {Tanimura}(1990)}]{tanimura90}%
  \BibitemOpen
  \bibfield  {author} {\bibinfo {author} {\bibfnamefont {Y.}~\bibnamefont
  {Tanimura}},\ }\href@noop {} {\bibfield  {journal} {\bibinfo  {journal}
  {Phys.~Rev.~A}\ }\textbf {\bibinfo {volume} {41}},\ \bibinfo {pages} {6676}
  (\bibinfo {year} {1990})}\BibitemShut {NoStop}%
\bibitem [{\citenamefont {Ozaki}(2007)}]{ozaki07}%
  \BibitemOpen
  \bibfield  {author} {\bibinfo {author} {\bibfnamefont {T.}~\bibnamefont
  {Ozaki}},\ }\href@noop {} {\bibfield  {journal} {\bibinfo  {journal}
  {Phys.~Rev.~B}\ }\textbf {\bibinfo {volume} {75}},\ \bibinfo {pages} {035123}
  (\bibinfo {year} {2007})}\BibitemShut {NoStop}%
\bibitem [{\citenamefont {Hu}\ \emph {et~al.}(2010)\citenamefont {Hu},
  \citenamefont {Xu},\ and\ \citenamefont {Yan}}]{hu10}%
  \BibitemOpen
  \bibfield  {author} {\bibinfo {author} {\bibfnamefont {J.}~\bibnamefont
  {Hu}}, \bibinfo {author} {\bibfnamefont {R.-X.}\ \bibnamefont {Xu}}, \ and\
  \bibinfo {author} {\bibfnamefont {Y.-J.}\ \bibnamefont {Yan}},\ }\href@noop
  {} {\bibfield  {journal} {\bibinfo  {journal} {J.~Chem.~Phys.}\ }\textbf
  {\bibinfo {volume} {133}},\ \bibinfo {pages} {101106} (\bibinfo {year}
  {2010})}\BibitemShut {NoStop}%
\bibitem [{\citenamefont {Hu}\ \emph {et~al.}(2011)\citenamefont {Hu},
  \citenamefont {Luo}, \citenamefont {Jiang}, \citenamefont {Xu},\ and\
  \citenamefont {Yan}}]{hu11}%
  \BibitemOpen
  \bibfield  {author} {\bibinfo {author} {\bibfnamefont {J.}~\bibnamefont
  {Hu}}, \bibinfo {author} {\bibfnamefont {M.}~\bibnamefont {Luo}}, \bibinfo
  {author} {\bibfnamefont {F.}~\bibnamefont {Jiang}}, \bibinfo {author}
  {\bibfnamefont {R.-X.}\ \bibnamefont {Xu}}, \ and\ \bibinfo {author}
  {\bibfnamefont {Y.-J.}\ \bibnamefont {Yan}},\ }\href@noop {} {\bibfield
  {journal} {\bibinfo  {journal} {J.~Chem.~Phys.}\ }\textbf {\bibinfo {volume}
  {134}},\ \bibinfo {pages} {244106} (\bibinfo {year} {2011})}\BibitemShut
  {NoStop}%
\bibitem [{\citenamefont {Ye}\ \emph {et~al.}(2017)\citenamefont {Ye},
  \citenamefont {Zhang}, \citenamefont {Wang}, \citenamefont {Zheng},\ and\
  \citenamefont {Yan}}]{ye17}%
  \BibitemOpen
  \bibfield  {author} {\bibinfo {author} {\bibfnamefont {L.}~\bibnamefont
  {Ye}}, \bibinfo {author} {\bibfnamefont {H.-D.}\ \bibnamefont {Zhang}},
  \bibinfo {author} {\bibfnamefont {Y.}~\bibnamefont {Wang}}, \bibinfo {author}
  {\bibfnamefont {X.}~\bibnamefont {Zheng}}, \ and\ \bibinfo {author}
  {\bibfnamefont {Y.}~\bibnamefont {Yan}},\ }\href@noop {} {\bibfield
  {journal} {\bibinfo  {journal} {J.~Chem.~Phys.}\ }\textbf {\bibinfo {volume}
  {147}},\ \bibinfo {pages} {074111} (\bibinfo {year} {2017})}\BibitemShut
  {NoStop}%
\bibitem [{\citenamefont {Tian}\ and\ \citenamefont {Chen}(2012)}]{tian12}%
  \BibitemOpen
  \bibfield  {author} {\bibinfo {author} {\bibfnamefont {H.}~\bibnamefont
  {Tian}}\ and\ \bibinfo {author} {\bibfnamefont {G.-H.}\ \bibnamefont
  {Chen}},\ }\href@noop {} {\bibfield  {journal} {\bibinfo  {journal}
  {J.~Chem.~Phys.}\ }\textbf {\bibinfo {volume} {137}},\ \bibinfo {pages}
  {204114} (\bibinfo {year} {2012})}\BibitemShut {NoStop}%
\bibitem [{\citenamefont {Popescu}\ \emph {et~al.}(2015)\citenamefont
  {Popescu}, \citenamefont {Rahman},\ and\ \citenamefont
  {Kleinekath{\"o}fer}}]{popescu15}%
  \BibitemOpen
  \bibfield  {author} {\bibinfo {author} {\bibfnamefont {B.}~\bibnamefont
  {Popescu}}, \bibinfo {author} {\bibfnamefont {H.}~\bibnamefont {Rahman}}, \
  and\ \bibinfo {author} {\bibfnamefont {U.}~\bibnamefont
  {Kleinekath{\"o}fer}},\ }\href@noop {} {\bibfield  {journal} {\bibinfo
  {journal} {J.~Chem.~Phys.}\ }\textbf {\bibinfo {volume} {142}},\ \bibinfo
  {pages} {154103} (\bibinfo {year} {2015})}\BibitemShut {NoStop}%
\bibitem [{\citenamefont {Nakamura}\ and\ \citenamefont
  {Tanimura}(2018)}]{nakamura18}%
  \BibitemOpen
  \bibfield  {author} {\bibinfo {author} {\bibfnamefont {K.}~\bibnamefont
  {Nakamura}}\ and\ \bibinfo {author} {\bibfnamefont {Y.}~\bibnamefont
  {Tanimura}},\ }\href@noop {} {\bibfield  {journal} {\bibinfo  {journal}
  {Phys.~Rev.~A}\ }\textbf {\bibinfo {volume} {98}},\ \bibinfo {pages} {012109}
  (\bibinfo {year} {2018})}\BibitemShut {NoStop}%
\bibitem [{\citenamefont {Rahman}\ and\ \citenamefont
  {Kleinekath{\"o}fer}(2019)}]{rahman19}%
  \BibitemOpen
  \bibfield  {author} {\bibinfo {author} {\bibfnamefont {H.}~\bibnamefont
  {Rahman}}\ and\ \bibinfo {author} {\bibfnamefont {U.}~\bibnamefont
  {Kleinekath{\"o}fer}},\ }\href@noop {} {\bibfield  {journal} {\bibinfo
  {journal} {J.~Chem.~Phys.}\ }\textbf {\bibinfo {volume} {150}},\ \bibinfo
  {pages} {244104} (\bibinfo {year} {2019})}\BibitemShut {NoStop}%
\bibitem [{\citenamefont {Tang}\ \emph {et~al.}(2015)\citenamefont {Tang},
  \citenamefont {Ouyang}, \citenamefont {Gong}, \citenamefont {Wang},\ and\
  \citenamefont {Wu}}]{tang15}%
  \BibitemOpen
  \bibfield  {author} {\bibinfo {author} {\bibfnamefont {Z.}~\bibnamefont
  {Tang}}, \bibinfo {author} {\bibfnamefont {X.}~\bibnamefont {Ouyang}},
  \bibinfo {author} {\bibfnamefont {Z.}~\bibnamefont {Gong}}, \bibinfo {author}
  {\bibfnamefont {H.}~\bibnamefont {Wang}}, \ and\ \bibinfo {author}
  {\bibfnamefont {J.}~\bibnamefont {Wu}},\ }\href@noop {} {\bibfield  {journal}
  {\bibinfo  {journal} {J.~Chem.~Phys.}\ }\textbf {\bibinfo {volume} {143}},\
  \bibinfo {pages} {224112} (\bibinfo {year} {2015})}\BibitemShut {NoStop}%
\bibitem [{\citenamefont {Ikeda}\ and\ \citenamefont
  {Scholes}(2020)}]{ikeda20}%
  \BibitemOpen
  \bibfield  {author} {\bibinfo {author} {\bibfnamefont {T.}~\bibnamefont
  {Ikeda}}\ and\ \bibinfo {author} {\bibfnamefont {G.~D.}\ \bibnamefont
  {Scholes}},\ }\href@noop {} {\bibfield  {journal} {\bibinfo  {journal}
  {J.~Chem.~Phys.}\ }\textbf {\bibinfo {volume} {152}},\ \bibinfo {pages}
  {204101} (\bibinfo {year} {2020})}\BibitemShut {NoStop}%
\bibitem [{\citenamefont {Erpenbeck}\ \emph {et~al.}(2018)\citenamefont
  {Erpenbeck}, \citenamefont {Hertlein}, \citenamefont {Schinabeck},\ and\
  \citenamefont {Thoss}}]{erpenbeck18}%
  \BibitemOpen
  \bibfield  {author} {\bibinfo {author} {\bibfnamefont {A.}~\bibnamefont
  {Erpenbeck}}, \bibinfo {author} {\bibfnamefont {C.}~\bibnamefont {Hertlein}},
  \bibinfo {author} {\bibfnamefont {C.}~\bibnamefont {Schinabeck}}, \ and\
  \bibinfo {author} {\bibfnamefont {M.}~\bibnamefont {Thoss}},\ }\href@noop {}
  {\bibfield  {journal} {\bibinfo  {journal} {J.~Chem.~Phys.}\ }\textbf
  {\bibinfo {volume} {149}},\ \bibinfo {pages} {064106} (\bibinfo {year}
  {2018})}\BibitemShut {NoStop}%
\bibitem [{\citenamefont {Cui}\ \emph {et~al.}(2019)\citenamefont {Cui},
  \citenamefont {Zhang}, \citenamefont {Zheng}, \citenamefont {Xu},\ and\
  \citenamefont {Yan}}]{cui19}%
  \BibitemOpen
  \bibfield  {author} {\bibinfo {author} {\bibfnamefont {L.}~\bibnamefont
  {Cui}}, \bibinfo {author} {\bibfnamefont {H.-D.}\ \bibnamefont {Zhang}},
  \bibinfo {author} {\bibfnamefont {X.}~\bibnamefont {Zheng}}, \bibinfo
  {author} {\bibfnamefont {R.-X.}\ \bibnamefont {Xu}}, \ and\ \bibinfo {author}
  {\bibfnamefont {Y.}~\bibnamefont {Yan}},\ }\href@noop {} {\bibfield
  {journal} {\bibinfo  {journal} {J.~Chem.~Phys.}\ }\textbf {\bibinfo {volume}
  {151}},\ \bibinfo {pages} {024110} (\bibinfo {year} {2019})}\BibitemShut
  {NoStop}%
\bibitem [{\citenamefont {Chen}\ \emph {et~al.}(2022)\citenamefont {Chen},
  \citenamefont {Wang}, \citenamefont {Zheng}, \citenamefont {Xu},\ and\
  \citenamefont {Yan}}]{chen22}%
  \BibitemOpen
  \bibfield  {author} {\bibinfo {author} {\bibfnamefont {Z.-H.}\ \bibnamefont
  {Chen}}, \bibinfo {author} {\bibfnamefont {Y.}~\bibnamefont {Wang}}, \bibinfo
  {author} {\bibfnamefont {X.}~\bibnamefont {Zheng}}, \bibinfo {author}
  {\bibfnamefont {R.-X.}\ \bibnamefont {Xu}}, \ and\ \bibinfo {author}
  {\bibfnamefont {Y.}~\bibnamefont {Yan}},\ }\href {\doibase 10.1063/5.0095961}
  {\bibfield  {journal} {\bibinfo  {journal} {J.~Chem.~Phys.}\ }\textbf
  {\bibinfo {volume} {156}},\ \bibinfo {pages} {221102} (\bibinfo {year}
  {2022})}\BibitemShut {NoStop}%
\bibitem [{\citenamefont {Li}\ \emph {et~al.}(2017)\citenamefont {Li},
  \citenamefont {Wei}, \citenamefont {Zheng}, \citenamefont {Yan},\ and\
  \citenamefont {Luo}}]{li17}%
  \BibitemOpen
  \bibfield  {author} {\bibinfo {author} {\bibfnamefont {Z.}~\bibnamefont
  {Li}}, \bibinfo {author} {\bibfnamefont {J.}~\bibnamefont {Wei}}, \bibinfo
  {author} {\bibfnamefont {X.}~\bibnamefont {Zheng}}, \bibinfo {author}
  {\bibfnamefont {Y.}~\bibnamefont {Yan}}, \ and\ \bibinfo {author}
  {\bibfnamefont {H.-G.}\ \bibnamefont {Luo}},\ }\href@noop {} {\bibfield
  {journal} {\bibinfo  {journal} {J. Phys.: Condens. Matter}\ }\textbf
  {\bibinfo {volume} {29}},\ \bibinfo {pages} {175601} (\bibinfo {year}
  {2017})}\BibitemShut {NoStop}%
\bibitem [{\citenamefont {Han}\ \emph {et~al.}(2018)\citenamefont {Han},
  \citenamefont {Zhang}, \citenamefont {Zheng},\ and\ \citenamefont
  {Yan}}]{han18}%
  \BibitemOpen
  \bibfield  {author} {\bibinfo {author} {\bibfnamefont {L.}~\bibnamefont
  {Han}}, \bibinfo {author} {\bibfnamefont {H.-D.}\ \bibnamefont {Zhang}},
  \bibinfo {author} {\bibfnamefont {X.}~\bibnamefont {Zheng}}, \ and\ \bibinfo
  {author} {\bibfnamefont {Y.}~\bibnamefont {Yan}},\ }\href@noop {} {\bibfield
  {journal} {\bibinfo  {journal} {J.~Chem.~Phys.}\ }\textbf {\bibinfo {volume}
  {148}},\ \bibinfo {pages} {234108} (\bibinfo {year} {2018})}\BibitemShut
  {NoStop}%
\bibitem [{\citenamefont {Zheng}\ \emph {et~al.}(2009)\citenamefont {Zheng},
  \citenamefont {Jin}, \citenamefont {Welack}, \citenamefont {Luo},\ and\
  \citenamefont {Yan}}]{zheng09b}%
  \BibitemOpen
  \bibfield  {author} {\bibinfo {author} {\bibfnamefont {X.}~\bibnamefont
  {Zheng}}, \bibinfo {author} {\bibfnamefont {J.-S.}\ \bibnamefont {Jin}},
  \bibinfo {author} {\bibfnamefont {S.}~\bibnamefont {Welack}}, \bibinfo
  {author} {\bibfnamefont {M.}~\bibnamefont {Luo}}, \ and\ \bibinfo {author}
  {\bibfnamefont {Y.-J.}\ \bibnamefont {Yan}},\ }\href@noop {} {\bibfield
  {journal} {\bibinfo  {journal} {J.~Chem.~Phys.}\ }\textbf {\bibinfo {volume}
  {130}},\ \bibinfo {pages} {164708} (\bibinfo {year} {2009})}\BibitemShut
  {NoStop}%
\bibitem [{\citenamefont {Xie}\ \emph {et~al.}(2013)\citenamefont {Xie},
  \citenamefont {Kwok}, \citenamefont {Zhang}, \citenamefont {Jiang},
  \citenamefont {Zheng}, \citenamefont {Yan},\ and\ \citenamefont
  {Chen}}]{xie13b}%
  \BibitemOpen
  \bibfield  {author} {\bibinfo {author} {\bibfnamefont {H.}~\bibnamefont
  {Xie}}, \bibinfo {author} {\bibfnamefont {Y.}~\bibnamefont {Kwok}}, \bibinfo
  {author} {\bibfnamefont {Y.}~\bibnamefont {Zhang}}, \bibinfo {author}
  {\bibfnamefont {F.}~\bibnamefont {Jiang}}, \bibinfo {author} {\bibfnamefont
  {X.}~\bibnamefont {Zheng}}, \bibinfo {author} {\bibfnamefont
  {Y.}~\bibnamefont {Yan}}, \ and\ \bibinfo {author} {\bibfnamefont
  {G.}~\bibnamefont {Chen}},\ }\href@noop {} {\bibfield  {journal} {\bibinfo
  {journal} {Phys. Status Solidi B}\ }\textbf {\bibinfo {volume} {250}},\
  \bibinfo {pages} {2481} (\bibinfo {year} {2013})}\BibitemShut {NoStop}%
\bibitem [{\citenamefont {Hou}\ \emph {et~al.}(2014)\citenamefont {Hou},
  \citenamefont {Wang}, \citenamefont {Zheng}, \citenamefont {Tong},
  \citenamefont {Wei},\ and\ \citenamefont {Yan}}]{hou14}%
  \BibitemOpen
  \bibfield  {author} {\bibinfo {author} {\bibfnamefont {D.}~\bibnamefont
  {Hou}}, \bibinfo {author} {\bibfnamefont {R.}~\bibnamefont {Wang}}, \bibinfo
  {author} {\bibfnamefont {X.}~\bibnamefont {Zheng}}, \bibinfo {author}
  {\bibfnamefont {N.}~\bibnamefont {Tong}}, \bibinfo {author} {\bibfnamefont
  {J.}~\bibnamefont {Wei}}, \ and\ \bibinfo {author} {\bibfnamefont
  {Y.}~\bibnamefont {Yan}},\ }\href@noop {} {\bibfield  {journal} {\bibinfo
  {journal} {Phys.~Rev.~B}\ }\textbf {\bibinfo {volume} {90}},\ \bibinfo
  {pages} {045141} (\bibinfo {year} {2014})}\BibitemShut {NoStop}%
\bibitem [{\citenamefont {Duan}\ \emph {et~al.}(2017)\citenamefont {Duan},
  \citenamefont {Tang}, \citenamefont {Cao},\ and\ \citenamefont
  {Wu}}]{duan17}%
  \BibitemOpen
  \bibfield  {author} {\bibinfo {author} {\bibfnamefont {C.}~\bibnamefont
  {Duan}}, \bibinfo {author} {\bibfnamefont {Z.}~\bibnamefont {Tang}}, \bibinfo
  {author} {\bibfnamefont {J.}~\bibnamefont {Cao}}, \ and\ \bibinfo {author}
  {\bibfnamefont {J.}~\bibnamefont {Wu}},\ }\href@noop {} {\bibfield  {journal}
  {\bibinfo  {journal} {Phys.~Rev.~B}\ }\textbf {\bibinfo {volume} {95}},\
  \bibinfo {pages} {214308} (\bibinfo {year} {2017})}\BibitemShut {NoStop}%
\bibitem [{\citenamefont {Xu}\ \emph {et~al.}(2022)\citenamefont {Xu},
  \citenamefont {Yan}, \citenamefont {Shi}, \citenamefont {Ankerhold},\ and\
  \citenamefont {Stockburger}}]{xu22}%
  \BibitemOpen
  \bibfield  {author} {\bibinfo {author} {\bibfnamefont {M.}~\bibnamefont
  {Xu}}, \bibinfo {author} {\bibfnamefont {Y.}~\bibnamefont {Yan}}, \bibinfo
  {author} {\bibfnamefont {Q.}~\bibnamefont {Shi}}, \bibinfo {author}
  {\bibfnamefont {J.}~\bibnamefont {Ankerhold}}, \ and\ \bibinfo {author}
  {\bibfnamefont {J.~T.}\ \bibnamefont {Stockburger}},\ }\href {\doibase
  10.1103/PhysRevLett.129.230601} {\bibfield  {journal} {\bibinfo  {journal}
  {Phys. Rev. Lett.}\ }\textbf {\bibinfo {volume} {129}},\ \bibinfo {pages}
  {230601} (\bibinfo {year} {2022})}\BibitemShut {NoStop}%
\bibitem [{\citenamefont {Nakatsukasa}\ \emph {et~al.}(2018)\citenamefont
  {Nakatsukasa}, \citenamefont {S{\`e}te},\ and\ \citenamefont
  {Trefethen}}]{nakatsukasa18}%
  \BibitemOpen
  \bibfield  {author} {\bibinfo {author} {\bibfnamefont {Y.}~\bibnamefont
  {Nakatsukasa}}, \bibinfo {author} {\bibfnamefont {O.}~\bibnamefont
  {S{\`e}te}}, \ and\ \bibinfo {author} {\bibfnamefont {L.~N.}\ \bibnamefont
  {Trefethen}},\ }\href@noop {} {\bibfield  {journal} {\bibinfo  {journal}
  {SIAM J SCI COMPUT}\ }\textbf {\bibinfo {volume} {40}},\ \bibinfo {pages}
  {A1494} (\bibinfo {year} {2018})}\BibitemShut {NoStop}%
\bibitem [{\citenamefont {Liu}\ \emph {et~al.}(2014)\citenamefont {Liu},
  \citenamefont {Zhu}, \citenamefont {Bai},\ and\ \citenamefont {Shi}}]{liu14}%
  \BibitemOpen
  \bibfield  {author} {\bibinfo {author} {\bibfnamefont {H.}~\bibnamefont
  {Liu}}, \bibinfo {author} {\bibfnamefont {L.}~\bibnamefont {Zhu}}, \bibinfo
  {author} {\bibfnamefont {S.}~\bibnamefont {Bai}}, \ and\ \bibinfo {author}
  {\bibfnamefont {Q.}~\bibnamefont {Shi}},\ }\href@noop {} {\bibfield
  {journal} {\bibinfo  {journal} {J. Chem. Phys.}\ }\textbf {\bibinfo {volume}
  {140}},\ \bibinfo {pages} {134106} (\bibinfo {year} {2014})}\BibitemShut
  {NoStop}%
\bibitem [{\citenamefont {Wang}\ and\ \citenamefont {Thoss}(2013)}]{wang13b}%
  \BibitemOpen
  \bibfield  {author} {\bibinfo {author} {\bibfnamefont {H.}~\bibnamefont
  {Wang}}\ and\ \bibinfo {author} {\bibfnamefont {M.}~\bibnamefont {Thoss}},\
  }\href@noop {} {\bibfield  {journal} {\bibinfo  {journal} {J.~Chem.~Phys.}\
  }\textbf {\bibinfo {volume} {138}},\ \bibinfo {pages} {134704} (\bibinfo
  {year} {2013})}\BibitemShut {NoStop}%
\bibitem [{\citenamefont {Xu}\ \emph {et~al.}(2017)\citenamefont {Xu},
  \citenamefont {Song}, \citenamefont {Song},\ and\ \citenamefont
  {Shi}}]{xu17}%
  \BibitemOpen
  \bibfield  {author} {\bibinfo {author} {\bibfnamefont {M.}~\bibnamefont
  {Xu}}, \bibinfo {author} {\bibfnamefont {L.}~\bibnamefont {Song}}, \bibinfo
  {author} {\bibfnamefont {K.}~\bibnamefont {Song}}, \ and\ \bibinfo {author}
  {\bibfnamefont {Q.}~\bibnamefont {Shi}},\ }\href@noop {} {\bibfield
  {journal} {\bibinfo  {journal} {J.~Chem.~Phys.}\ }\textbf {\bibinfo {volume}
  {146}},\ \bibinfo {pages} {064102} (\bibinfo {year} {2017})}\BibitemShut
  {NoStop}%
\bibitem [{\citenamefont {Shi}\ \emph {et~al.}(2009)\citenamefont {Shi},
  \citenamefont {Chen}, \citenamefont {Nan}, \citenamefont {Xu},\ and\
  \citenamefont {Yan}}]{shi09b}%
  \BibitemOpen
  \bibfield  {author} {\bibinfo {author} {\bibfnamefont {Q.}~\bibnamefont
  {Shi}}, \bibinfo {author} {\bibfnamefont {L.-P.}\ \bibnamefont {Chen}},
  \bibinfo {author} {\bibfnamefont {G.-J.}\ \bibnamefont {Nan}}, \bibinfo
  {author} {\bibfnamefont {R.-X.}\ \bibnamefont {Xu}}, \ and\ \bibinfo {author}
  {\bibfnamefont {Y.-J.}\ \bibnamefont {Yan}},\ }\href@noop {} {\bibfield
  {journal} {\bibinfo  {journal} {J. Chem. Phys.}\ }\textbf {\bibinfo {volume}
  {130}},\ \bibinfo {pages} {084105} (\bibinfo {year} {2009})}\BibitemShut
  {NoStop}%
\bibitem [{\citenamefont {Ke}\ \emph {et~al.}(2022{\natexlab{a}})\citenamefont
  {Ke}, \citenamefont {Borrelli},\ and\ \citenamefont {Thoss}}]{ke22a}%
  \BibitemOpen
  \bibfield  {author} {\bibinfo {author} {\bibfnamefont {Y.}~\bibnamefont
  {Ke}}, \bibinfo {author} {\bibfnamefont {R.}~\bibnamefont {Borrelli}}, \ and\
  \bibinfo {author} {\bibfnamefont {M.}~\bibnamefont {Thoss}},\ }\href@noop {}
  {\bibfield  {journal} {\bibinfo  {journal} {J.~Chem.~Phys.}\ }\textbf
  {\bibinfo {volume} {156}},\ \bibinfo {pages} {194102} (\bibinfo {year}
  {2022}{\natexlab{a}})}\BibitemShut {NoStop}%
\bibitem [{\citenamefont {Hewson}(1997)}]{hewson97}%
  \BibitemOpen
  \bibfield  {author} {\bibinfo {author} {\bibfnamefont {A.~C.}\ \bibnamefont
  {Hewson}},\ }\href@noop {} {\emph {\bibinfo {title} {The Kondo problem to
  heavy fermions}}}\ (\bibinfo  {publisher} {Cambridge university press},\
  \bibinfo {year} {1997})\BibitemShut {NoStop}%
\bibitem [{\citenamefont {Mahan}(2000)}]{mahan00}%
  \BibitemOpen
  \bibfield  {author} {\bibinfo {author} {\bibfnamefont {G.~D.}\ \bibnamefont
  {Mahan}},\ }\href@noop {} {\emph {\bibinfo {title} {Many-Particle Physics}}}\
  (\bibinfo  {publisher} {Kluwer Academic/Plenum},\ \bibinfo {address} {New
  York},\ \bibinfo {year} {2000})\BibitemShut {NoStop}%
\bibitem [{\citenamefont {Tanimura}(2014)}]{tanimura14}%
  \BibitemOpen
  \bibfield  {author} {\bibinfo {author} {\bibfnamefont {Y.}~\bibnamefont
  {Tanimura}},\ }\href@noop {} {\bibfield  {journal} {\bibinfo  {journal}
  {J.~Chem.~Phys.}\ }\textbf {\bibinfo {volume} {141}},\ \bibinfo {pages}
  {044114} (\bibinfo {year} {2014})}\BibitemShut {NoStop}%
\bibitem [{\citenamefont {Song}\ and\ \citenamefont {Shi}(2015)}]{song15b}%
  \BibitemOpen
  \bibfield  {author} {\bibinfo {author} {\bibfnamefont {L.}~\bibnamefont
  {Song}}\ and\ \bibinfo {author} {\bibfnamefont {Q.}~\bibnamefont {Shi}},\
  }\href@noop {} {\bibfield  {journal} {\bibinfo  {journal} {J.~Chem.~Phys.}\
  }\textbf {\bibinfo {volume} {143}},\ \bibinfo {pages} {194106} (\bibinfo
  {year} {2015})}\BibitemShut {NoStop}%
\bibitem [{\citenamefont {Xing}\ \emph {et~al.}(2022)\citenamefont {Xing},
  \citenamefont {Li}, \citenamefont {Yan}, \citenamefont {Bai},\ and\
  \citenamefont {Shi}}]{xing22}%
  \BibitemOpen
  \bibfield  {author} {\bibinfo {author} {\bibfnamefont {T.}~\bibnamefont
  {Xing}}, \bibinfo {author} {\bibfnamefont {T.}~\bibnamefont {Li}}, \bibinfo
  {author} {\bibfnamefont {Y.}~\bibnamefont {Yan}}, \bibinfo {author}
  {\bibfnamefont {S.}~\bibnamefont {Bai}}, \ and\ \bibinfo {author}
  {\bibfnamefont {Q.}~\bibnamefont {Shi}},\ }\href@noop {} {\bibfield
  {journal} {\bibinfo  {journal} {J.~Chem.~Phys.}\ }\textbf {\bibinfo {volume}
  {156}},\ \bibinfo {pages} {244102} (\bibinfo {year} {2022})}\BibitemShut
  {NoStop}%
\bibitem [{\citenamefont {Ke}\ \emph {et~al.}(2022{\natexlab{b}})\citenamefont
  {Ke}, \citenamefont {Kaspar}, \citenamefont {Erpenbecka}, \citenamefont
  {Peskin},\ and\ \citenamefont {Thoss}}]{ke22b}%
  \BibitemOpen
  \bibfield  {author} {\bibinfo {author} {\bibfnamefont {Y.}~\bibnamefont
  {Ke}}, \bibinfo {author} {\bibfnamefont {C.}~\bibnamefont {Kaspar}}, \bibinfo
  {author} {\bibfnamefont {A.}~\bibnamefont {Erpenbecka}}, \bibinfo {author}
  {\bibfnamefont {U.}~\bibnamefont {Peskin}}, \ and\ \bibinfo {author}
  {\bibfnamefont {M.}~\bibnamefont {Thoss}},\ }\href@noop {} {\bibfield
  {journal} {\bibinfo  {journal} {J.~Chem.~Phys.}\ }\textbf {\bibinfo {volume}
  {157}},\ \bibinfo {pages} {034103} (\bibinfo {year}
  {2022}{\natexlab{b}})}\BibitemShut {NoStop}%
\bibitem [{\citenamefont {Ye}\ \emph {et~al.}(2016)\citenamefont {Ye},
  \citenamefont {Wang}, \citenamefont {Hou}, \citenamefont {Xu}, \citenamefont
  {Zheng},\ and\ \citenamefont {Yan}}]{ye16}%
  \BibitemOpen
  \bibfield  {author} {\bibinfo {author} {\bibfnamefont {L.}~\bibnamefont
  {Ye}}, \bibinfo {author} {\bibfnamefont {X.}~\bibnamefont {Wang}}, \bibinfo
  {author} {\bibfnamefont {D.}~\bibnamefont {Hou}}, \bibinfo {author}
  {\bibfnamefont {R.-X.}\ \bibnamefont {Xu}}, \bibinfo {author} {\bibfnamefont
  {X.}~\bibnamefont {Zheng}}, \ and\ \bibinfo {author} {\bibfnamefont
  {Y.}~\bibnamefont {Yan}},\ }\href@noop {} {\bibfield  {journal} {\bibinfo
  {journal} {Wiley Interdiscip. Rev.: Comput. Mol. Sci.}\ }\textbf {\bibinfo
  {volume} {6}},\ \bibinfo {pages} {608} (\bibinfo {year} {2016})}\BibitemShut
  {NoStop}%
\bibitem [{\citenamefont {Kaspar}\ and\ \citenamefont
  {Thoss}(2021)}]{kaspar21}%
  \BibitemOpen
  \bibfield  {author} {\bibinfo {author} {\bibfnamefont {C.}~\bibnamefont
  {Kaspar}}\ and\ \bibinfo {author} {\bibfnamefont {M.}~\bibnamefont {Thoss}},\
  }\href@noop {} {\bibfield  {journal} {\bibinfo  {journal} {J.~Phys.~Chem.~A}\
  }\textbf {\bibinfo {volume} {125}},\ \bibinfo {pages} {5190} (\bibinfo {year}
  {2021})}\BibitemShut {NoStop}%
\bibitem [{\citenamefont {Hofreither}(2021)}]{hofreither21}%
  \BibitemOpen
  \bibfield  {author} {\bibinfo {author} {\bibfnamefont {C.}~\bibnamefont
  {Hofreither}},\ }\href@noop {} {\bibfield  {journal} {\bibinfo  {journal}
  {Numer Algorithms}\ }\textbf {\bibinfo {volume} {88}},\ \bibinfo {pages}
  {365} (\bibinfo {year} {2021})}\BibitemShut {NoStop}%
\bibitem [{\citenamefont {Barthel}\ \emph {et~al.}(2009)\citenamefont
  {Barthel}, \citenamefont {Schollw{\"o}ck},\ and\ \citenamefont
  {White}}]{barthel09}%
  \BibitemOpen
  \bibfield  {author} {\bibinfo {author} {\bibfnamefont {T.}~\bibnamefont
  {Barthel}}, \bibinfo {author} {\bibfnamefont {U.}~\bibnamefont
  {Schollw{\"o}ck}}, \ and\ \bibinfo {author} {\bibfnamefont {S.~R.}\
  \bibnamefont {White}},\ }\href@noop {} {\bibfield  {journal} {\bibinfo
  {journal} {Phys.~Rev.~B}\ }\textbf {\bibinfo {volume} {79}},\ \bibinfo
  {pages} {245101} (\bibinfo {year} {2009})}\BibitemShut {NoStop}%
\bibitem [{\citenamefont {Wolf}\ \emph
  {et~al.}(2015{\natexlab{a}})\citenamefont {Wolf}, \citenamefont {Go},
  \citenamefont {McCulloch}, \citenamefont {Millis},\ and\ \citenamefont
  {Schollw{\"o}ck}}]{wolf15}%
  \BibitemOpen
  \bibfield  {author} {\bibinfo {author} {\bibfnamefont {F.~A.}\ \bibnamefont
  {Wolf}}, \bibinfo {author} {\bibfnamefont {A.}~\bibnamefont {Go}}, \bibinfo
  {author} {\bibfnamefont {I.~P.}\ \bibnamefont {McCulloch}}, \bibinfo {author}
  {\bibfnamefont {A.~J.}\ \bibnamefont {Millis}}, \ and\ \bibinfo {author}
  {\bibfnamefont {U.}~\bibnamefont {Schollw{\"o}ck}},\ }\href@noop {}
  {\bibfield  {journal} {\bibinfo  {journal} {Phys.~Rev.~X}\ }\textbf {\bibinfo
  {volume} {5}},\ \bibinfo {pages} {041032} (\bibinfo {year}
  {2015}{\natexlab{a}})}\BibitemShut {NoStop}%
\bibitem [{\citenamefont {Wolf}\ \emph
  {et~al.}(2015{\natexlab{b}})\citenamefont {Wolf}, \citenamefont {Justiniano},
  \citenamefont {McCulloch},\ and\ \citenamefont {Schollw{\"o}ck}}]{wolf15b}%
  \BibitemOpen
  \bibfield  {author} {\bibinfo {author} {\bibfnamefont {F.~A.}\ \bibnamefont
  {Wolf}}, \bibinfo {author} {\bibfnamefont {J.~A.}\ \bibnamefont
  {Justiniano}}, \bibinfo {author} {\bibfnamefont {I.~P.}\ \bibnamefont
  {McCulloch}}, \ and\ \bibinfo {author} {\bibfnamefont {U.}~\bibnamefont
  {Schollw{\"o}ck}},\ }\href@noop {} {\bibfield  {journal} {\bibinfo  {journal}
  {Phys.~Rev.~B}\ }\textbf {\bibinfo {volume} {91}},\ \bibinfo {pages} {115144}
  (\bibinfo {year} {2015}{\natexlab{b}})}\BibitemShut {NoStop}%
\bibitem [{\citenamefont {White}\ and\ \citenamefont
  {Affleck}(2008)}]{white08}%
  \BibitemOpen
  \bibfield  {author} {\bibinfo {author} {\bibfnamefont {S.~R.}\ \bibnamefont
  {White}}\ and\ \bibinfo {author} {\bibfnamefont {I.}~\bibnamefont
  {Affleck}},\ }\href@noop {} {\bibfield  {journal} {\bibinfo  {journal}
  {Phys.~Rev.~B}\ }\textbf {\bibinfo {volume} {77}},\ \bibinfo {pages} {134437}
  (\bibinfo {year} {2008})}\BibitemShut {NoStop}%
\bibitem [{\citenamefont {Ganahl}\ \emph {et~al.}(2015)\citenamefont {Ganahl},
  \citenamefont {Aichhorn}, \citenamefont {Evertz}, \citenamefont
  {Thunstr{\"o}m}, \citenamefont {Held},\ and\ \citenamefont
  {Verstraete}}]{ganahl15}%
  \BibitemOpen
  \bibfield  {author} {\bibinfo {author} {\bibfnamefont {M.}~\bibnamefont
  {Ganahl}}, \bibinfo {author} {\bibfnamefont {M.}~\bibnamefont {Aichhorn}},
  \bibinfo {author} {\bibfnamefont {H.~G.}\ \bibnamefont {Evertz}}, \bibinfo
  {author} {\bibfnamefont {P.}~\bibnamefont {Thunstr{\"o}m}}, \bibinfo {author}
  {\bibfnamefont {K.}~\bibnamefont {Held}}, \ and\ \bibinfo {author}
  {\bibfnamefont {F.}~\bibnamefont {Verstraete}},\ }\href@noop {} {\bibfield
  {journal} {\bibinfo  {journal} {Phys.~Rev.~B}\ }\textbf {\bibinfo {volume}
  {92}},\ \bibinfo {pages} {155132} (\bibinfo {year} {2015})}\BibitemShut
  {NoStop}%
\bibitem [{\citenamefont {Langreth}(1966)}]{langreth66}%
  \BibitemOpen
  \bibfield  {author} {\bibinfo {author} {\bibfnamefont {D.~C.}\ \bibnamefont
  {Langreth}},\ }\href@noop {} {\bibfield  {journal} {\bibinfo  {journal}
  {Phys. Rev.}\ }\textbf {\bibinfo {volume} {150}},\ \bibinfo {pages} {516}
  (\bibinfo {year} {1966})}\BibitemShut {NoStop}%
\bibitem [{\citenamefont {Martinek}\ \emph {et~al.}(2003)\citenamefont
  {Martinek}, \citenamefont {Sindel}, \citenamefont {Borda}, \citenamefont
  {Barna{\'s}}, \citenamefont {K{\"o}nig}, \citenamefont {Sch{\"o}n},\ and\
  \citenamefont {Von~Delft}}]{martinek03}%
  \BibitemOpen
  \bibfield  {author} {\bibinfo {author} {\bibfnamefont {J.}~\bibnamefont
  {Martinek}}, \bibinfo {author} {\bibfnamefont {M.}~\bibnamefont {Sindel}},
  \bibinfo {author} {\bibfnamefont {L.}~\bibnamefont {Borda}}, \bibinfo
  {author} {\bibfnamefont {J.}~\bibnamefont {Barna{\'s}}}, \bibinfo {author}
  {\bibfnamefont {J.}~\bibnamefont {K{\"o}nig}}, \bibinfo {author}
  {\bibfnamefont {G.}~\bibnamefont {Sch{\"o}n}}, \ and\ \bibinfo {author}
  {\bibfnamefont {J.}~\bibnamefont {Von~Delft}},\ }\href@noop {} {\bibfield
  {journal} {\bibinfo  {journal} {Phys.~Rev.~Lett.}\ }\textbf {\bibinfo
  {volume} {91}},\ \bibinfo {pages} {247202} (\bibinfo {year}
  {2003})}\BibitemShut {NoStop}%
\bibitem [{\citenamefont {Zhang}\ \emph {et~al.}(2022)\citenamefont {Zhang},
  \citenamefont {Ding}, \citenamefont {Zhang}, \citenamefont {Zheng},\ and\
  \citenamefont {Yan}}]{zhang22}%
  \BibitemOpen
  \bibfield  {author} {\bibinfo {author} {\bibfnamefont {D.}~\bibnamefont
  {Zhang}}, \bibinfo {author} {\bibfnamefont {X.}~\bibnamefont {Ding}},
  \bibinfo {author} {\bibfnamefont {H.-D.}\ \bibnamefont {Zhang}}, \bibinfo
  {author} {\bibfnamefont {X.}~\bibnamefont {Zheng}}, \ and\ \bibinfo {author}
  {\bibfnamefont {Y.}~\bibnamefont {Yan}},\ }\href@noop {} {\bibfield
  {journal} {\bibinfo  {journal} {Chin. J.~Chem.~Phys.}\ }\textbf {\bibinfo
  {volume} {34}},\ \bibinfo {pages} {905} (\bibinfo {year} {2022})}\BibitemShut
  {NoStop}%
\bibitem [{\citenamefont {Ding}\ \emph {et~al.}(2022)\citenamefont {Ding},
  \citenamefont {Zhang}, \citenamefont {Ye}, \citenamefont {Zheng},\ and\
  \citenamefont {Yan}}]{ding22}%
  \BibitemOpen
  \bibfield  {author} {\bibinfo {author} {\bibfnamefont {X.}~\bibnamefont
  {Ding}}, \bibinfo {author} {\bibfnamefont {D.}~\bibnamefont {Zhang}},
  \bibinfo {author} {\bibfnamefont {L.}~\bibnamefont {Ye}}, \bibinfo {author}
  {\bibfnamefont {X.}~\bibnamefont {Zheng}}, \ and\ \bibinfo {author}
  {\bibfnamefont {Y.}~\bibnamefont {Yan}},\ }\href@noop {} {\bibfield
  {journal} {\bibinfo  {journal} {J.~Chem.~Phys.}\ }\textbf {\bibinfo {volume}
  {157}},\ \bibinfo {pages} {224107} (\bibinfo {year} {2022})}\BibitemShut
  {NoStop}%
\bibitem [{\citenamefont {Wolf}\ \emph {et~al.}(2014)\citenamefont {Wolf},
  \citenamefont {McCulloch},\ and\ \citenamefont {Schollw{\"o}ck}}]{wolf14}%
  \BibitemOpen
  \bibfield  {author} {\bibinfo {author} {\bibfnamefont {F.~A.}\ \bibnamefont
  {Wolf}}, \bibinfo {author} {\bibfnamefont {I.~P.}\ \bibnamefont {McCulloch}},
  \ and\ \bibinfo {author} {\bibfnamefont {U.}~\bibnamefont {Schollw{\"o}ck}},\
  }\href@noop {} {\bibfield  {journal} {\bibinfo  {journal} {Phys.~Rev.~B}\
  }\textbf {\bibinfo {volume} {90}},\ \bibinfo {pages} {235131} (\bibinfo
  {year} {2014})}\BibitemShut {NoStop}%
\bibitem [{\citenamefont {Karski}\ \emph {et~al.}(2008)\citenamefont {Karski},
  \citenamefont {Raas},\ and\ \citenamefont {Uhrig}}]{karski08}%
  \BibitemOpen
  \bibfield  {author} {\bibinfo {author} {\bibfnamefont {M.}~\bibnamefont
  {Karski}}, \bibinfo {author} {\bibfnamefont {C.}~\bibnamefont {Raas}}, \ and\
  \bibinfo {author} {\bibfnamefont {G.~S.}\ \bibnamefont {Uhrig}},\ }\href@noop
  {} {\bibfield  {journal} {\bibinfo  {journal} {Phys.~Rev.~B}\ }\textbf
  {\bibinfo {volume} {77}},\ \bibinfo {pages} {075116} (\bibinfo {year}
  {2008})}\BibitemShut {NoStop}%
\bibitem [{\citenamefont {Dorda}\ \emph {et~al.}(2014)\citenamefont {Dorda},
  \citenamefont {Nuss}, \citenamefont {von~der Linden},\ and\ \citenamefont
  {Arrigoni}}]{dorda14}%
  \BibitemOpen
  \bibfield  {author} {\bibinfo {author} {\bibfnamefont {A.}~\bibnamefont
  {Dorda}}, \bibinfo {author} {\bibfnamefont {M.}~\bibnamefont {Nuss}},
  \bibinfo {author} {\bibfnamefont {W.}~\bibnamefont {von~der Linden}}, \ and\
  \bibinfo {author} {\bibfnamefont {E.}~\bibnamefont {Arrigoni}},\ }\href@noop
  {} {\bibfield  {journal} {\bibinfo  {journal} {Phys.~Rev.~B}\ }\textbf
  {\bibinfo {volume} {89}},\ \bibinfo {pages} {165105} (\bibinfo {year}
  {2014})}\BibitemShut {NoStop}%
\bibitem [{\citenamefont {W{\'o}jtowicz}\ \emph {et~al.}(2021)\citenamefont
  {W{\'o}jtowicz}, \citenamefont {Elenewski}, \citenamefont {Rams},\ and\
  \citenamefont {Zwolak}}]{wojtowicz21}%
  \BibitemOpen
  \bibfield  {author} {\bibinfo {author} {\bibfnamefont {G.}~\bibnamefont
  {W{\'o}jtowicz}}, \bibinfo {author} {\bibfnamefont {J.~E.}\ \bibnamefont
  {Elenewski}}, \bibinfo {author} {\bibfnamefont {M.~M.}\ \bibnamefont {Rams}},
  \ and\ \bibinfo {author} {\bibfnamefont {M.}~\bibnamefont {Zwolak}},\
  }\href@noop {} {\bibfield  {journal} {\bibinfo  {journal} {Phys.~Rev.~B}\
  }\textbf {\bibinfo {volume} {104}},\ \bibinfo {pages} {165131} (\bibinfo
  {year} {2021})}\BibitemShut {NoStop}%
\bibitem [{\citenamefont {Cohen}\ and\ \citenamefont {Rabani}(2011)}]{cohen11}%
  \BibitemOpen
  \bibfield  {author} {\bibinfo {author} {\bibfnamefont {G.}~\bibnamefont
  {Cohen}}\ and\ \bibinfo {author} {\bibfnamefont {E.}~\bibnamefont {Rabani}},\
  }\href@noop {} {\bibfield  {journal} {\bibinfo  {journal} {Phys.~Rev.~B}\
  }\textbf {\bibinfo {volume} {84}},\ \bibinfo {pages} {075150} (\bibinfo
  {year} {2011})}\BibitemShut {NoStop}%
\bibitem [{\citenamefont {Cohen}\ \emph {et~al.}(2013)\citenamefont {Cohen},
  \citenamefont {Wilner},\ and\ \citenamefont {Rabani}}]{cohen13b}%
  \BibitemOpen
  \bibfield  {author} {\bibinfo {author} {\bibfnamefont {G.}~\bibnamefont
  {Cohen}}, \bibinfo {author} {\bibfnamefont {E.~Y.}\ \bibnamefont {Wilner}}, \
  and\ \bibinfo {author} {\bibfnamefont {E.}~\bibnamefont {Rabani}},\
  }\href@noop {} {\bibfield  {journal} {\bibinfo  {journal} {New.~J.~Phys.}\
  }\textbf {\bibinfo {volume} {15}},\ \bibinfo {pages} {073018} (\bibinfo
  {year} {2013})}\BibitemShut {NoStop}%
\bibitem [{\citenamefont {Yang}\ and\ \citenamefont {White}(2020)}]{yang20}%
  \BibitemOpen
  \bibfield  {author} {\bibinfo {author} {\bibfnamefont {M.}~\bibnamefont
  {Yang}}\ and\ \bibinfo {author} {\bibfnamefont {S.~R.}\ \bibnamefont
  {White}},\ }\href@noop {} {\bibfield  {journal} {\bibinfo  {journal}
  {Phys.~Rev.~B}\ }\textbf {\bibinfo {volume} {102}},\ \bibinfo {pages}
  {094315} (\bibinfo {year} {2020})}\BibitemShut {NoStop}%
\bibitem [{\citenamefont {Borrelli}\ and\ \citenamefont
  {Dolgov}(2021)}]{borrelli21}%
  \BibitemOpen
  \bibfield  {author} {\bibinfo {author} {\bibfnamefont {R.}~\bibnamefont
  {Borrelli}}\ and\ \bibinfo {author} {\bibfnamefont {S.}~\bibnamefont
  {Dolgov}},\ }\href@noop {} {\bibfield  {journal} {\bibinfo  {journal}
  {J.~Phys.~Chem.~B}\ }\textbf {\bibinfo {volume} {125}},\ \bibinfo {pages}
  {5397} (\bibinfo {year} {2021})}\BibitemShut {NoStop}%
\bibitem [{\citenamefont {Yan}\ \emph {et~al.}(2021)\citenamefont {Yan},
  \citenamefont {Liu}, \citenamefont {Xing},\ and\ \citenamefont
  {Shi}}]{yan21}%
  \BibitemOpen
  \bibfield  {author} {\bibinfo {author} {\bibfnamefont {Y.}~\bibnamefont
  {Yan}}, \bibinfo {author} {\bibfnamefont {Y.}~\bibnamefont {Liu}}, \bibinfo
  {author} {\bibfnamefont {T.}~\bibnamefont {Xing}}, \ and\ \bibinfo {author}
  {\bibfnamefont {Q.}~\bibnamefont {Shi}},\ }\href@noop {} {\bibfield
  {journal} {\bibinfo  {journal} {Wiley Interdiscip. Rev.: Comput. Mol. Sci.}\
  }\textbf {\bibinfo {volume} {11}},\ \bibinfo {pages} {e1498} (\bibinfo {year}
  {2021})}\BibitemShut {NoStop}%
\bibitem [{\citenamefont {Suess}\ \emph {et~al.}(2014)\citenamefont {Suess},
  \citenamefont {Eisfeld},\ and\ \citenamefont {Strunz}}]{suess14}%
  \BibitemOpen
  \bibfield  {author} {\bibinfo {author} {\bibfnamefont {D.}~\bibnamefont
  {Suess}}, \bibinfo {author} {\bibfnamefont {A.}~\bibnamefont {Eisfeld}}, \
  and\ \bibinfo {author} {\bibfnamefont {W.~T.}\ \bibnamefont {Strunz}},\
  }\href@noop {} {\bibfield  {journal} {\bibinfo  {journal} {Phys.~Rev.~Lett.}\
  }\textbf {\bibinfo {volume} {113}},\ \bibinfo {eid} {150403} (\bibinfo {year}
  {2014})}\BibitemShut {NoStop}%
\bibitem [{\citenamefont {Beach}\ \emph {et~al.}(2000)\citenamefont {Beach},
  \citenamefont {Gooding},\ and\ \citenamefont {Marsiglio}}]{beach00}%
  \BibitemOpen
  \bibfield  {author} {\bibinfo {author} {\bibfnamefont {K.}~\bibnamefont
  {Beach}}, \bibinfo {author} {\bibfnamefont {R.}~\bibnamefont {Gooding}}, \
  and\ \bibinfo {author} {\bibfnamefont {F.}~\bibnamefont {Marsiglio}},\
  }\href@noop {} {\bibfield  {journal} {\bibinfo  {journal} {Phys.~Rev.~B}\
  }\textbf {\bibinfo {volume} {61}},\ \bibinfo {pages} {5147} (\bibinfo {year}
  {2000})}\BibitemShut {NoStop}%
\bibitem [{\citenamefont {Croy}\ and\ \citenamefont {Saalmann}(2009)}]{croy09}%
  \BibitemOpen
  \bibfield  {author} {\bibinfo {author} {\bibfnamefont {A.}~\bibnamefont
  {Croy}}\ and\ \bibinfo {author} {\bibfnamefont {U.}~\bibnamefont
  {Saalmann}},\ }\href@noop {} {\bibfield  {journal} {\bibinfo  {journal}
  {Phys.~Rev.~B}\ }\textbf {\bibinfo {volume} {80}},\ \bibinfo {pages} {245311}
  (\bibinfo {year} {2009})}\BibitemShut {NoStop}%
\bibitem [{\citenamefont {Gu}\ \emph {et~al.}(2020)\citenamefont {Gu},
  \citenamefont {Chen}, \citenamefont {Wang},\ and\ \citenamefont
  {Zhang}}]{gu20}%
  \BibitemOpen
  \bibfield  {author} {\bibinfo {author} {\bibfnamefont {J.}~\bibnamefont
  {Gu}}, \bibinfo {author} {\bibfnamefont {J.}~\bibnamefont {Chen}}, \bibinfo
  {author} {\bibfnamefont {Y.}~\bibnamefont {Wang}}, \ and\ \bibinfo {author}
  {\bibfnamefont {X.-G.}\ \bibnamefont {Zhang}},\ }\href@noop {} {\bibfield
  {journal} {\bibinfo  {journal} {Comput. Phys. Commun.}\ }\textbf {\bibinfo
  {volume} {253}},\ \bibinfo {pages} {107178} (\bibinfo {year}
  {2020})}\BibitemShut {NoStop}%
\bibitem [{\citenamefont {Arrigoni}\ \emph {et~al.}(2013)\citenamefont
  {Arrigoni}, \citenamefont {Knap},\ and\ \citenamefont {Von
  Der~Linden}}]{arrigoni13}%
  \BibitemOpen
  \bibfield  {author} {\bibinfo {author} {\bibfnamefont {E.}~\bibnamefont
  {Arrigoni}}, \bibinfo {author} {\bibfnamefont {M.}~\bibnamefont {Knap}}, \
  and\ \bibinfo {author} {\bibfnamefont {W.}~\bibnamefont {Von Der~Linden}},\
  }\href@noop {} {\bibfield  {journal} {\bibinfo  {journal} {Phys.~Rev.~Lett.}\
  }\textbf {\bibinfo {volume} {110}},\ \bibinfo {pages} {086403} (\bibinfo
  {year} {2013})}\BibitemShut {NoStop}%
\end{thebibliography}%

\pagebreak

\pagebreak

\pagebreak

\pagebreak

\pagebreak

\pagebreak

\pagebreak

\pagebreak

\pagebreak

\pagebreak

\pagebreak

\pagebreak

\pagebreak

\pagebreak

\end{document}